\documentclass[12pt]{article}
\usepackage{jheppub}
\usepackage{ytableau}
\usepackage{comment}
\usepackage[vcentermath]{youngtab}
\usepackage{tikz}

\newcommand\be{\begin{equation}}
\newcommand\ee{\end{equation}}

\newcommand\Tr{\mathrm{Tr}}

\preprint{
}

\title{Ortho-symplectic interface line operators in $\mathcal{N}=4$ SYM theories}
\abstract{
We study the line operators at the half-BPS interfaces in 4d $\mathcal{N}=4$ super Yang-Mills (SYM) theories of orthogonal and symplectic gauge groups. 
For the NS5-type interfaces, line operators are labeled by pairs of electric or magnetic charges under the gauge groups on either side of the interface, 
whereas for the D5-type interfaces they are labeled by electric or magnetic charges under the single gauge group preserved by the interface. 
We propose dualities between the Wilson lines and vortex lines at the interfaces under the action of S-duality. 
As highly non-trivial evidence of the dualities on the junctions of the line operators and the interfaces, 
we find matching of the line defect half-indices for the dual pairs. 
}
\author[a]{Yasuyuki Hatsuda}
\author[b]{and Tadashi Okazaki}

\affiliation[a]{Department of Physics, Rikkyo University, Toshima, Tokyo 171-8501, Japan}
\affiliation[b]{School of Physics and Shing-Tung Yau Center, Southeast University,\\
Yifu Architecture Building, No.2 Sipailou, Xuanwu district, Nanjing, Jiangsu, 210096, China}

\emailAdd{yhatsuda@rikkyo.ac.jp}
\emailAdd{tokazaki@seu.edu.cn}
\begin{document}
%%%%%%%%%%%%%%%%%%%%%%%%%%%%%%%%%%%%%%%%%%%%
%%%%%%%%%%%%%%%%%%%%%%%%%%%%%%%%%%%%%%%%%%%%
\maketitle

%%%%%%%%%%%%%%%%%%%%%%%%%%%%%%%%%%%%%%%%%%%%%%%%%
%%%%%%%%%%%%%%%%%%%%%%%%%%%%%%%%%%%%%%%%%%%%%%%%%
\section{Introduction and summary}
\label{sec_intro}
%%%%%%%%%%%%%%%%%%%%%%%%%%%%%%%%%%%%%%%%%%%%%%%%%
%%%%%%%%%%%%%%%%%%%%%%%%%%%%%%%%%%%%%%%%%%%%%%%%%
%line
Line defect operators provide a sharp probe of the global structure and non-perturbative regime of quantum field theories. 
Wilson lines \cite{Wilson:1974sk} and 't Hooft lines \cite{tHooft:1977nqb,Kapustin:2005py} in 4d gauge theories are charged under electric and magnetic 1-form symmetries respectively. 
Consequently, line operators furnish a powerful diagnostic of global forms of the gauge groups and dualities beyond the data of local operators \cite{Aharony:2013hda}.
This structure becomes particularly intricate for orthogonal and symplectic gauge groups. 
In contrast to the unitary case, 4d gauge theories based on $\mathfrak{so}(N)$ or $\mathfrak{usp}(2N)$ gauge algebras 
admit several inequivalent global realizations of gauge groups, such as $Spin(N)$, $SO(N)$, $O(N)^{\pm}$, $USp(2N)$, $USp(2N)/\mathbb{Z}_2$, $USp(2N)'$, 
distinguished by discrete electric and magnetic 1-form symmetries and discrete theta angles. 

%what we do
In this paper, we study the line defect operators at the half-BPS interfaces in 4d $\mathcal{N}=4$ super Yang-Mills (SYM) theories with orthogonal and symplectic gauge groups 
by analyzing the supersymmetric line defect half-indices (see \cite{Hatsuda:2025zvi} for unitary cases). 
The half-BPS interfaces provide a more refined setting in which these global structures are sharply exposed 
(see \cite{Gaiotto:2008sa,Gaiotto:2008sd,Gaiotto:2008ak} for the details of the half-BPS boundary conditions and interfaces in $\mathcal{N}=4$ SYM theories). 
The interface separates two 4d $\mathcal{N}=4$ SYM theories of gauge groups $G_1$ and $G_2$ and supports 3d degrees of freedom. 
These interfaces admit brane constructions in Type IIB string theory. 
In particular, two types of the half-BPS interfaces arise, 
corresponding to the NS5- and D5-branes in Type IIB string theory. 
For the NS5-type interface the orthogonal and symplectic gauge groups of the two SYM theories are fully preserved, 
whereas for the D5-type interface only the gauge group with the smaller rank is preserved. 
Electric line operators supported on such an interface are identified with the Wilson line operators charged under the preserved gauge group, 
whereas magnetic line operators admit an interpretation as vortex line defects 
that induce singular configurations for the 3d matter fields localized on the interface. 

%duality
In what follows, we propose and test dualities between the vortex lines and the Wilson lines, 
both of which are inserted along the interfaces in 4d $\mathcal{N}=4$ SYM theories with orthogonal and symplectic gauge groups. 
The dualities we propose fall into two classes.

\begin{enumerate}

\item \textbf{NS5-type interface Wilson lines and D5-type interface vortex lines}
%NS5-Wilson D5-vortex

The Wilson lines supported on an NS5-type interface are generally labeled by pairs of representations $\mathcal{R}_1$ and $\mathcal{R}_2$ 
of the orthogonal and symplectic gauge groups on the two sides of the interface. 
In contrast, vortex lines for the dual D5-type interface are naturally characterized by the Abelianized vortex numbers 
associated with the preserved gauge group of the smaller rank. 
In particular, for the D5-type interface involving an orthogonal gauge group 
one can introduce the minuscule vortex line labeled by the fundamental coweight. 
We propose that this operator is dual to the Wilson line transforming as $(\tiny \yng(1);\yng(1))$, 
a pair of the fundamental representations under both orthogonal and symplectic gauge groups. 
The dualities are summarized as follows: 
\begin{align}
\label{Table_dual}
\begin{array}{c|c||c|c}
\textrm{NS5-type interface}&\textrm{representation $\mathcal{R}$}&\textrm{D5-type interface}&\textrm{magnetic charge $k$} \\ \hline 
\mathfrak{so}(2N)|\mathfrak{usp}(2M)&(\tiny\yng(1);\tiny\yng(1))&\mathfrak{so}(2N)|\mathfrak{so}(2M+1)&(1,0^{\min (N,M)-1}) \\
\mathfrak{o}(2N)^{\pm}|\mathfrak{usp}(2M)&(\tiny\yng(1);\tiny\yng(1))&\mathfrak{o}(2N)^{\pm}|\mathfrak{so}(2M+1)&(1,0^{\min (N,M)-1}) \\
\end{array}. 
\end{align}
A particularly simple situation arises for the D5-type $\mathfrak{so}(2)|\mathfrak{so}(2M+1)$ interface. 
Since the preserved gauge algebra $\mathfrak{so}(2)$ is Abelian, the vortex lines do not receive monopole bubbling contributions. 
This allows the S-duality map between the Wilson lines and vortex lines to be generalized to higher electric and magnetic charges respectively
\begin{align}
\label{Table_dual_so2}
\begin{array}{c|c||c|c}
\textrm{NS5-type interface}&\textrm{representation $\mathcal{R}$}&\textrm{D5-type interface}&\textrm{magnetic charge $k$} \\ \hline 
\mathfrak{so}(2)|\mathfrak{usp}(2M)&(n;(n))&\mathfrak{so}(2)|\mathfrak{so}(2M+1)&n \\
\end{array}, 
\end{align}
where the representation $(n;(n))$ stands for the $\mathfrak{so}(2)$ electric charge $n$ and the rank-$n$ $\mathfrak{usp}(2M)$ symmetric representation, the rank being understood as $|n|$ for $n<0$. 
For this Abelian family, the equality of the line defect half-indices is proved analytically for arbitrary $M\ge 1$ and $n\in \mathbb{Z}$ in section \ref{sec_DCD5V_Sduality}. 
Remarkably, the situation differs from that in ordinary 4d $\mathcal{N}=4$ $USp(2M)$ SYM theory, 
where the rank-$n$ symmetric Wilson line is not S-dual to a simple minuscule 't Hooft line of the dual $SO(2M+1)$ SYM theory. 
At the NS5-type $\mathfrak{so}(2)|\mathfrak{usp}(2M)$ interface, however, it is mapped to a magnetic line defect that does not receive monopole bubbling contributions.

%obstruction of spinor Wilson line at NS-type interface
For the D5-type $\mathfrak{usp}(2N)|\mathfrak{usp}(2M)'$ interface preserving the symplectic gauge algebra, 
the minuscule dominant coweight corresponds to the highest weight of the spinor representation of the Langlands dual $\mathfrak{so}(2N+1)$. 
However, there will be an obstruction to the insertion of a single spinor Wilson line at the NS5-type half-BPS interfaces 
between orthogonal and symplectic gauge theories due to the center symmetry. 
Correspondingly, we find that the line defect half-indices involving a single Wilson line in the spinor representation vanish
\begin{align}
\langle W_{\textrm{sp};\mathcal{R}}\rangle_{\mathcal{N}}^{\mathfrak{so}(2N+1)|\mathfrak{usp}(2M)'}(t;q)&=0, \\
\langle W_{\textrm{sp};\mathcal{R}}\rangle_{\mathcal{N}}^{\mathfrak{so}(2N)|\mathfrak{usp}(2M)}(t;q)&=0, 
\end{align}
where $\mathcal{R}$ stands for an irreducible representation of $\mathfrak{usp}(2M)$. 
The vortex lines associated with the integer coweight of symplectic gauge groups are generically non-minuscule. 
Accordingly, the associated vortex line defect half-index is expected to involve complicated monopole bubbling effects. 
Nevertheless, in the special case with $N=0$, which is dual to the NS5-type $\mathfrak{so}(1)|\mathfrak{usp}(2M)'$ interface, 
we obtain an explicit formula for the vortex line defect half-index. 

\item  \textbf{NS5-type interface vortex lines and D5-type interface Wilson lines}
%NS5-vortex D5-Wilson

On the other hand, the vortex lines at the NS5-type interface are labeled by the Abelianized vortex numbers for both gauge groups, 
namely the orthogonal and the symplectic gauge factors. 
For the orthogonal gauge group, there exist magnetic line operators associated with the fundamental weights 
corresponding to minuscule representations, for which monopole bubbling is absent. 
Such operators are expected to be dual, under S-duality, 
to electric line operators realized as Wilson lines in the fundamental representation of the Langlands dual gauge group at the D5-type interface. 
We propose the following dualities: 
\begin{align}
\label{Table_dual2}
\begin{array}{c|c||c|c}
\textrm{NS5-type interface}&\textrm{magnetic charge $k$} &\textrm{D5-type interface}&\textrm{representation $\mathcal{R}$} \\ \hline 
\mathfrak{so}(2N+1)|\mathfrak{usp}(2M)'&(1,0^{N-1};0)&\mathfrak{usp}(2N)|\mathfrak{usp}(2M)'&(\tiny\yng(1)) \\
\mathfrak{so}(2N)|\mathfrak{usp}(2M)&(1,0^{N-1};0)&\mathfrak{so}(2N)|\mathfrak{so}(2M+1)&(\tiny\yng(1)) \\
\mathfrak{o}(2N)^{\pm}|\mathfrak{usp}(2M)&(1,0^{N-1};0)&\mathfrak{o}(2N)^{\pm}|\mathfrak{so}(2M+1)&(\tiny\yng(1)) \\
\end{array}. 
\end{align}
Here we take the ranks to satisfy $N\le M$. 
For the NS5-type $\mathfrak{so}(2)|\mathfrak{usp}(2M)$ interface, 
the preserved gauge algebra $\mathfrak{so}(2)$ is Abelian, so the vortex lines are free from the monopole bubbling effects. 
Consequently, the S-duality map extends naturally to Wilson lines and vortex lines of arbitrary charge $n\in \mathbb{Z}$
\begin{align}
\label{Table_dual2_so2}
\begin{array}{c|c||c|c}
\textrm{NS5-type interface}&\textrm{magnetic charge $k$} &\textrm{D5-type interface}&\textrm{electric charge} \\ \hline 
\mathfrak{so}(2)|\mathfrak{usp}(2M)&(n;0)&\mathfrak{so}(2)|\mathfrak{so}(2M+1)&n \\
\end{array}. 
\end{align}
For this Abelian family, the equality of the line defect half-indices is proved analytically for arbitrary $M\ge 1$ and $n\in\mathbb{Z}$ in section \ref{sec_DCD5W_Sduality}.

\end{enumerate}

%line defect Witten index
We also propose several conjectures motivated by the exact results obtained for the line defect half-indices. 
In \cite{Hatsuda:2025zvi}, we found that 
the normalized Wilson line defect half-index of the D5-type $U(N)|U(M)$ interface is realized, 
in the Higgs limit, as the principal specialization of the Schur function associated with the corresponding representation. 
For ortho-symplectic interfaces, it remains an open question whether an analogous description in terms of principal specializations of certain symmetric functions exists. 
Nevertheless, we find that the $\mathfrak{q}\rightarrow 1$ limit, 
which we refer to as the \textit{line defect Witten index}, reproduces the dimension of the corresponding orthogonal or symplectic group representation. 
We are therefore led to conjecture the following general formula for the Wilson line defect Witten index of the D5-type interface $\mathfrak{g}_1|\mathfrak{g}_2$:
\begin{align}
{\langle \mathcal{W}_{\mathcal{R}_1}\cdots \mathcal{W}_{\mathcal{R}_k} \rangle_{\mathcal{D}}^{\mathfrak{g}_1|\mathfrak{g}_2}}^{(H)}(1)
&=\prod_{j=1}^k \chi_{\mathcal{R}_j}^{\mathfrak{g}_{\min}}(1). 
\end{align}
Here $\mathfrak{g}_{\min}$ denotes the gauge algebra whose rank is the smaller of those of $\mathfrak{g}_1$ and $\mathfrak{g}_2$. 
$\chi_{\mathcal{R}}^{\mathfrak{g}}$ is the character of the representation $\mathcal{R}$ of gauge algebra $\mathfrak{g}$. 

While the Coulomb limit results reduce to the half-BPS line defect indices of ordinary 4d $\mathcal{N}=4$ SYM theory, 
we conjecture that the corresponding line defect Witten index is given by the following general formula: 
\begin{align}
{\langle \mathcal{W}_{\mathcal{R}_1}\cdots \mathcal{W}_{\mathcal{R}_k} \rangle_{\mathcal{D}}^{\mathfrak{g}_1|\mathfrak{g}_2}}^{(C)}(1)
&=\oint 
\prod_{i=1}^{\mathrm{rank}(\mathfrak{g}_{\min})}
\frac{ds_i}{2\pi is_i}
\prod_{j=1}^k 
\chi_{\mathcal{R}_j}^{\mathfrak{g}_{\min}}(s). 
\end{align}

These formulae admit a simple physical interpretation, 
which also allows us to determine the line defect Witten indices of the Wilson lines at the NS5-type interfaces. 
The line defect Witten index is obtained by integrating the character over the residual gauge group $H_0$ 
which remains unbroken at a generic point of the Coulomb or Higgs branch 
\begin{align}
\label{WI_residual_intro}
\langle \mathcal{W}_{\mathcal{R}}\rangle(1)
&=\int_{H_0}d\mu_{H_0}\, \chi_{\mathcal{R}}\big|_{H_0}
=\dim \mathcal{R}^{H_0}, 
\end{align}
that is, it counts the $H_0$-invariant components of $\mathcal{R}$ which can terminate on the interface. 
Analogous formulae for the line defect Witten indices of 3d $\mathcal{N}\ge 4$ supersymmetric gauge theories, 
together with the same characterization in terms of the residual gauge group at a generic point of the branch, 
have been proposed in \cite{Hayashi:2026ngm}. 
For the D5-type interfaces $H_0$ is the maximal torus of the preserved gauge group in the Coulomb limit, 
while it is trivial in the Higgs limit, 
which reproduces the above conjectures.

For the NS5-type interfaces both vector multiplets obey the Neumann boundary condition and survive the Coulomb limit, 
so that $H_0$ is the product of the maximal tori of the two gauge groups and 
\begin{align}
\label{WI_NS5_C_intro}
{\langle \mathcal{W}_{\mathcal{R}_1;\mathcal{R}_2} \rangle_{\mathcal{N}}^{\mathfrak{g}_1|\mathfrak{g}_2}}^{(C)}(1)
&=\oint \prod_{i=1}^{\mathrm{rank}(\mathfrak{g}_1)}\frac{ds_i^{(1)}}{2\pi is_i^{(1)}}
\prod_{j=1}^{\mathrm{rank}(\mathfrak{g}_2)}\frac{ds_j^{(2)}}{2\pi is_j^{(2)}}
\chi_{\mathcal{R}_1}^{\mathfrak{g}_1}(s^{(1)})\chi_{\mathcal{R}_2}^{\mathfrak{g}_2}(s^{(2)}), 
\end{align}
which is the product of the zero weight multiplicities. 
This follows from Kostant's theorem \cite{MR158024}, and we have checked it in low-rank examples. 

In the Higgs limit $H_0=U(1)^{r}\times H_0'$ with $r=\min(N,M)$, 
where $U(1)^{r}$ is embedded diagonally into both gauge groups 
and $H_0'$ is the orthogonal or symplectic gauge group of rank $|N-M|$ on the side with the larger rank (see Table \ref{Table_residual}). 
A subtlety arises for the $\mathfrak{so}(2N+1)|\mathfrak{usp}(2M)'$ interface with $N<M$, 
where a half-hypermultiplet $S$ in the fundamental representation of the residual $USp(2(M-N))$ gauge group stays massless. 
It generates no gauge invariant, but it contributes to the covariants and therefore has to be kept in the residual integral: 
\begin{align}
\label{WI_NS5_H_intro}
&{\langle \mathcal{W}_{\mathcal{R}_1;\mathcal{R}_2} \rangle_{\mathcal{N}}^{\mathfrak{g}_1|\mathfrak{g}_2}}^{(H)}(1)
\nonumber\\
&=\lim_{\mathfrak{q}\rightarrow 1}
\oint \prod_{i=1}^{r}\frac{du_i}{2\pi iu_i}\int_{H_0'}d\mu_{H_0'}(h)\, 
\chi_{\mathcal{R}_1}^{\mathfrak{g}_1}(u,h)\chi_{\mathcal{R}_2}^{\mathfrak{g}_2}(u,h)\,
\mathrm{PE}\left[\mathfrak{q}\chi_{S}(h)\right], 
\end{align}
where $u_i$ with $i=1,\cdots,r$ are the fugacities of the diagonal $U(1)^{r}$, 
which enter the holonomies of the two gauge groups as $s^{(1)}_i=s^{(2)}_i=u_i$, 
and $h$ stands for the fugacities of $H_0'$, which fill the remaining entries of the holonomy of the gauge group containing it, 
the character of the other gauge group being independent of $h$. 
Here $\int d\mu_{H_0'}$ is the normalized Haar integral and $\chi_{S}$ is the character of $S$. 
For all the other cases $S$ is empty, the plethystic exponential is trivial 
and (\ref{WI_NS5_H_intro}) reduces to (\ref{WI_residual_intro}). 
This formula follows from the observation that in the Higgs limit 
the normalized line defect half-index is the ratio of the Hilbert series of a module of covariants to that of the ring of invariants, 
so that its $\mathfrak{q}\rightarrow 1$ limit is the rank of the module, 
which is computed on a generic fiber of the Higgs branch. 
Unlike in the Coulomb limit, the residual gauge group and the massless charged half-hypermultiplet have to be determined case by case, 
and we have confirmed the resulting formulae against the exact Higgs limit matrix integrals 
for a number of pairs of irreducible representations in low-rank cases. 
The details of both limits are given in Appendix \ref{app_residual}. 

%%%%%%%%%%%%%%%%%%%%%%%%%%%%%%%%%%%%%%%%%%%%%%%%%
\subsection{Structure}
%%%%%%%%%%%%%%%%%%%%%%%%%%%%%%%%%%%%%%%%%%%%%%%%%
The paper is organized as follows. 
In section \ref{sec_intline} we briefly review the half-BPS interfaces and line operators in 4d $\mathcal{N}=4$ SYM theories 
and introduce the conventions and the definition of the line defect half-indices. 
In section \ref{sec_BC1} and section \ref{sec_BC2} 
we examine the line operators in the NS5-type $\mathfrak{so}(2N+1)|\mathfrak{usp}(2M)'$ interface 
and the S-dual D5-type $\mathfrak{usp}(2N)|\mathfrak{usp}(2M)'$ interface respectively. 
The action of S-duality is given in section \ref{sec_BC2}.
Subsequently, we analyze the line operators in the NS5-type $\mathfrak{so}(2N)|\mathfrak{usp}(2M)$ interface 
and the S-dual D5-type $\mathfrak{so}(2N)|\mathfrak{so}(2M+1)$ interface in section \ref{sec_DC1} and section \ref{sec_DC2}. 
We propose S-duality of the corresponding interface line operators in section \ref{sec_DC2}. 
In Appendix \ref{app_koornwinder} we prove the closed-form expressions for the Wilson line defect half-indices 
of the NS5-type $\mathfrak{so}(1)|\mathfrak{usp}(2M)'$ interface 
by identifying the corresponding matrix integrals with Koornwinder integrals of type $BC_M$. 
In Appendix \ref{app_expansion} we exhibit the $q$-series expansions of the line defect half-indices. 
In Appendix \ref{app_residual} we derive the line defect Witten indices from the residual gauge groups on the Coulomb and Higgs branches. 

%%%%%%%%%%%%%%%%%%%%%%%%%%%%%%%%%%%%%%%%%%%%%%%%%
\subsection{Future works}
%%%%%%%%%%%%%%%%%%%%%%%%%%%%%%%%%%%%%%%%%%%%%%%%%

\begin{itemize}

\item In this work we focus on the line operators of minuscule charge at the interfaces. 
For the non-minuscule magnetic charges we encounter the monopole bubbling effect \cite{Kapustin:2006pk}. 
For the orthogonal and symplectic gauge theories it would be significant to figure out 
how the bubbling patterns depend on the global forms of the gauge group and discrete theta angles.  
The line defect half-indices of the line operators with non-minuscule charges will be addressed by taking into account 
our results and the monopole bubbling indices \cite{Ito:2011ea,Hayashi:2020ofu,Brennan:2018yuj,Brennan:2018rcn,Hayashi:2019rpw} 
which contain the monopole bubbling contributions. 

\item Gravity duals of the line operators at the interfaces will provide us with more refined examples of the AdS/CFT correspondence 
(see e.g. \cite{Nagasaki:2011ue,Estes:2012nx,Nagasaki:2013hwa,Karch:2022rvr,Kristjansen:2024dnm} for the relevant studies). 
In the large gauge rank limit, the interface background dominates and the line operator can be viewed as a probe string/brane. 
In particular, it would be intriguing to figure out the gravity interpretation of the monopole bubbling for the non-minuscule line operators. 
It would also be significant to extract the finite-rank corrections to the line defect half-indices 
in the form of the giant graviton expansions \cite{Imamura:2021ytr,Gaiotto:2021xce,Murthy:2022ien,Hatsuda:2024uwt,Beccaria:2024oif,Imamura:2024zvw}, 
both for the ortho-symplectic interfaces studied here and for the unitary interfaces \cite{Hatsuda:2025zvi}. 
For the former the branes can wrap the torsion cycles of $\mathbb{RP}^5$ associated with the Pfaffian-like operators and the spinor Wilson lines \cite{Witten:1998xy}, 
so that the expansion is expected to be structurally different from the unitary case, 
as observed for the indices of the orientifold and S-fold theories \cite{Arai:2019xmp,Fujiwara:2023bdc,Imamura:2026dhk}. 

\item The half-BPS interface with a single NS5- or D5-brane can be generalized by introducing multiple 5-branes \cite{Gaiotto:2008ak}. 
In such a case, the interface may involve additional 3d degrees of freedom and extra global symmetries 
and the line operators can end on different segments of D3-branes between 5-branes. 
We hope to report dualities of the configurations by keeping track of global data through several interfaces. 

\end{itemize}
%%%%%%%%%%%%%%%%%%%%%%%%%%%%%%%%%%%%%%%%%%%%%%%%%
%%%%%%%%%%%%%%%%%%%%%%%%%%%%%%%%%%%%%%%%%%%%%%%%%
\section{Interface line operators}
\label{sec_intline}
%%%%%%%%%%%%%%%%%%%%%%%%%%%%%%%%%%%%%%%%%%%%%%%%%
%%%%%%%%%%%%%%%%%%%%%%%%%%%%%%%%%%%%%%%%%%%%%%%%%

%%%%%%%%%%%%%%%%%%%%%%%%%%%%%%%%%%%%%%%%%%%%%%%%%
\subsection{Field theory setup}
\label{sec_fieldsetup}
%%%%%%%%%%%%%%%%%%%%%%%%%%%%%%%%%%%%%%%%%%%%%%%%%

%Wilson line
The Wilson lines are electric order-type line operators \cite{Wilson:1974sk}. 
They are defined as traces of the holonomy matrices of gauge connections along the line operators. 
The half-BPS Wilson line in 4d $\mathcal{N}=4$ SYM theory of gauge group $G$ 
can be constructed by coupling to a scalar field \cite{Maldacena:1998im}
\begin{align}
\label{W_op}
W_{\mathcal{R}}
&={\Tr}_{\mathcal{R}} 
\mathcal{P}
\exp\left[
i\int \left(
A_{\mu}\dot{x}^{\mu}+\Phi|\dot{x}|
\right)ds
\right], 
\end{align}
where the Wilson line is supported on the curve $x^{\mu}$ $=$ $x^{\mu}(s)$ and $|\dot{x}|ds$ is the proper length element along the curve.  
The holonomy of the generalized connection is evaluated in the representation $\mathcal{R}$ of gauge group $G$. 
$\Phi$ is one of the six real adjoint scalar fields. 
It breaks the $SO(6)_R$ R-symmetry rotating six scalar fields to $SO(5)$. 

%vortex line
The 't Hooft lines are magnetic disorder operators \cite{tHooft:1977nqb,Kapustin:2005py} in 4d gauge theories. 
They are defined by specifying singular boundary conditions for the gauge and matter fields around the line operators. 
The half-BPS 't Hooft line in 4d $\mathcal{N}=4$ SYM theory of gauge group $G$ is defined by prescribing the Dirac monopole singularity along the line 
\begin{align}
\label{BPSeq}
F&\sim \frac{m}{2} \sin\theta d\theta \wedge d\phi, 
\qquad 
\Phi\sim \frac{m}{2r}, 
\end{align}
where $(\theta,\phi)$ are the spherical coordinates on a 2-sphere that surrounds the line in the transverse $\mathbb{R}^3$ 
and $r$ is the distance from the line in the three transverse directions. 
$F$ is the field strength of the gauge connection carrying monopole flux
\begin{align}
\frac{1}{2\pi} \int_{S^2}F&=m
\end{align}
and $\Phi$ is one of the six real adjoint scalar fields. 
The $SO(6)_R$ R-symmetry rotating six scalar fields is broken to $SO(5)$.
The configuration is labeled by $m$ $\in$ $\mathrm{Hom}(U(1),G)$, that is, a magnetic charge. 
Two magnetic charges which are related by the Weyl transformation give rise to the same magnetic line operator
\begin{align}
m&\sim w(m), \qquad w\in \mathrm{Weyl}(G). 
\end{align}
Hence the magnetic lines are classified by the Weyl orbits of coweights. 
%minuscule vortex
A magnetic charge is called minuscule if \cite{MR2109105}
\begin{align}
\langle \alpha, m\rangle&\in \{0,\pm 1\}
\end{align}
for all roots $\alpha$ in the root system of $G$. 
In this case, the Weyl orbit of $m$ contains no smaller non-zero charges 
so that there is no way to decompose it as
\begin{align}
\label{m_split}
m&=m'+\sum_{\alpha_i\in \Delta}\alpha_i^{\vee}, 
\end{align}
where $m'<m$ and $\Delta$ is the set of simple roots. 
Physically, the decomposition (\ref{m_split}) implies that 
the magnetic charge $m$ is screened by emitting smooth monopoles associated with the simple root directions 
where the associated simple coroots $\alpha_i^{\vee}$ $=$ $\frac{2\alpha_i}{(\alpha_i,\alpha_i)}$ can be viewed as magnetic charges of the smooth BPS monopoles. 
This phenomenon is known as the monopole bubbling effect \cite{Kapustin:2006pk}. 
The minuscule magnetic charges $m$ of gauge groups $G$ are given as follows: 
\begin{align}
\label{Table_minuscule}
\begin{array}{c|c}
G&m \\ \hline 
U(N)&(1^{k},0^{N-k}) \\
USp(2N)/\mathbb{Z}_2&(\frac12^N) \\
SO(2N+1)&(1,0^{N-1}) \\
SO(2N)/\mathbb{Z}_2&(\frac12^N) \\ 
SO(2N)&(1,0^{N-1}) \\ 
O(2N)&(1,0^{N-1}) \\ 
\end{array}.
\end{align}

%interface
Let us further introduce the half-BPS interface between 4d $\mathcal{N}=4$ SYM theories \cite{Gaiotto:2008ak}. 
The configuration preserves 3d $\mathcal{N}=4$ supersymmetry and $SU(2)_C$ $\times$ $SU(2)_H$ R-symmetry. 
At the interface that preserves eight supercharges, 
the half-hypermultiplet is required to transform in a quaternionic representation. 

For the half-hypermultiplet transforming as the bifundamental representation under a pair of gauge groups, 
a quaternionic representation is possible if exactly one factor is pseudoreal, namely for the real-pseudoreal combination, i.e. orthogonal-symplectic. 

On the other hand, when one has real-real or pseudoreal-pseudoreal combinations, 
i.e. orthogonal-orthogonal or symplectic-symplectic, 
the interface can only preserve a single gauge factor as its diagonal subgroup in such a way that 
the matter multiplet transforms as the fundamental representation. 
Furthermore, when the ranks of the two gauge groups are different, some of the 4d bulk fields obey the Nahm pole boundary condition. 
For the symplectic-symplectic interface preserving a single gauge factor of symplectic gauge group, 
a single half-hypermultiplet is allowed only if we turn on the discrete theta angle $\theta=\pi$, 
which we denote by $\mathfrak{usp}|\mathfrak{usp}'$. 

There are four basic types of the half-BPS interfaces between orthogonal and symplectic gauge theories
\begin{enumerate}
\item $\mathfrak{so}(2N+1)|\mathfrak{usp}(2M)'$ 
\item $\mathfrak{so}(2N)|\mathfrak{usp}(2M)$ 
\item $\mathfrak{usp}(2N)|\mathfrak{usp}(2M)'$ 
\item $\mathfrak{so}(2N)|\mathfrak{so}(2M+1)$
\end{enumerate}
As reviewed in section \ref{sec_brane} (see \cite{Gaiotto:2008sd,Gaiotto:2008ak} for the details), 
these interfaces can be engineered in Type IIB brane configurations. 
Under the action of S-duality, 
the first $\mathfrak{so}(2N+1)|\mathfrak{usp}(2M)'$ interface maps to 
the third $\mathfrak{usp}(2N)|\mathfrak{usp}(2M)'$ interface 
while the second $\mathfrak{so}(2N)|\mathfrak{usp}(2M)$ interface maps to 
the fourth $\mathfrak{so}(2N)|\mathfrak{so}(2M+1)$ interface. 

%interface line
When we introduce the Wilson lines at the above interfaces, they are labeled by representations of the preserved gauge group. 
As in the half-BPS Wilson lines (\ref{W_op}) in the 4d bulk theory, the holonomy matrices are given by modified connections 
involving ordinary gauge connections and the adjoint scalar field $\Phi$. 
They preserve four supercharges and break the $SU(2)_{C}$ R-symmetry down to $U(1)_{C}$ while keeping the $SU(2)_H$ R-symmetry. 
We can also consider the magnetic lines at the interface. 
In this case, the magnetic charge is obtained from a flux of the interface gauge group through a small disk $D$ 
surrounding the line and the 3d matter fields are also required to obey singular boundary conditions along the line. 
They behave as the vortex lines labeled by the Abelianized vortex numbers
\begin{align}
k=(k_1,\cdots,k_{\mathrm{rank}(H)})\in \mathbb{Z}^{\mathrm{rank}(H)}, 
\end{align}
where $H$ is the gauge group preserved by the interface. 
Unlike the Wilson lines, the vortex lines break the $SU(2)_{H}$ R-symmetry down to $U(1)_{H}$, whereas they preserve the $SU(2)_C$ R-symmetry. 

%%%%%%%%%%%%%%%%%%%%%%%%%%%%%%%%%%%%%%%%%%%%%%%%%
\subsection{Brane setup}
\label{sec_brane}
%%%%%%%%%%%%%%%%%%%%%%%%%%%%%%%%%%%%%%%%%%%%%%%%%

%orthogonal/symplectic gauge
Let us consider the setup in Type IIB string theory. 
To obtain 4d $\mathcal{N}=4$ SYM theories with orthogonal or symplectic gauge groups, we consider an O3-plane parallel to D3-branes. 
There are four types of O3-planes, which are distinguished by discrete RR and NSNS 2-form fluxes (torsions $\theta_{RR}$ and $\theta_{NS}$)
\begin{align}
\label{O3_SYM}
\begin{array}{c|c|c|c|c}
&\mathfrak{so}(2N+1)&\mathfrak{usp}(2N)&\mathfrak{so}(2N)&\mathfrak{usp}(2N)'\\ \hline 
\theta_{RR}&1/2&0&0&1/2 \\
\theta_{NS}&0&1/2&0&1/2 \\
\textrm{D3-brane charge}&1/4&1/4&-1/4&1/4 \\
\textrm{orientifold}&\widetilde{\textrm{O3}}^{-}&\textrm{O3}^+&\textrm{O3}^{-}&\widetilde{\textrm{O3}}^+ \\
\textrm{$S$ operation}&\textrm{O3}^+&\widetilde{\textrm{O3}}^{-}&\textrm{O3}^{-}&\widetilde{\textrm{O3}}^+ \\
\end{array}.
\end{align}
These O3-planes transform into each other under S-duality of Type IIB string theory. 
We introduce D3-branes as well as a certain O3-plane that extend along the directions $0126$. 

%interface
To construct the half-BPS interfaces, we introduce a half NS5-brane or half D5-brane sitting at $x^6=0$ 
in the presence of $N$ D3-branes in $x^6<0$ and $M$ D3-branes in $x^6>0$. 
Since the half NS5-brane carries NSNS flux that jumps across the interface, 
the O3-plane changes the NSNS discrete torsion $\theta_{NS}$ by one unit while leaving $\theta_{RR}$ invariant. 
Hence the half NS5-brane can engineer the following two types of the $\mathfrak{so}|\mathfrak{usp}$ interfaces
\begin{enumerate}
\item $\mathfrak{so}(2N+1)|\mathfrak{usp}(2M)'$ ($\widetilde{\textrm{O3}}^-$ and $\widetilde{\textrm{O3}}^+$)
\item $\mathfrak{so}(2N)|\mathfrak{usp}(2M)$ (O3$^-$ and O3$^+$)
\end{enumerate}
The NS5-type interfaces can preserve full gauge algebras and they support the bifundamental half-hypermultiplet 
arising from open strings across the half NS5-brane. 
On the other hand, a half D5-brane placed at $x^6=0$ gives rise to an interface, 
across which the gauge algebras are of the type $\mathfrak{usp}|\mathfrak{usp}$ or $\mathfrak{so}|\mathfrak{so}$
\begin{enumerate}
\item $\mathfrak{usp}(2N)|\mathfrak{usp}(2M)'$ (O3$^+$ and $\widetilde{\textrm{O3}}^+$)
\item $\mathfrak{so}(2N)|\mathfrak{so}(2M+1)$ (O3$^-$ and $\widetilde{\textrm{O3}}^-$)
\end{enumerate}
In this case the gauge algebra is effectively broken down to its diagonal subalgebra 
and the open strings between D3-branes and a half D5-brane lead to a fundamental hypermultiplet. 

%line operator
%Wilson
Furthermore, we can realize the line operators at the interface by introducing additional strings stretched along the $x^0$ time direction. 
When we have a single F1-string ending on D3-branes, 
it can generate the Wilson line in the fundamental representation of gauge group in $\mathcal{N}=4$ SYM theory \cite{Maldacena:1998im,Rey:1998ik}. 
When $k$ coincident F1-strings along the directions $05$ terminate on the D3-branes and another probe D3-brane, 
one finds the Wilson lines in the $k$-fold symmetric tensor product of the fundamental representations 
\cite{Drukker:2005kx,Gomis:2006sb,Gomis:2006im,Rodriguez-Gomez:2006fmx,Yamaguchi:2007ps}. 
Moreover, one can introduce new brane species to realize the Wilson lines. 
While keeping the amount of supersymmetry, we can introduce a D5-brane along the $034789$ directions 
that has world-volume electric flux that dissolves $k$ units of F1-string charge. 
We denote it by $\widetilde{\textrm{D5}}$. 
It realizes the Wilson lines in the rank-$k$ antisymmetric tensor product of the fundamental representation \cite{Yamaguchi:2006tq,Gomis:2006sb,Rodriguez-Gomez:2006fmx,Hartnoll:2006hr}. 
For the theories with orthogonal gauge group, one can also consider the Wilson line operator in the spinor representation. 
To realize the spinor Wilson line, we need to take into account the possible compact cycle $\mathbb{RP}^4$ in the background of $\widetilde{\textrm{O3}}^-$- and O3$^-$-planes 
which can be wrapped by a D5-brane \cite{Witten:1998xy} 
(also see \cite{Fiol:2014fla,Giombi:2020kvo,Okuyama:2022lbn,Zhang:2023yus,Hatsuda:2025jze} for more recent analyses). 
The brane configuration is summarized as follows: 
\begin{align}
\begin{array}{c|cccccccccc}
&0&1&2&3&4&5&6&7&8&9 \\ \hline 
\textrm{D3}&\circ&\circ&\circ&&&&\circ&&& \\
\textrm{O3}&\circ&\circ&\circ&&&&\circ&&& \\
\textrm{NS5}&\circ&\circ&\circ&\circ&\circ&\circ&&&& \\
\textrm{D5}&\circ&\circ&\circ&&&&&\circ&\circ&\circ \\
\textrm{F1}&\circ&&&&&\circ&&&& \\
\widetilde{\textrm{D5}}&\circ&&&\circ&\circ&&&\circ&\circ&\circ \\
\end{array}.
\end{align}

%vortex
The vortex line operators can be realized by D1-strings along the directions $09$ which are the S-dual of the F1-strings 
and the NS5-brane ($\widetilde{\textrm{NS5}}$) along the directions $034578$ which is the S-dual of the $\widetilde{\textrm{D5}}$-brane. 
It is expected that 
the vortex lines with the minuscule magnetic charges that are free from the monopole bubbling effect can terminate cleanly on the D3-branes. 
In the following we focus on the minuscule vortex lines. 
The brane configuration is summarized as follows: 
\begin{align}
\begin{array}{c|cccccccccc}
&0&1&2&3&4&5&6&7&8&9 \\ \hline 
\textrm{D3}&\circ&\circ&\circ&&&&\circ&&& \\
\textrm{O3}&\circ&\circ&\circ&&&&\circ&&& \\
\textrm{NS5}&\circ&\circ&\circ&\circ&\circ&\circ&&&& \\
\textrm{D5}&\circ&\circ&\circ&&&&&\circ&\circ&\circ \\
\textrm{D1}&\circ&&&&&&&&&\circ \\
\widetilde{\textrm{NS5}}&\circ&&&\circ&\circ&\circ&&\circ&\circ& \\
\end{array}.
\end{align}

%%%%%%%%%%%%%%%%%%%%%%%%%%%%%%%%%%%%%%%%%%%%%%%%%
%%%%%%%%%%%%%%%%%%%%%%%%%%%%%%%%%%%%%%%%%%%%%%%%%
\subsection{Line defect half-indices}
\label{sec_hind}
%%%%%%%%%%%%%%%%%%%%%%%%%%%%%%%%%%%%%%%%%%%%%%%%%
%%%%%%%%%%%%%%%%%%%%%%%%%%%%%%%%%%%%%%%%%%%%%%%%%
%DEF
To count the BPS local operators at the junction of the line operators $L_j$ with $j=1,\cdots,k$ and the interface $\mathcal{B}$ 
between theory $\mathcal{T}_1$ and theory $\mathcal{T}_2$, we analyze the line defect half-indices. 
They can be defined as a trace over the cohomology $\mathcal{H}$ of the preserved supercharges 
that belong to the subalgebra of the 3d $\mathcal{N}=4$ superalgebra. 
They are weighted by fugacities for global symmetries\footnote{See \cite{Gaiotto:2019jvo,Okazaki:2019ony} for the convention. Also see \cite{Dimofte:2011py,Gang:2012yr,Gang:2012ff} for other conventions. }
\begin{align}
\label{DEF_hind}
\langle L_1\cdots L_k \rangle_{\mathcal{B}}^{\mathcal{T}_1|\mathcal{T}_2}(t,z;q)
&=\Tr_{\mathcal{H}} (-1)^F q^{J+\frac{H+C}{4}} t^{H-C}z^f, 
\end{align}
where $F$ is the fermion number operator, $J$ the spin, $C$ and $H$ the Cartan generators of the $SU(2)_C$ and $SU(2)_H$ R-symmetries 
and $f$ the Cartan generators of the other global symmetries. 
The line defect half-indices can be computed as certain matrix integrals. 
We define
\begin{align}
\label{qpoch_def}
(a;q)_{0}&:=1,\qquad
(a;q)_{n}:=\prod_{k=0}^{n-1}(1-aq^{k}),\qquad 
(q)_{n}:=\prod_{k=1}^{n}(1-q^{k}),
\nonumber \\
(a;q)_{\infty}&:=\prod_{k=0}^{\infty}(1-aq^{k}),\qquad 
(q)_{\infty}:=\prod_{k=1}^{\infty} (1-q^k), 
\end{align}
with $a, q \in \mathbb{C}$ and $|q|<1$. 
For simplicity we also make use of the following shorthand notations: 
\begin{align}
(x^\pm; q)_\infty&=(x;q)_\infty (x^{-1};q)_\infty, \\
(x^\pm y^\pm;q)_\infty&=(xy;q)_\infty (xy^{-1};q)_\infty (x^{-1}y;q)_\infty (x^{-1}y^{-1};q)_\infty.
\end{align}

%properties
The line defect half-indices are crucial for testing dualities of line operators supported at the interfaces. 
In the following sections, we obtain strong evidence for S-duality among ortho-symplectic interface line operators 
by comparing the indices of expected dual configurations, e.g. Wilson lines at an NS5-type interface and vortex lines at a D5-type interface. 

%Coulomb and Higgs limits
As the half-index (\ref{DEF_hind}) can be viewed as a generalization of the supersymmetric index of 3d $\mathcal{N}=4$ supersymmetric field theories, 
we can take the two special fugacity limits which isolate different kinds of protected operators, i.e. Coulomb and Higgs branch operators. 
We shall refer to them as the Coulomb and Higgs limits \cite{Razamat:2014pta}
\begin{align}
\label{C_lim}
{\langle L_1\cdots L_k\rangle_{\mathcal{B}}^{\mathcal{T}_1|\mathcal{T}_2}}^{(C)}(z;\mathfrak{q})
&=\lim_{
\begin{smallmatrix}
\textrm{$\mathfrak{q}:=q^{1/4}t^{-1}$: fixed}\\
q\rightarrow 0\\
\end{smallmatrix}
}
\langle L_1\cdots L_k\rangle_{\mathcal{B}}^{\mathcal{T}_1|\mathcal{T}_2}(t,z;q), \\
\label{H_lim}
{\langle L_1\cdots L_k\rangle_{\mathcal{B}}^{\mathcal{T}_1|\mathcal{T}_2}}^{(H)}(z;\mathfrak{q})
&=\lim_{
\begin{smallmatrix}
\textrm{$\mathfrak{q}:=q^{1/4}t$: fixed}\\
q\rightarrow 0\\
\end{smallmatrix}
}
\langle L_1\cdots L_k\rangle_{\mathcal{B}}^{\mathcal{T}_1|\mathcal{T}_2}(t,z;q). 
\end{align}
%normalized
Also we define the normalized line defect half-indices by
\begin{align}
\langle \mathcal{L}_1\cdots \mathcal{L}_k\rangle_{\mathcal{B}}^{\mathcal{T}_1|\mathcal{T}_2}(t,z;q)
&=\frac{\langle L_1\cdots L_k\rangle_{\mathcal{B}}^{\mathcal{T}_1|\mathcal{T}_2}(t,z;q)}
{\langle 1 \rangle_{\mathcal{B}}^{\mathcal{T}_1|\mathcal{T}_2}(t,z;q)}, 
\end{align}
where we use calligraphic letters for the inserted line operators in the normalized objects. 
Here $\langle 1 \rangle_{\mathcal{B}}^{\mathcal{T}_1|\mathcal{T}_2}(t,z;q)$ in the denominator is the half-index of the interface $\mathcal{B}$ 
between $\mathcal{T}_1$ and $\mathcal{T}_2$ without any insertion of line operators, 
which we also denote by $\mathbb{II}_{\mathcal{B}}^{\mathcal{T}_1|\mathcal{T}_2}(t,z;q)$ \cite{Gaiotto:2019jvo,Hatsuda:2024uwt,Hatsuda:2024lcc}. 
The normalized line defect indices focus on contributions from the line defect itself. 

From a physical viewpoint, the normalized line defect Coulomb/Higgs indices are expected to become integers in the $\mathfrak{q}\rightarrow 1$ limit. 
We call the resulting quantities the \textit{line defect Witten indices}. 
The evaluation of the line defect Witten index seems considerably more subtle from the matrix-integral perspective. 
In what follows, we present a collection of exact results and use them to formulate several general conjectures. 
As we show in Appendix \ref{app_residual}, 
the line defect Witten indices are nevertheless controlled by the residual gauge group $H_0$ at a generic point of the Coulomb or Higgs branch 
through (\ref{WI_residual_intro}). 
In the Coulomb limit the Hall-Littlewood measure of the vector multiplet tends to unity as $\mathfrak{q}\rightarrow 1$ 
and the matrix integral localizes onto the maximal torus. 
In the Higgs limit the vector multiplet with the Neumann boundary condition only yields the Haar measure 
and the matrix integral becomes a Molien-Weyl integral, 
so that in the cases at hand the normalized line defect Higgs index is a polynomial in $\mathfrak{q}$ 
whose value at $\mathfrak{q}=1$ is the rank of the module of covariants over the ring of invariants. 

%%%%%%%%%%%%%%%%%%%%%%%%%%%%%%%%%%%%%%%%%%%%%%%%%
%%%%%%%%%%%%%%%%%%%%%%%%%%%%%%%%%%%%%%%%%%%%%%%%%
\section{$\mathfrak{so}(2N+1)|\mathfrak{usp}(2M)'$}
\label{sec_BC1}
\begin{figure}
\centering
\scalebox{1.2}{
\begin{tikzpicture}
\draw[thick,solid,black] (-3.5,1) -- (-3.5,3) node[above] {\textrm{1/2 NS5}};
\draw[thick,dashed,black] (3.5,1) -- (3.5,3) node[above] {\textrm{1/2 D5}};
\draw[thick,dotted,blue] (-5.5,2) node[below] {$\widetilde{\textrm{O3}}^-$} -- (-3.5,2);
\draw[thick,dotted,red] (-3.5,2) -- (-1.5,2) node[below] {$\widetilde{\textrm{O3}}^+$};
\draw[thick,dotted,red] (1.5,2) node[below] {$\textrm{O3}^+$} -- (3.5,2);
\draw[thick,dotted,red] (3.5,2) -- (5.5,2) node[below] {$\widetilde{\textrm{O3}}^+$};
\draw[thick,solid,yellow] (-5.5,2.2) node[above] {\textcolor{black}{\textrm{D3}}} -- (-3.5,2.2);
\draw[thick,solid,yellow] (-5.5,2.1)  -- (-3.5,2.1);
\draw[thick,solid,yellow] (-3.5,2.3)  -- (-1.5,2.3) node[above] {\textcolor{black}{\textrm{D3}}};
\draw[thick,solid,yellow] (-3.5,2.2)  -- (-1.5,2.2);
\draw[thick,solid,yellow] (-3.5,2.1)  -- (-1.5,2.1);
\draw[thick,solid,yellow] (1.5,2.2) node[above] {\textcolor{black}{\textrm{D3}}} -- (3.5,2.2);
\draw[thick,solid,yellow] (1.5,2.1)  -- (3.5,2.1);
\draw[thick,solid,yellow] (3.5,2.3)  -- (5.5,2.3) node[above] {\textcolor{black}{\textrm{D3}}};
\draw[thick,solid,yellow] (3.5,2.2)  -- (5.5,2.2);
\draw[thick,solid,yellow] (3.5,2.1)  -- (5.5,2.1);
\end{tikzpicture}
}
\caption{\textrm{
The S-dual brane configurations for the NS5-type $\mathfrak{so}(2N+1)|\mathfrak{usp}(2M)'$ interface and D5-type $\mathfrak{usp}(2N)|\mathfrak{usp}(2M)'$ interface.  
}}
\label{fig_so2N+1|usp2M'}
\end{figure}
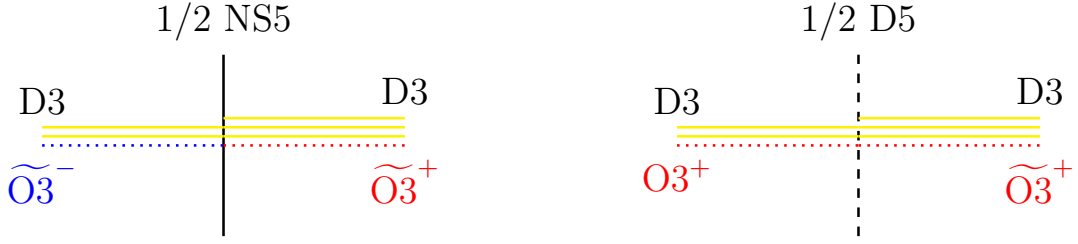

%%%%%%%%%%%%%%%%%%%%%%%%%%%%%%%%%%%%%%%%%%%%%%%%%
%%%%%%%%%%%%%%%%%%%%%%%%%%%%%%%%%%%%%%%%%%%%%%%%%
%so(2N+1)|usp(2m)' interface
For the NS5-type interface between gauge theories of type $B$ and $C$, 
the gauge group is the product of the gauge groups associated with the pair of Lie algebras $\mathfrak{so}(2N+1)$ and $\mathfrak{usp}(2M)$. 
It supports a 3d $\mathcal{N}=4$ half-hypermultiplet transforming in the bifundamental representation of the gauge group. 
It can be realized by $N$ D3-branes in the presence of an $\widetilde{\textrm{O3}}^-$-plane in $x^6<0$ terminating on a half NS5-brane from one side 
and $M$ D3-branes together with an $\widetilde{\textrm{O3}}^+$-plane in $x^6>0$ ending on it from the other side (see Figure \ref{fig_so2N+1|usp2M'}). 
%
%
%
%
%
%%%%%%%%%%%%%%%%%%%%%%%%%%%%%%%%%%%%%%%%%%%%%%%%%
\subsection{Wilson lines}
\label{sec_BCNS5W}
%%%%%%%%%%%%%%%%%%%%%%%%%%%%%%%%%%%%%%%%%%%%%%%%%
%Wilson line
The half-BPS Wilson line $W_{\mathcal{R}_1;\mathcal{R}_2}$ supported on the NS5-type $\mathfrak{so}(2N+1)|\mathfrak{usp}(2M)'$ interface is labeled 
by a pair of representations $(\mathcal{R}_1,\mathcal{R}_2)$ under the gauge algebras $\mathfrak{so}(2N+1)$ and $\mathfrak{usp}(2M)$. 

%half-index
The line defect half-index of the NS5-type $\mathfrak{so}(2N+1)|\mathfrak{usp}(2M)'$ interface involving the Wilson line operator $W_{\mathcal{R}_1;\mathcal{R}_2}$ takes the form
\begin{align}
\label{h_so2N+1|usp2M_W}
&
\langle W_{\mathcal{R}_1;\mathcal{R}_2} \rangle_{\mathcal{N}}^{\mathfrak{so}(2N+1)|\mathfrak{usp}(2M)'}(t;q)
\nonumber\\
&
=\frac{1}{2^N N!}
\frac{(q)_{\infty}^N}{(q^{\frac12}t^{-2};q)_{\infty}^N}
\oint \prod_{i=1}^N \frac{ds_i^{(1)}}{2\pi is_i^{(1)}}
\frac{(s_i^{(1)\pm};q)_{\infty}}{(q^{\frac12}t^{-2}s_i^{(1)\pm};q)_{\infty}}
\prod_{i<j}
\frac{(s_i^{(1)\pm}s_j^{(1)\pm};q)_{\infty}}
{(q^{\frac12}t^{-2}s_i^{(1)\pm}s_j^{(1)\pm};q)_{\infty}}
\nonumber\\
&\times 
\frac{1}{2^M M!}
\frac{(q)_{\infty}^M}{(q^{\frac12}t^{-2};q)_{\infty}^M}
\oint \prod_{i=1}^M \frac{ds_i^{(2)}}{2\pi is_i^{(2)}}
\frac{(s_i^{(2)\pm 2};q)_{\infty}}{(q^{\frac12}t^{-2}s_i^{(2)\pm 2};q)_{\infty}}
\prod_{i< j}
\frac{(s_i^{(2)\pm}s_j^{(2)\pm};q)_{\infty}}
{(q^{\frac12}t^{-2}s_i^{(2)\pm}s_j^{(2)\pm};q)_{\infty}}
\nonumber\\
&\times 
\prod_{i=1}^N\prod_{j=1}^M
\frac{(q^{\frac34}t^{-1}s_i^{(1)\pm}s_j^{(2)\pm};q)_{\infty}}
{(q^{\frac14}ts_i^{(1)\pm}s_{j}^{(2)\pm};q)_{\infty}}
\prod_{j=1}^M 
\frac{(q^{\frac34}t^{-1}s_j^{(2)\pm};q)_{\infty}}{(q^{\frac14}ts_{j}^{(2)\pm};q)_{\infty}}
\chi_{\mathcal{R}_1}^{\mathfrak{so}(2N+1)}(s^{(1)})\chi_{\mathcal{R}_2}^{\mathfrak{usp}(2M)}(s^{(2)}), 
\end{align}
where $\chi_{\mathcal{R}_1}^{\mathfrak{so}(2N+1)}(s^{(1)})$ and $\chi_{\mathcal{R}_2}^{\mathfrak{usp}(2M)}(s^{(2)})$ are 
the character of the representation $\mathcal{R}_1$ of $\mathfrak{so}(2N+1)$ 
and that of the representation $\mathcal{R}_2$ of $\mathfrak{usp}(2M)$ respectively. 

%so(2N+1)
%spinor
The gauge groups associated with the Lie algebra $\mathfrak{so}(2N+1)$ are globally distinguished. 
For $Spin(2N+1)$ there exist the Wilson lines transforming in the spinor representation. 
The character of the spinor representation of $\mathfrak{so}(2N+1)$ is given by
\begin{align}
\label{ch_spin2N+1}
\chi_{\textrm{sp}}^{\mathfrak{so}(2N+1)}(s)
&=\prod_{i=1}^{N}(s_i^{\frac12}+s_i^{-\frac12}). 
\end{align}
For $Spin(2N+1)$ one has an electric $\mathbb{Z}_2$ 1-form symmetry associated with the non-trivial center $\mathbb{Z}_2$. 
The gauge invariant singlet that contributes to the line defect half-index must carry a trivial center charge $+1$. 
The BPS local operators appearing in the integrand of the half-index (\ref{h_so2N+1|usp2M_W}) 
without a spinor Wilson line are invariant under the $\mathbb{Z}_2$ center symmetry. 
On the other hand, the spinor representation carries non-trivial $\mathbb{Z}_2$ center charge $-1$. 
Hence we have a selection rule according to the total $\mathbb{Z}_2$ center charge 
\begin{align}
\label{h_so2N+1|usp2M_Wsp}
\langle W_{\textrm{sp};\mathcal{R}_2} \rangle_{\mathcal{N}}^{\mathfrak{so}(2N+1)|\mathfrak{usp}(2M)'}(t;q)
&=0. 
\end{align}
Crucially, this topological obstruction can be bypassed when we introduce an even number of spinor Wilson lines 
or consider a tensor product of even powers of the spinor representation. 

More generally, the $\mathfrak{so}(2N+1)$ character of the irreducible representation  
with the highest weight corresponding to the Young diagram $\lambda$ is \cite{MR1153249}
\begin{align}
\label{ch_so2N+1_irrep}
\chi_{\lambda}^{\mathfrak{so}(2N+1)}(s)
&=\frac{\det (s_j^{\lambda_i+N-i+1/2}-s_j^{-(\lambda_i+N-i+1/2)} )}
{\det(s_j^{N-i+1/2}-s_j^{-(N-i+1/2)})}. 
\end{align}
This character carries exclusively integer weights, so that the corresponding Wilson line can combine with the BPS local operators to form gauge invariants 
with trivial $\mathbb{Z}_2$ center charge. 

%usp(2M)
We next consider the representation $\mathcal{R}_2$ of $\mathfrak{usp}(2M)$. 
All its irreducible representations are labeled by the Young diagrams. 
The character of the irreducible representation labeled by the Young diagram $\lambda$ is given by \cite{MR1153249}
\begin{align}
\label{ch_usp2N_irrep}
\chi_{\lambda}^{\mathfrak{usp}(2M)}(s)
&=\frac{\det(s_j^{\lambda_i+M-i+1}-s_j^{-\lambda_i-M+i-1})}{\det(s_j^{M-i+1}-s_j^{-M+i-1})}. 
\end{align}

%duality for SO(1)|USp(2M)'
As a first example, let us consider the case with $N=0$, namely the NS5-type $SO(1)|USp(2M)'$ interface. 
In this configuration the $SO(1)$ gauge theory itself is trivial, implying that no dynamical bulk gauge fields or Wilson lines for the gauge group exist. 
However, at the interface there exists a half-hypermultiplet in the fundamental representation of $USp(2M)$ 
as well as 4d $\mathcal{N}=4$ $USp(2M)'$ SYM theory obeying the Neumann boundary condition $\mathcal{N}$. 
In the absence of the Wilson line the S-dual configuration is the D5-type $0|USp(2M)'$ interface, 
or equivalently the D5-type boundary condition of 4d $\mathcal{N}=4$ $USp(2M)'$ SYM theory, 
where the bulk $USp(2M)'$ gauge group is completely broken due to the Nahm pole boundary condition. 
In fact, the corresponding interface half-index is equal to the $USp(2M)$ Nahm pole half-index, as discussed in \cite{Hatsuda:2024lcc}. 
%Wilson
When we introduce the Wilson line in the fundamental representation of $USp(2M)$, 
the line defect half-index (\ref{h_so2N+1|usp2M_W}) reduces to
\begin{align}
\label{h_so1|usp2M_W}
&
\langle W_{0;\tiny \yng(1)} \rangle_{\mathcal{N}}^{\mathfrak{so}(1)|\mathfrak{usp}(2M)'}(t;q)
\nonumber\\
&=\frac{1}{2^M M!}
\frac{(q)_{\infty}^M}{(q^{\frac12}t^{-2};q)_{\infty}^M}
\oint \prod_{i=1}^{M}
\frac{ds_i}{2\pi is_i}
\frac{(s_i^{\pm2};q)_{\infty}}{(q^{\frac12}t^{-2}s_i^{\pm2};q)_{\infty}}
\prod_{i<j}
\frac{(s_i^{\pm}s_j^{\pm};q)_{\infty}}
{(q^{\frac12}t^{-2}s_i^{\pm}s_j^{\pm};q)_{\infty}}
\nonumber\\
&\times \prod_{i=1}^M \frac{(q^{\frac34}t^{-1}s_i^{\pm};q)_{\infty}}
{(q^{\frac14}ts_i^{\pm};q)_{\infty}}
\sum_{i=1}^M (s_i+s_i^{-1}). 
\end{align}
The fundamental representation of $USp(2M)$ is a minuscule representation. 
Under S-duality, the fundamental Wilson line is expected to map to a magnetic vortex line characterized 
by a minuscule magnetic weight in the dual $USp(2M)'$ gauge theory \cite{Kapustin:2006pk}. 
We address this dual vortex line defect half-index in section \ref{sec_BCD5V}. 
Remarkably, the matrix integral (\ref{h_so1|usp2M_W}) can be rewritten in terms of a simple infinite product expression,
as we show in (\ref{h_so1|usp2M_W_closed}) and prove in Appendix \ref{app_koornwinder}. 

%%%%%%%%%%%%%%%%%%%%%%%%%%%%%%%%%%%%%%%%%%%%%%%%%
\subsubsection{Coulomb limit}
%%%%%%%%%%%%%%%%%%%%%%%%%%%%%%%%%%%%%%%%%%%%%%%%%
%Coulomb limits
In the Coulomb limit the line defect half-index (\ref{h_so2N+1|usp2M_W}) is factorized as
\begin{align}
\label{hC_so2N+1|usp2M_W}
{\langle W_{\mathcal{R}_1;\mathcal{R}_2} \rangle_{\mathcal{N}}^{\mathfrak{so}(2N+1)|\mathfrak{usp}(2M)'}}^{(C)}(\mathfrak{q})
&=\langle W_{\mathcal{R}_1} \rangle_{\textrm{$\frac12$BPS}}^{\mathfrak{so}(2N+1)}(\mathfrak{q})
\langle W_{\mathcal{R}_2} \rangle_{\textrm{$\frac12$BPS}}^{\mathfrak{usp}(2M)}(\mathfrak{q}), 
\end{align}
where $\langle W_{\mathcal{R}_1} \rangle_{\textrm{$\frac12$BPS}}^{\mathfrak{so}(2N+1)}(\mathfrak{q})$ 
is the half-BPS line defect index of the Wilson line $W_{\mathcal{R}_1}$ in $\mathcal{N}=4$ $\mathfrak{so}(2N+1)$ SYM theory 
and $\langle W_{\mathcal{R}_2} \rangle_{\textrm{$\frac12$BPS}}^{\mathfrak{usp}(2M)}(\mathfrak{q})$ 
is that of the Wilson line $W_{\mathcal{R}_2}$ in $\mathcal{N}=4$ $\mathfrak{usp}(2M)$ SYM theory \cite{Hatsuda:2025jze}. 
In the absence of line operator insertions, the interface Coulomb half-index takes the form \cite{Hatsuda:2024lcc}
\begin{align}
{\mathbb{II}_{\mathcal{N}}^{\mathfrak{so}(2N+1)|\mathfrak{usp}(2M)'}}^{(C)}(\mathfrak{q})
&=\prod_{n=1}^{N}\frac{1}{1-\mathfrak{q}^{4n}}
\prod_{n=1}^{M}\frac{1}{1-\mathfrak{q}^{4n}}. 
\end{align}
Thus the normalized Wilson line defect half-index also factorizes in a similar manner 
in terms of the normalized half-BPS line defect indices 
\begin{align}
{\langle \mathcal{W}_{\mathcal{R}_1;\mathcal{R}_2} \rangle_{\mathcal{N}}^{\mathfrak{so}(2N+1)|\mathfrak{usp}(2M)'}}^{(C)}(\mathfrak{q})
&=\langle \mathcal{W}_{\mathcal{R}_1} \rangle_{\textrm{$\frac12$BPS}}^{\mathfrak{so}(2N+1)}(\mathfrak{q})
\langle \mathcal{W}_{\mathcal{R}_2} \rangle_{\textrm{$\frac12$BPS}}^{\mathfrak{usp}(2M)}(\mathfrak{q}). 
\end{align}
The normalized half-BPS line defect indices can be evaluated in closed form. 
In the Coulomb limit the vector multiplet contributes the Hall-Littlewood measure 
$\prod_{\alpha}(1-s^{\alpha})/(1-\mathfrak{q}^2 s^{\alpha})$, 
which is the Haar measure dressed by the plethystic exponential of $\mathfrak{q}^2\chi_{\mathrm{adj}}$. 
Hence the normalized half-BPS index of the Wilson line in the irreducible representation $\lambda$ of $\mathfrak{g}$ 
is the graded multiplicity of $\lambda$ in the harmonic polynomials on $\mathfrak{g}$ \cite{MR158024}, 
namely the generating function of the generalized exponents of $\lambda$, 
which coincides with Lusztig's $q$-analogue of the zero weight multiplicity $K_{\lambda,0}$ \cite{MR737932,MR593631}
\begin{align}
\label{halfBPS_K}
\langle \mathcal{W}_{\lambda} \rangle_{\textrm{$\frac12$BPS}}^{\mathfrak{g}}(\mathfrak{q})
&=K_{\lambda,0}(\mathfrak{q}^2). 
\end{align}
For instance, 
\begin{align}
\langle \mathcal{W}_{\tiny \yng(1)} \rangle_{\textrm{$\frac12$BPS}}^{\mathfrak{so}(2N+1)}(\mathfrak{q})&=\mathfrak{q}^{2N}, &
\langle \mathcal{W}_{\tiny \yng(1)} \rangle_{\textrm{$\frac12$BPS}}^{\mathfrak{usp}(2M)}(\mathfrak{q})&=0. 
\end{align}
In the normalized index the $s$-independent prefactor cancels out. 
As $\mathfrak{q}\rightarrow 1$ the Hall-Littlewood measure tends to unity while remaining bounded on the unit circle, 
so that the contributions of the W-bosons cancel against those of the off-diagonal adjoint scalar fields 
and the matrix integral localizes onto the maximal torus $T^N\times T^M$ of the gauge group, 
which is the residual gauge group at a generic point of the Coulomb branch. 
Consequently, the line defect Witten index is given by (\ref{WI_NS5_C_intro}) 
\begin{align}
\label{WI_so2N+1|usp2M_C}
&{\langle \mathcal{W}_{\mathcal{R}_1;\mathcal{R}_2} \rangle_{\mathcal{N}}^{\mathfrak{so}(2N+1)|\mathfrak{usp}(2M)'}}^{(C)}(1)
\nonumber\\
&=\oint \prod_{i=1}^{N}\frac{ds_i^{(1)}}{2\pi is_i^{(1)}}\prod_{j=1}^{M}\frac{ds_j^{(2)}}{2\pi is_j^{(2)}}
\chi_{\mathcal{R}_1}^{\mathfrak{so}(2N+1)}(s^{(1)})\chi_{\mathcal{R}_2}^{\mathfrak{usp}(2M)}(s^{(2)})
\nonumber\\
&=\dim \mathcal{R}_1[0]\cdot \dim \mathcal{R}_2[0], 
\end{align}
where $\mathcal{R}[0]$ denotes the zero weight subspace of $\mathcal{R}$. 
Unlike for the D5-type interfaces, the torus has rank $N+M$ rather than the rank of $\mathfrak{g}_{\min}$. 
For example, 
${\langle \mathcal{W}_{\tiny \yng(1);0} \rangle_{\mathcal{N}}^{\mathfrak{so}(2N+1)|\mathfrak{usp}(2M)'}}^{(C)}(1)=1$, 
whereas 
${\langle \mathcal{W}_{0;\tiny \yng(1)} \rangle_{\mathcal{N}}^{\mathfrak{so}(2N+1)|\mathfrak{usp}(2M)'}}^{(C)}(1)$ $=$ 
${\langle \mathcal{W}_{\tiny \yng(1);\yng(1)} \rangle_{\mathcal{N}}^{\mathfrak{so}(2N+1)|\mathfrak{usp}(2M)'}}^{(C)}(1)$ $=$ $0$ 
since the fundamental representation of $\mathfrak{usp}(2M)$ has no zero weight. 

%%%%%%%%%%%%%%%%%%%%%%%%%%%%%%%%%%%%%%%%%%%%%%%%%
\subsubsection{Higgs limit}
%%%%%%%%%%%%%%%%%%%%%%%%%%%%%%%%%%%%%%%%%%%%%%%%%
%Higgs limits
In the Higgs limit the expression (\ref{h_so2N+1|usp2M_W}) is not simply factorized. 
It becomes the following matrix integral: 
\begin{align}
\label{hH_so2N+1|usp2M_W}
&
{\langle W_{\mathcal{R}_1;\mathcal{R}_2} \rangle_{\mathcal{N}}^{\mathfrak{so}(2N+1)|\mathfrak{usp}(2M)'}}^{(H)}(\mathfrak{q})
\nonumber\\
&=\frac{1}{2^N N!}\oint \prod_{i=1}^N \frac{ds_i^{(1)}}{2\pi is_i^{(1)}}
(1-s_i^{(1)\pm})
\prod_{i<j}
(1-s_i^{(1)\pm}s_j^{(1)\pm})
\nonumber\\
&\times 
\frac{1}{2^M M!}\oint \prod_{i=1}^M \frac{ds_i^{(2)}}{2\pi is_i^{(2)}}
(1-s_i^{(2)\pm 2})
\prod_{i<j}(1-s_i^{(2)\pm}s_j^{(2)\pm})
\nonumber\\
&\times 
\prod_{i=1}^{N}\prod_{j=1}^{M}
\frac{1}{1-\mathfrak{q}s_i^{(1)\pm}s_j^{(2)\pm}}
\prod_{j=1}^M \frac{1}{1-\mathfrak{q}s_j^{(2)\pm}}
 \chi_{\mathcal{R}_1}^{\mathfrak{so}(2N+1)}(s^{(1)})
 \chi_{\mathcal{R}_2}^{\mathfrak{usp}(2M)}(s^{(2)}). 
\end{align}
Before introducing line operators, the corresponding interface Higgs half-index is given by \cite{Hatsuda:2024lcc}
\begin{align}
{\mathbb{II}_{\mathcal{N}}^{\mathfrak{so}(2N+1)|\mathfrak{usp}(2M)'}}^{(H)}(\mathfrak{q})
&=\prod_{n=1}^{\min(N,M)}\frac{1}{1-\mathfrak{q}^{4n}}. 
\end{align}

%fund
For the insertion of a Wilson line transforming in the fundamental representation, 
we obtain the following normalized line defect Higgs half-indices for $N\ge 1$:
\begin{align}
{\langle \mathcal{W}_{\tiny \yng(1);0} \rangle_{\mathcal{N}}^{\mathfrak{so}(2N+1)|\mathfrak{usp}(2M)'}}^{(H)}(\mathfrak{q})
&=
\begin{cases}
\mathfrak{q}^{2N}&\textrm{for $N\le M$}\cr
0&\textrm{for $N> M$}\cr
\end{cases}, \\
{\langle \mathcal{W}_{0;\tiny \yng(1)} \rangle_{\mathcal{N}}^{\mathfrak{so}(2N+1)|\mathfrak{usp}(2M)'}}^{(H)}(\mathfrak{q})
&=
\begin{cases}
\mathfrak{q}^{2M-1}&\textrm{for $N< M$}\cr
0&\textrm{for $N\ge M$}\cr
\end{cases}, \\
{\langle \mathcal{W}_{\tiny \yng(1);\yng(1)} \rangle_{\mathcal{N}}^{\mathfrak{so}(2N+1)|\mathfrak{usp}(2M)'}}^{(H)}(\mathfrak{q})
&=
\begin{cases}
\mathfrak{q}+\mathfrak{q}^3+\mathfrak{q}^5+\cdots+\mathfrak{q}^{4N+1}&\textrm{for $N< M$} \cr
\mathfrak{q}+\mathfrak{q}^3+\mathfrak{q}^5+\cdots+\mathfrak{q}^{4M-1}&\textrm{for $N\ge M$} \cr
\end{cases}. 
\end{align}
The second expression is not valid for $N=0$. 
Indeed, for the NS5-type $\mathfrak{so}(1)|\mathfrak{usp}(2M)'$ interface the closed-form result (\ref{h_so1|usp2M_W_closed}) 
gives in the Higgs limit 
\begin{align}
{\langle \mathcal{W}_{0;\tiny \yng(1)} \rangle_{\mathcal{N}}^{\mathfrak{so}(1)|\mathfrak{usp}(2M)'}}^{(H)}(\mathfrak{q})
&=\mathfrak{q}
\end{align}
for any $M$, which is consistent with the last expression as the fundamental representation of $\mathfrak{so}(1)$ is trivial. 

The line defect Witten indices of these results can be understood from the residual gauge group 
at a generic point of the Higgs branch (see Appendix \ref{app_residual} for details). 
For a generic D-flat configuration of the bifundamental half-hypermultiplet $X$, 
the moment maps $XX^{\dagger}\in \mathfrak{so}(2N+1)$ and $X^{\dagger}X\in \mathfrak{usp}(2M)$ share $\min(N,M)$ pairs of non-zero eigenvalues. 
The unbroken gauge group consists of $U(1)^{\min(N,M)}$, 
embedded diagonally as $s_i^{(1)}=s_i^{(2)}=u_i$ with $i=1,\cdots,\min(N,M)$, 
together with $SO(2(N-M)+1)$ for $N\ge M$ or $USp(2(M-N))$ for $N<M$ acting on the zero eigenspace. 
In the brane picture, $\min(N,M)$ D3-branes reconnect across the half NS5-brane and move away along the $789$ directions, 
leaving behind the NS5-type $\mathfrak{so}(2(N-M)+1)|\mathfrak{usp}(0)'$ or $\mathfrak{so}(1)|\mathfrak{usp}(2(M-N))'$ interface. 
For $N\ge M$ the action of the gauge group on the half-hypermultiplet is stable, 
that is, its generic orbits are closed, and we find 
\begin{align}
\label{WI_so2N+1|usp2M_H1}
&{\langle \mathcal{W}_{\mathcal{R}_1;\mathcal{R}_2} \rangle_{\mathcal{N}}^{\mathfrak{so}(2N+1)|\mathfrak{usp}(2M)'}}^{(H)}(1)
\nonumber\\
&=\oint \prod_{i=1}^{M}\frac{du_i}{2\pi iu_i}\int_{SO(2(N-M)+1)}dg\, 
\chi_{\mathcal{R}_1}^{\mathfrak{so}(2N+1)}(u,g)\chi_{\mathcal{R}_2}^{\mathfrak{usp}(2M)}(u)
\nonumber\\
&=\dim (\mathcal{R}_1\otimes \mathcal{R}_2)^{U(1)^M\times SO(2(N-M)+1)}, 
\end{align}
where $u_i$ are the fugacities of the diagonal $U(1)^M$ and $g_i$ those of the residual $SO(2(N-M)+1)$, 
so that the characters are evaluated at 
$s^{(1)}=(u_1,\cdots,u_M,g_1,\cdots,g_{N-M})$ and $s^{(2)}=(u_1,\cdots,u_M)$, 
and $\int dg$ is the normalized Haar integral. 
For $N<M$ the residual $\mathfrak{so}(1)|\mathfrak{usp}(2(M-N))'$ interface supports a half-hypermultiplet 
in the fundamental representation of $USp(2(M-N))$, which cannot acquire a D-flat expectation value. 
Although it does not generate any gauge invariant, it contributes to the covariants 
and must be kept in the residual integral 
\begin{align}
\label{WI_so2N+1|usp2M_H2}
&{\langle \mathcal{W}_{\mathcal{R}_1;\mathcal{R}_2} \rangle_{\mathcal{N}}^{\mathfrak{so}(2N+1)|\mathfrak{usp}(2M)'}}^{(H)}(1)
\nonumber\\
&=\lim_{\mathfrak{q}\rightarrow 1}
\oint \prod_{i=1}^{N}\frac{du_i}{2\pi iu_i}\int_{USp(2(M-N))}dh\, 
\chi_{\mathcal{R}_1}^{\mathfrak{so}(2N+1)}(u)\chi_{\mathcal{R}_2}^{\mathfrak{usp}(2M)}(u,h)
\prod_{j=1}^{M-N}\frac{1}{1-\mathfrak{q}h_j^{\pm}}
\nonumber\\
&=\sum_{n=0}^{\infty}
\oint \prod_{i=1}^{N}\frac{du_i}{2\pi iu_i}\int_{USp(2(M-N))}dh\, 
\chi_{\mathcal{R}_1}^{\mathfrak{so}(2N+1)}(u)\chi_{\mathcal{R}_2}^{\mathfrak{usp}(2M)}(u,h)
\chi_{(n)}^{\mathfrak{usp}(2(M-N))}(h). 
\end{align}
Here $u_i$ are again the fugacities of the diagonal $U(1)^N$ and $h_j$ those of the residual $USp(2(M-N))$, 
so that this time the characters are evaluated at 
$s^{(1)}=(u_1,\cdots,u_N)$ and $s^{(2)}=(u_1,\cdots,u_N,h_1,\cdots,h_{M-N})$, 
while $\prod_{j}(1-\mathfrak{q}h_j^{\pm})^{-1}$ is the plethystic exponential of the character 
$\chi_{S}(h)=\sum_{j=1}^{M-N}(h_j+h_j^{-1})$ of the residual half-hypermultiplet. 
In the second line we have used the fact that 
the polynomial ring on $\mathbb{C}^{2K}$ decomposes as $\bigoplus_{n\ge 0}\mathrm{Sym}^n\mathbb{C}^{2K}$ under $USp(2K)$ without multiplicities. 
Equivalently, the Higgs limit of the Wilson line defect half-index for the NS5-type $\mathfrak{so}(1)|\mathfrak{usp}(2K)'$ interface is 
\begin{align}
\label{hH_so1|usp2M_Wgen}
{\langle \mathcal{W}_{0;\lambda} \rangle_{\mathcal{N}}^{\mathfrak{so}(1)|\mathfrak{usp}(2K)'}}^{(H)}(\mathfrak{q})
&=
\begin{cases}
\mathfrak{q}^n&\textrm{for $\lambda=(n)$}\cr
0&\textrm{otherwise}\cr
\end{cases}, 
\end{align}
which generalizes the above result for the fundamental representation. 

The formulae (\ref{WI_so2N+1|usp2M_H1}) and (\ref{WI_so2N+1|usp2M_H2}) reproduce the $\mathfrak{q}\rightarrow 1$ limits of the above results. 
For $\mathcal{W}_{\tiny \yng(1);0}$ the $u$-neutral part of the fundamental representation of $\mathfrak{so}(2N+1)$ is 
the vector representation of $SO(2(N-M)+1)$ for $N\ge M$, which contains an invariant only for $N=M$, 
and the singlet for $N<M$, 
so that the Witten index is $1$ for $N\le M$ and $0$ for $N>M$. 
For $\mathcal{W}_{0;\tiny \yng(1)}$ the $u$-neutral part is absent for $N\ge M$, 
while for $N<M$ it is the fundamental representation of $USp(2(M-N))$ which pairs with the $n=1$ term, 
so that the Witten index is $0$ for $N\ge M$ and $1$ for $N<M$. 
For $\mathcal{W}_{\tiny \yng(1);\yng(1)}$ there are $2\min(N,M)$ $u$-neutral components $u_i^{\pm1}\otimes u_i^{\mp1}$ 
and an additional contribution from the $n=1$ term for $N<M$, 
which gives $2N+1$ for $N<M$ and $2M$ for $N\ge M$. 
Notably, for $\mathcal{W}_{0;\tiny \yng(1)}$ with $N<M$ the Witten index originates entirely from the residual half-hypermultiplet. 
The integral over the compact residual gauge group alone would give $0$. 
We have checked (\ref{WI_so2N+1|usp2M_H1}) and (\ref{WI_so2N+1|usp2M_H2}) against the exact matrix integral (\ref{hH_so2N+1|usp2M_W}) 
for various representations (see Appendix \ref{app_residual}). 
For example, we find 
\begin{align}
{\langle \mathcal{W}_{0;\tiny \yng(2)} \rangle_{\mathcal{N}}^{\mathfrak{so}(3)|\mathfrak{usp}(4)'}}^{(H)}(\mathfrak{q})
&=\mathfrak{q}^2+\mathfrak{q}^6, \\
{\langle \mathcal{W}_{\tiny \yng(1);\yng(2,1)} \rangle_{\mathcal{N}}^{\mathfrak{so}(3)|\mathfrak{usp}(4)'}}^{(H)}(\mathfrak{q})
&=\mathfrak{q}^3+2\mathfrak{q}^5+2\mathfrak{q}^7+\mathfrak{q}^9, 
\end{align}
whose Witten indices $2$ and $6$ follow from (\ref{WI_so2N+1|usp2M_H2}), 
whereas the compact residual gauge group alone would give $1$ and $2$. 

%%%%%%%%%%%%%%%%%%%%%%%%%%%%%%%%%%%%%%%%%%%%%%%%%
\subsection{Vortex lines}
\label{sec_BCNS5V}
%%%%%%%%%%%%%%%%%%%%%%%%%%%%%%%%%%%%%%%%%%%%%%%%%
%vortex
Let us consider the half-BPS vortex lines at the NS5-type $\mathfrak{so}(2N+1)|\mathfrak{usp}(2M)'$ interface 
which can be labeled by a pair $(k^{(1)},k^{(2)})$ of the Abelianized vortex numbers. 

%so(2N+1)
In 4d gauge theory the 't Hooft line is labeled by a coweight. 
It is free from the monopole bubbling effect if the magnetic charge is associated with the minuscule weight \cite{Kapustin:2006pk}. 
For the gauge algebra $\mathfrak{so}(2N+1)$, coweights coincide with weights of $\mathfrak{usp}(2N)$. 
The minuscule coweight of $\mathfrak{so}(2N+1)$ corresponds to the fundamental weight $(1,0^{N-1})$ of $\mathfrak{usp}(2N)$. 
In this case, the 't Hooft line is primitive and irreducible. 
In Type IIB string theory the minuscule 't Hooft line will be realized as a single D1-brane ending on a stack of D3-branes in the presence of an $\widetilde{\textrm{O3}}^-$-plane. 

Similarly, the vortex number $k^{(1)}$ $\in$ $\mathbb{Z}^N$ lies in the cocharacter lattice. 
Under the action of S-duality the $\mathfrak{so}(2N+1)$ gauge algebra of the NS5-type interface 
maps to the Langlands dual $\mathfrak{usp}(2N)$ gauge algebra in the dual D5-type interface which is kept for $N\le M$. 
In this case the minuscule vortex line at the interface 
with the fundamental coweight $(1,0^{N-1})$ of $SO(2N+1)$ gauge group 
is expected to be S-dual to the Wilson line transforming in the fundamental representation of the $USp(2N)$ gauge group. 

%half-index
We propose that 
the line defect half-index of the NS5-type $\mathfrak{so}(2N+1)|\mathfrak{usp}(2M)'$ interface 
involving the minuscule vortex line with $k^{(1)}=(1,0^{N-1})$ and $k^{(2)}=0$ for $N\le M$ is given by
\begin{align}
\label{h_so2N+1|usp2M_V}
&
\langle V_{(1,0^{N-1});0}\rangle_{\mathcal{N}}^{\mathfrak{so}(2N+1)|\mathfrak{usp}(2M)'}(t;q)
\nonumber\\
&=q^{\frac14+\frac{M-N}{2}}t^{-1-2(M-N)}
\frac{1}{2^{N-1}(N-1)!}
\frac{(q)_{\infty}^{N}}{(q^{\frac12}t^{-2};q)_{\infty}^N}
\nonumber\\
&\times 
\oint \prod_{i=1}^{N}
\frac{ds_i^{(1)}}{2\pi is_i^{(1)}}
\frac{(q^{\frac12}s_1^{(1)\pm};q)_{\infty}}{(qt^{-2}s_1^{(1)\pm};q)_{\infty}}
\prod_{i=2}^N \frac{(s_i^{(1)\pm};q)_{\infty}}{(q^{\frac12}t^{-2}s_i^{(1)\pm};q)_{\infty}}
\nonumber\\
&\times 
\prod_{i=2}^{N}
\frac{(q^{\frac12}s_1^{(1)\pm}s_i^{(1)\pm};q)_{\infty}}
{(qt^{-2}s_1^{(1)\pm}s_i^{(1)\pm};q)_{\infty}}
\prod_{2\le i<j\le N}
\frac{(s_i^{(1)\pm}s_j^{(1)\pm};q)_{\infty}}
{(q^{\frac12}t^{-2}s_i^{(1)\pm}s_j^{(1)\pm};q)_{\infty}}
\nonumber\\
&\times 
\frac{1}{2^M M!}
\frac{(q)_{\infty}^{M}}{(q^{\frac12}t^{-2};q)_{\infty}^M}
\oint \prod_{i=1}^{M}
\frac{ds_i^{(2)}}{2\pi is_i^{(2)}}
\frac{(s_i^{(2)\pm 2};q)_{\infty}}
{(q^{\frac12}t^{-2}s_i^{(2)\pm 2};q)_{\infty}}
\prod_{i<j}
\frac{(s_i^{(2)\pm}s_j^{(2)\pm};q)_{\infty}}
{(q^{\frac12}t^{-2}s_i^{(2)\pm}s_j^{(2)\pm};q)_{\infty}}
\nonumber\\
&\times 
\prod_{j=1}^M
\frac{(q^{\frac54}t^{-1}s_1^{(1)\pm}s_j^{(2)\pm};q)_{\infty}}
{(q^{\frac34}ts_1^{(1)\pm}s_j^{(2)\pm};q)_{\infty}} 
\prod_{i=2}^N
\prod_{j=1}^M
\frac{(q^{\frac34}t^{-1}s_i^{(1)\pm}s_j^{(2)\pm};q)_{\infty}}
{(q^{\frac14}ts_i^{(1)\pm}s_j^{(2)\pm};q)_{\infty}}
\prod_{j=1}^M
\frac{(q^{\frac34}t^{-1}s_j^{(2)\pm};q)_{\infty}}
{(q^{\frac14}ts_j^{(2)\pm};q)_{\infty}}. 
\end{align}
Here the overall factor $q^{\frac14+\frac{M-N}{2}}t^{-1-2(M-N)}$ is the contribution from the background bare monopole operator. 
This matrix integral is a generalized version of the ordinary interface half-index matrix integral, 
in which a spin shift is incorporated into the $\mathfrak{so}(2N+1)$ gauge fugacity $s_i^{(1)}$ for $i=1$, 
corresponding to a singular configuration associated with the vortex number. 

%%%%%%%%%%%%%%%%%%%%%%%%%%%%%%%%%%%%%%%%%%%%%%%%%
%ADDED
Let us spell out these two ingredients in more detail as we will make use of them repeatedly in what follows. 
In the background of the vortex line of magnetic charge $k=(k^{(1)},k^{(2)})$, 
the singular profile of the fields shifts the spin $J$ of a field of weight $\rho$ 
in such a way that its fugacity is multiplied by $q^{\frac{|\rho(k)|}{2}}$, 
both in the numerator and in the denominator of the corresponding $q$-Pochhammer factors. 
For $k^{(1)}=(1,0^{N-1})$ and $k^{(2)}=0$ the fields charged under $k$ are 
the components of the 4d $\mathfrak{so}(2N+1)$ vector multiplet associated with the roots $\pm e_1$ and $\pm e_1\pm e_i$ with $i\neq 1$, 
together with the 3d bifundamental half-hypermultiplet of weights $\pm e_1 \pm f_j$, 
all of which obey $|\rho(k)|=1$ and are therefore shifted by $q^{\frac12}$, 
whereas the weights $\pm f_j$ of the bifundamental half-hypermultiplet, 
which descend from the zero weight of the vector representation of $\mathfrak{so}(2N+1)$, are left intact. 
In addition, the background bare monopole operator carries the $SU(2)_C$ R-charge 
\begin{align}
\label{bare_monopole_D}
\Delta(k)&=\frac12\sum_{\rho}|\rho(k)|-\sum_{\alpha>0}|\alpha(k)|, 
\end{align}
where $\rho$ runs over the weights of the 3d hypermultiplet localized at the interface 
and $\alpha$ over the positive roots of the 4d gauge algebras preserved at the interface. 
As the half-index (\ref{DEF_hind}) weights such a Coulomb branch type operator by $q^{\frac{C}{4}}t^{-C}$, 
the bare monopole operator contributes the overall factor $q^{\frac{\Delta(k)}{4}}t^{-\Delta(k)}$. 
In the case at hand the $4M$ weights $\pm e_1\pm f_j$ of the bifundamental half-hypermultiplet 
and the $2N-1$ positive roots $e_1$ and $e_1\pm e_i$ with $i\neq 1$ of $\mathfrak{so}(2N+1)$ are charged under $k$, so that 
\begin{align}
\label{Delta_so2N+1|usp2M}
\Delta\bigl((1,0^{N-1});0\bigr)&=\frac12\cdot 4M-(2N-1)=1+2(M-N), 
\end{align}
which precisely accounts for the non-trivial prefactor $q^{\frac14+\frac{M-N}{2}}t^{-1-2(M-N)}$ in (\ref{h_so2N+1|usp2M_V}). 
It is also consistent with the Coulomb limit (\ref{hC_so2N+1|usp2M_V}), 
whose leading power $\mathfrak{q}^{1+2(M-N)}$ is nothing but $\mathfrak{q}^{\Delta}$ 
as the Coulomb limit is defined by $\mathfrak{q}=q^{\frac14}t^{-1}$. 
Finally, the factor $\frac{1}{2^{N-1}(N-1)!}$ in (\ref{h_so2N+1|usp2M_V}) 
is the inverse of the order of the Weyl group of the residual $U(1)\times SO(2N-1)$ gauge symmetry 
which is preserved in the vortex line background. 
%%%%%%%%%%%%%%%%%%%%%%%%%%%%%%%%%%%%%%%%%%%%%%%%%

We will see in section \ref{sec_BCD5W} that 
the matrix integral (\ref{h_so2N+1|usp2M_V}) agrees with the dual D5-type interface Wilson line defect half-index. 

%%%%%%%%%%%%%%%%%%%%%%%%%%%%%%%%%%%%%%%%%%%%%%%%%
\subsubsection{Coulomb limit}
%%%%%%%%%%%%%%%%%%%%%%%%%%%%%%%%%%%%%%%%%%%%%%%%%
%Coulomb limit
In the Coulomb limit the line defect half-index (\ref{h_so2N+1|usp2M_V}) is evaluated as
\begin{align}
\label{hC_so2N+1|usp2M_V}
{\langle V_{(1,0^{N-1});0}\rangle_{\mathcal{N}}^{\mathfrak{so}(2N+1)|\mathfrak{usp}(2M)'}}^{(C)}(\mathfrak{q})
&=\frac{\mathfrak{q}^{1+2(M-N)}}{1-\mathfrak{q}^2}
\mathcal{I}_{\textrm{$\frac12$BPS}}^{\mathfrak{so}(2N+1)}(\mathfrak{q})
\mathcal{I}_{\textrm{$\frac12$BPS}}^{\mathfrak{usp}(2M)}(\mathfrak{q})
\nonumber\\
&=\frac{\mathfrak{q}^{1+2(M-N)}}{1-\mathfrak{q}^2}\prod_{n=1}^{N-1}\frac{1}{1-\mathfrak{q}^{4n}}\prod_{n=1}^{M}\frac{1}{1-\mathfrak{q}^{4n}},
\end{align}
where $\mathcal{I}_{\textrm{$\frac12$BPS}}^{\mathfrak{so}(2N+1)}(\mathfrak{q})$ and 
$\mathcal{I}_{\textrm{$\frac12$BPS}}^{\mathfrak{usp}(2M)}(\mathfrak{q})$ are the half-BPS indices 
of 4d $\mathcal{N}=4$ $SO(2N+1)$ and $USp(2M)$ gauge theories \cite{Caputa:2013vla,Hatsuda:2024lcc}. 
The normalized line defect half-index is
\begin{align}
\label{NhC_so2N+1|usp2M_V}
{\langle \mathcal{V}_{(1,0^{N-1});0}\rangle_{\mathcal{N}}^{\mathfrak{so}(2N+1)|\mathfrak{usp}(2M)'}}^{(C)}(\mathfrak{q})
&=\mathfrak{q}^{1+2(M-N)}(1+\mathfrak{q}^{2}+\mathfrak{q}^4+\cdots+\mathfrak{q}^{4N-2}). 
\end{align}
The line defect Witten index is
\begin{align}
{\langle \mathcal{V}_{(1,0^{N-1});0}\rangle_{\mathcal{N}}^{\mathfrak{so}(2N+1)|\mathfrak{usp}(2M)'}}^{(C)}(1)
&=2N. 
\end{align}

%%%%%%%%%%%%%%%%%%%%%%%%%%%%%%%%%%%%%%%%%%%%%%%%%
\subsubsection{Higgs limit}
%%%%%%%%%%%%%%%%%%%%%%%%%%%%%%%%%%%%%%%%%%%%%%%%%
%Higgs limit
On the other hand, the expression (\ref{h_so2N+1|usp2M_V}) vanishes in the Higgs limit
\begin{align}
\label{hH_so2N+1|usp2M_V}
{\langle V_{(1,0^{N-1});0}\rangle_{\mathcal{N}}^{\mathfrak{so}(2N+1)|\mathfrak{usp}(2M)'}}^{(H)}(\mathfrak{q})
&=0. 
\end{align}
This implies that the corresponding vortex line does not form a gauge invariant operator 
together with the bifundamental half-hypermultiplet localized on the interface. 

%%%%%%%%%%%%%%%%%%%%%%%%%%%%%%%%%%%%%%%%%%%%%%%%%
%%%%%%%%%%%%%%%%%%%%%%%%%%%%%%%%%%%%%%%%%%%%%%%%%
\section{$\mathfrak{usp}(2N)|\mathfrak{usp}(2M)'$}
\label{sec_BC2}
%usp(2N)|usp(2m)' interface
%%%%%%%%%%%%%%%%%%%%%%%%%%%%%%%%%%%%%%%%%%%%%%%%%
%%%%%%%%%%%%%%%%%%%%%%%%%%%%%%%%%%%%%%%%%%%%%%%%%
The NS5-type $\mathfrak{so}(2N+1)|\mathfrak{usp}(2M)'$ interface is S-dual to the D5-type $\mathfrak{usp}(2N)|\mathfrak{usp}(2M)'$ interface. 
When $N=M$, the $\mathfrak{usp}(2N)\oplus \mathfrak{usp}(2N)$ gauge algebra breaks down to $\mathfrak{usp}(2N)$ at the interface. 
The interface involves the half-hypermultiplet transforming in the fundamental representation under the $USp(2N)$ gauge group. 
For $N\neq M$, the $\mathfrak{usp}(2N)\oplus \mathfrak{usp}(2M)$ gauge algebra reduces to $\mathfrak{usp}(2 \min(N,M))$ and there is a Nahm pole of rank $|N-M|$. 
Though there is no ordinary half-hypermultiplet, there remain extra matter field degrees of freedom transforming in the fundamental representation at the interface \cite{Gaiotto:2019jvo,Hatsuda:2024lcc}. 
The D5-type interface is realized as a half D5-brane on which 
$N$ D3-branes in the presence of an O3$^+$-plane in $x^6<0$ and $M$ D3-branes together with an $\widetilde{\textrm{O3}}^+$-plane in $x^6>0$ end (see Figure \ref{fig_so2N+1|usp2M'}).

%%%%%%%%%%%%%%%%%%%%%%%%%%%%%%%%%%%%%%%%%%%%%%%%%
\subsection{Wilson lines}
\label{sec_BCD5W}
%%%%%%%%%%%%%%%%%%%%%%%%%%%%%%%%%%%%%%%%%%%%%%%%%
%Wilson
The half-BPS Wilson lines at the D5-type $\mathfrak{usp}(2N)|\mathfrak{usp}(2M)'$ interface are labeled by 
a representation under the gauge group $USp(2 \min(N,M))$. 

%half-index
The line defect half-index of the Wilson lines in the representation $\mathcal{R}_j$ with $j=1,\cdots, k$  
for the D5-type $\mathfrak{usp}(2N)|\mathfrak{usp}(2M)'$ interface is given by the following matrix integral: 
\begin{align}
\label{h_usp2N|usp2M_W}
&
\langle W_{\mathcal{R}_1}\cdots W_{\mathcal{R}_k} \rangle_{\mathcal{D}}^{\mathfrak{usp}(2N)|\mathfrak{usp}(2M)'}(t;q)
\nonumber\\
&=\frac{1}{2^{\min(N,M)}\min(N,M)!}
\frac{(q)_{\infty}^{2\min(N,M)}}{(q^{\frac12}t^{\pm2};q)_{\infty}^{\min(N,M)}}
\oint \prod_{i=1}^{\min(N,M)}
\frac{ds_i}{2\pi is_i}
\frac{(s_i^{\pm2};q)_{\infty} (qs_i^{\pm 2};q)_{\infty}}
{(q^{\frac12}t^2 s_i^{\pm2};q)_{\infty}(q^{\frac12}t^{-2} s_i^{\pm2};q)_{\infty}}
\nonumber\\
&\times 
\prod_{i<j}
\frac{(s_i^{\pm}s_j^{\pm};q)_{\infty}(qs_i^{\pm}s_j^{\pm};q)_{\infty}}
{(q^{\frac12}t^2s_i^{\pm}s_j^{\pm};q)_{\infty} (q^{\frac12}t^{-2}s_i^{\pm}s_j^{\pm};q)_{\infty}}
\prod_{n=1}^{|N-M|}
\frac{(q^{n+\frac12}t^{4n-2};q)_{\infty}}
{(q^n t^{4n};q)_{\infty}}
\nonumber\\
&\times 
\prod_{i=1}^{\min(N,M)}
\frac{(q^{\frac34+\frac{|N-M|}{2}}t^{-1+2|N-M|}s_i^{\pm};q)_{\infty}}
{(q^{\frac14+\frac{|N-M|}{2}}t^{1+2|N-M|}s_i^{\pm};q)_{\infty}}
\prod_{j=1}^{k}\chi_{\mathcal{R}_j}^{\mathfrak{usp}(2 \min (N,M))}(s). 
\end{align}

%%%%%%%%%%%%%%%%%%%%%%%%%%%%%%%%%%%%%%%%%%%%%%%%%
\subsubsection{S-duality}
%%%%%%%%%%%%%%%%%%%%%%%%%%%%%%%%%%%%%%%%%%%%%%%%%
%S-dual
We are now ready to discuss S-duality of the line operators at the interface. 
Under S-duality the Wilson line in the fundamental representation of 4d $\mathcal{N}=4$ $USp(2N)$ SYM theory 
maps to the minuscule vortex line with $k^{(1)}=(1,0^{N-1})$ for 4d $\mathcal{N}=4$ $SO(2N+1)$ SYM theory. 

In the D5-type interface, the $\mathfrak{usp}(2N)$ gauge algebra survives only for $N\le M$ and is broken when $N>M$.
It is therefore natural to expect that, when $N\le M$, the $\mathfrak{usp}(2N)$ fundamental Wilson line 
at the D5-type $\mathfrak{usp}(2N)|\mathfrak{usp}(2M)'$ interface is mapped under S-duality to the minuscule vortex line with $k^{(1)}=(1,0^{N-1})$ 
at the NS5-type $\mathfrak{so}(2N+1)|\mathfrak{usp}(2M)'$ interface.
We conjecture that 
the two matrix integrals (\ref{h_so2N+1|usp2M_V}) and (\ref{h_usp2N|usp2M_W}) are equivalent 
\begin{align}
\langle V_{(1,0^{N-1});0}\rangle_{\mathcal{N}}^{\mathfrak{so}(2N+1)|\mathfrak{usp}(2M)'}(t;q)
&=\langle W_{\tiny \yng(1)}\rangle_{\mathcal{D}}^{\mathfrak{usp}(2N)|\mathfrak{usp}(2M)'}(t^{-1};q)
\end{align}
for $N\le M$. 
We have explicitly confirmed the agreement of the line defect half-indices for various low-rank examples. 
The $q$-series expansions are collected in Appendix \ref{app_expansion}. 

%N>M
For $N>M$, the $USp(2N)$ gauge group of the D5-type $\mathfrak{usp}(2N)|\mathfrak{usp}(2M)'$ interface is broken, 
and therefore the expressions (\ref{h_so2N+1|usp2M_V}) and (\ref{h_usp2N|usp2M_W}) no longer coincide.

%%%%%%%%%%%%%%%%%%%%%%%%%%%%%%%%%%%%%%%%%%%%%%%%%
\subsubsection{Coulomb limit}
%%%%%%%%%%%%%%%%%%%%%%%%%%%%%%%%%%%%%%%%%%%%%%%%%
%Coulomb limits
Let us examine the Coulomb limit of the line defect half-index (\ref{h_usp2N|usp2M_W}). 
It is shown to coincide with the half-BPS limit of the line defect index for $USp(2\min (N,M))$ gauge theory
\begin{align}
\label{hC_usp2N|usp2M_W}
{\langle W_{\mathcal{R}_1}\cdots W_{\mathcal{R}_k} \rangle_{\mathcal{D}}^{\mathfrak{usp}(2N)|\mathfrak{usp}(2M)'}}^{(C)}(\mathfrak{q})
&=\langle W_{\mathcal{R}_1}\cdots W_{\mathcal{R}_k}\rangle_{\textrm{$\frac12$BPS}}^{\mathfrak{usp}(2\min (N,M))}(\mathfrak{q}).
\end{align}
In particular, when we consider the one-point function of the Wilson line $W_{\mathcal{R}}$ in a non-trivial irreducible representation $\mathcal{R}$ 
without zero weight, e.g. the fundamental representation, 
the line defect half-index (\ref{hC_usp2N|usp2M_W}) vanishes \cite{Hatsuda:2025jze}. 
More generally, the normalized one-point function is given by (\ref{halfBPS_K}), 
which is non-vanishing if and only if $\mathcal{R}$ contains the zero weight. 
For example, for the adjoint representation of $\mathfrak{usp}(2)$ one finds 
$\langle \mathcal{W}_{\tiny \yng(2)} \rangle_{\textrm{$\frac12$BPS}}^{\mathfrak{usp}(2)}(\mathfrak{q})=\mathfrak{q}^2$. 
For $N\le M$, the vanishing of the one-point function of the fundamental Wilson line is consistent with the fact that 
the Higgs limit of the proposed S-dual minuscule vortex line defect half-index for the $\mathfrak{so}(2N+1)|\mathfrak{usp}(2M)'$ interface also vanishes 
(see (\ref{hH_so2N+1|usp2M_V})).

%Witten ind
The line defect Witten index effectively follows from the half-BPS limit of the line defect index for $USp(2\min (N,M))$ gauge theory. 
We propose that the line defect Witten index can be extracted as the constant term in the product of characters
\begin{align}
{\langle \mathcal{W}_{\mathcal{R}_1}\cdots \mathcal{W}_{\mathcal{R}_k} \rangle_{\mathcal{D}}^{\mathfrak{usp}(2N)|\mathfrak{usp}(2M)'}}^{(C)}(1)
&=\oint \prod_{i=1}^{\min (N,M)} \frac{ds_i}{2\pi is_i} 
\prod_{j=1}^k \chi_{\mathcal{R}_j}^{\mathfrak{usp}(2\min (N,M))}(s). 
\end{align}
This follows from (\ref{halfBPS_K}) and Kostant's theorem \cite{MR158024} 
that the multiplicity of $\mathcal{R}$ in the harmonic polynomials on $\mathfrak{g}$ is $\dim \mathcal{R}[0]$, i.e. $K_{\lambda,0}(1)=\dim \mathcal{R}[0]$. 
Physically, the integral localizes onto the maximal torus, 
which is the residual gauge group at a generic point of the Coulomb branch (see Appendix \ref{app_residual}). 

%%%%%%%%%%%%%%%%%%%%%%%%%%%%%%%%%%%%%%%%%%%%%%%%%
\subsubsection{Higgs limit}
%%%%%%%%%%%%%%%%%%%%%%%%%%%%%%%%%%%%%%%%%%%%%%%%%
In contrast to the Coulomb limit, 
the Higgs limit is not equivalent to the line defect index of 4d $\mathcal{N}=4$ $USp(2 \min(N,M))$ SYM theory. 
In the Higgs limit the matrix integral (\ref{h_usp2N|usp2M_W}) reduces to 
\begin{align}
\label{hH_usp2N|usp2M_W}
&
{\langle W_{\mathcal{R}_1}\cdots W_{\mathcal{R}_k} \rangle_{\mathcal{D}}^{\mathfrak{usp}(2N)|\mathfrak{usp}(2M)'}}^{(H)}(\mathfrak{q})
\nonumber\\
&=\frac{1}{2^{\min(N,M)}\min(N,M)!}
\frac{1}{(1-\mathfrak{q}^2)^{\min(N,M)}}
\oint \prod_{i=1}^{\min(N,M)}
\frac{ds_i}{2\pi is_i}
\frac{(1-s_i^{\pm2})}{(1-\mathfrak{q}^2s_i^{\pm2})}
\nonumber\\
&\times 
\prod_{i<j}\frac{(1-s_i^{\pm}s_j^{\pm})}
{(1-\mathfrak{q}^2s_i^{\pm}s_j^{\pm})}
\prod_{n=1}^{|N-M|}
\frac{1}{1-\mathfrak{q}^{4n}}
\prod_{i=1}^{\min(N,M)}
\frac{1}{1-\mathfrak{q}^{1+2|N-M|}s_i^{\pm}}
\prod_{j=1}^{k}\chi_{\mathcal{R}_j}^{\mathfrak{usp}(2\min (N,M))}(s). 
\end{align}
In the absence of the line operator the Higgs half-index is \cite{Hatsuda:2024lcc}
\begin{align}
{\mathbb{II}_{\mathcal{D}}^{\mathfrak{usp}(2N)|\mathfrak{usp}(2M)'}}^{(H)}(\mathfrak{q})
&=\prod_{n=1}^{N}\frac{1}{1-\mathfrak{q}^{4n}}\prod_{m=1}^{M}\frac{1}{1-\mathfrak{q}^{4m}}. 
\end{align}
As expected, this coincides with the Coulomb limit of the interface half-index 
of the S-dual NS5-type $\mathfrak{so}(2N+1)|\mathfrak{usp}(2M)'$ interface given in section \ref{sec_BCNS5W}, 
since the exchange $t\rightarrow t^{-1}$ under S-duality exchanges the Coulomb and Higgs limits. 
Conversely, $\prod_{n=1}^{\min (N,M)}(1-\mathfrak{q}^{4n})^{-1}$ is the Coulomb limit of the D5-type interface half-index, 
which is dual to the Higgs limit of the NS5-type one. 
Let us perform several explicit computations for lower-rank gauge groups. 
For the $\mathfrak{usp}(2)|\mathfrak{usp}(2)'$ interface, the insertion of a Wilson line transforming in the rank-$k$ symmetric representation yields the following result:
\begin{align}
{\langle \mathcal{W}_{(k)} \rangle_{\mathcal{D}}^{\mathfrak{usp}(2)|\mathfrak{usp}(2)'}}^{(H)}(\mathfrak{q})
&=\left\lceil \frac{k+1}{2}\right\rceil  \mathfrak{q}^k+\left\lfloor \frac{k+1}{2}\right\rfloor \mathfrak{q}^{k+2}. 
\end{align}
For example, 
\begin{align}
{\langle \mathcal{W}_{\tiny \yng(1)} \rangle_{\mathcal{D}}^{\mathfrak{usp}(2)|\mathfrak{usp}(2)'}}^{(H)}(\mathfrak{q})
&=\mathfrak{q}+\mathfrak{q}^3, \\
{\langle \mathcal{W}_{\tiny \yng(2)} \rangle_{\mathcal{D}}^{\mathfrak{usp}(2)|\mathfrak{usp}(2)'}}^{(H)}(\mathfrak{q})
&=2\mathfrak{q}^2+\mathfrak{q}^4, \\
{\langle \mathcal{W}_{\tiny \yng(3)} \rangle_{\mathcal{D}}^{\mathfrak{usp}(2)|\mathfrak{usp}(2)'}}^{(H)}(\mathfrak{q})
&=2\mathfrak{q}^3+2\mathfrak{q}^5, \\
{\langle \mathcal{W}_{\tiny \yng(4)} \rangle_{\mathcal{D}}^{\mathfrak{usp}(2)|\mathfrak{usp}(2)'}}^{(H)}(\mathfrak{q})
&=3\mathfrak{q}^4+2\mathfrak{q}^6. 
\end{align}

%Witten ind
We conjecture that the line defect Witten index is given by
\begin{align}
{\langle \mathcal{W}_{\mathcal{R}_1}\cdots \mathcal{W}_{\mathcal{R}_k} \rangle_{\mathcal{D}}^{\mathfrak{usp}(2N)|\mathfrak{usp}(2M)'}}^{(H)}(1)
&=\prod_{j=1}^k \chi_{\mathcal{R}_j}^{\mathfrak{usp}(2\min (N,M))}(1), 
\end{align}
where 
\begin{align}
\chi_{\mathcal{R}}^{\mathfrak{usp}(2n)}(1)
&=\prod_{1\le i<j\le n}\frac{\lambda_i-\lambda_j+j-i}{j-i}
\prod_{1\le i\le j\le n}
\frac{\lambda_i+\lambda_j+2n+2-i-j}{2n+2-i-j}
\end{align}
is the dimension of the $USp(2n)$ irreducible representation labeled by the partition $\lambda$. 
This conjecture is naturally understood in terms of the residual gauge group. 
The Higgs limit (\ref{hH_usp2N|usp2M_W}) is a Molien-Weyl integral for $USp(2\min(N,M))$ 
acting on the adjoint representation with weight $\mathfrak{q}^2$ and on the fundamental representation with weight $\mathfrak{q}^{1+2|N-M|}$. 
At a generic point the adjoint scalar breaks the gauge group to the maximal torus and the fundamental scalar breaks it completely. 
Since the action is stable, the residual gauge group is trivial and (\ref{WI_residual_intro}) gives $\dim \mathcal{R}$ (see Appendix \ref{app_residual}). 
We have also confirmed the conjecture for non-minuscule representations, e.g. 
\begin{align}
{\langle \mathcal{W}_{\tiny \yng(1,1)} \rangle_{\mathcal{D}}^{\mathfrak{usp}(4)|\mathfrak{usp}(4)'}}^{(H)}(\mathfrak{q})
&=2\mathfrak{q}^4+\mathfrak{q}^6+\mathfrak{q}^8+\mathfrak{q}^{10}, \\
{\langle \mathcal{W}_{\tiny \yng(2)} \rangle_{\mathcal{D}}^{\mathfrak{usp}(4)|\mathfrak{usp}(4)'}}^{(H)}(\mathfrak{q})
&=2\mathfrak{q}^2+\mathfrak{q}^4+3\mathfrak{q}^6+2\mathfrak{q}^8+\mathfrak{q}^{10}+\mathfrak{q}^{12}, 
\end{align}
whose values at $\mathfrak{q}=1$ are $5$ and $10$. 

%%%%%%%%%%%%%%%%%%%%%%%%%%%%%%%%%%%%%%%%%%%%%%%%%
\subsection{Vortex lines}
\label{sec_BCD5V}
%%%%%%%%%%%%%%%%%%%%%%%%%%%%%%%%%%%%%%%%%%%%%%%%%
%vortex
The D5-type $\mathfrak{usp}(2N)|\mathfrak{usp}(2M)'$ interface preserves the symplectic gauge algebra. 
In gauge theories with symplectic gauge groups, magnetic charges are generally associated with non-minuscule weights 
unless one considers vortex lines carrying half-integer spinorial magnetic charges. 
Therefore the vortex line defect half-index is expected to involve intricate monopole bubbling contributions, making its general structure rather complicated. 
In the following we instead propose the closed-form expression in the distinguished case with $N=0$, 
corresponding to the dual NS5-type interface with the $SO(1)$ theory, 
for which we find a remarkably simple but non-trivial expression for the vortex line defect half-index.

%%%%%%%%%%%%%%%%%%%%%%%%%%%%%%%%%%%%%%%%%%%%%%%%%
\subsubsection{S-duality}
%%%%%%%%%%%%%%%%%%%%%%%%%%%%%%%%%%%%%%%%%%%%%%%%%
When $N=0$, the D5-type interface itself is expected to be described by   
the $USp(2M)'$ SYM theory defined on a half-space obeying the Nahm pole boundary condition 
as a dual configuration of the NS5-type $SO(1)|USp(2M)'$ interface. 
Indeed, in the absence of line defect insertions, 
the NS5-type $SO(1)|USp(2M)'$ interface half-index agrees with the $USp(2M)'$ Nahm pole half-index \cite{Hatsuda:2024lcc}.\footnote{In \cite{Hatsuda:2024lcc} the NS5-type $SO(1)|USp(2M)'$ interface half-index is evaluated as the modified Neumann half-index for the $USp(2M)'$ SYM theory. }
Motivated by this duality dictionary, 
we investigate how the $USp(2M)'$ Nahm pole boundary condition is modified 
when the $USp(2M)'$ fundamental Wilson line is inserted on the dual $SO(1)|USp(2M)'$ interface side. 
We find that the line defect half-index (\ref{h_so1|usp2M_W}) agrees with the following expression: 
\begin{align}
\label{h_so1|usp2M_W_closed}
\langle W_{0;\tiny \yng(1)} \rangle_{\mathcal{N}}^{\mathfrak{so}(1)|\mathfrak{usp}(2M)'}(t;q)
&=
q^{\frac14}t 
\frac{1-q^{M}t^{-4M}}{1-q^{M+\frac12}t^{-4M+2}}
\mathbb{II}_{\textrm{Nahm}}^{USp(2M)'}(t;q), 
\end{align}
where 
\begin{align}
\mathbb{II}_{\textrm{Nahm}}^{USp(2M)'}(t;q)
&=\prod_{k=1}^{M}\frac{(q^{\frac{2k+1}{2}} t^{-4k+2};q)_{\infty}}{(q^k t^{-4k};q)_{\infty}}
\end{align}
is the half-index of the $USp(2M)'$ Nahm pole boundary condition \cite{Hatsuda:2024lcc}. 
Physically, the additional rational function factor in (\ref{h_so1|usp2M_W_closed}) will capture the non-perturbative vortex degrees of freedom trapped at the D5-type interface. 

The expression (\ref{h_so1|usp2M_W_closed}) can be established rigorously. 
As we show in Appendix \ref{app_koornwinder}, 
the matrix integral (\ref{h_so1|usp2M_W}) is a Koornwinder integral of type $BC_M$ 
for a specific choice of the Askey-Wilson parameters, 
so that Gustafson's evaluation formula and the orthogonality of the Koornwinder polynomials 
determine the half-index for an arbitrary representation $\lambda$ of $\mathfrak{usp}(2M)$ 
in the closed form (\ref{app_master2}). 

It is worth noting that 
there are also interesting cases in which the half-index with a non-minuscule Wilson line can be computed exactly. 
For example, for the rank-two antisymmetric representation, we find
\begin{align}
\label{h_so1|usp2M_Wasym2_closed}
\langle W_{0;\tiny \yng(1,1)} \rangle_{\mathcal{N}}^{\mathfrak{so}(1)|\mathfrak{usp}(2M)'}(t;q)
&=qt^{-4}\frac{1-q^{M-1}t^{-4M+4}}{1-qt^{-4}}
\frac{1-q^{\frac12}t^2}{1-q^{M+\frac12}t^{-4M+2}}
\mathbb{II}_{\textrm{Nahm}}^{USp(2M)'}(t;q). 
\end{align}
It would be interesting to find the closed-form expression for a general representation, 
which by (\ref{app_master2}) amounts to the evaluation of a single Koornwinder expansion coefficient. 
We leave it as interesting future work to explore more general vortex line defect half-indices in the D5-type $\mathfrak{usp}(2N)|\mathfrak{usp}(2M)'$ interface 
by taking into account the monopole bubbling indices \cite{Ito:2011ea,Hayashi:2020ofu,Brennan:2018yuj,Brennan:2018rcn,Hayashi:2019rpw}.

%%%%%%%%%%%%%%%%%%%%%%%%%%%%%%%%%%%%%%%%%%%%%%%%%
%%%%%%%%%%%%%%%%%%%%%%%%%%%%%%%%%%%%%%%%%%%%%%%%%
\section{$\mathfrak{so}(2N)|\mathfrak{usp}(2M)$}
\label{sec_DC1}
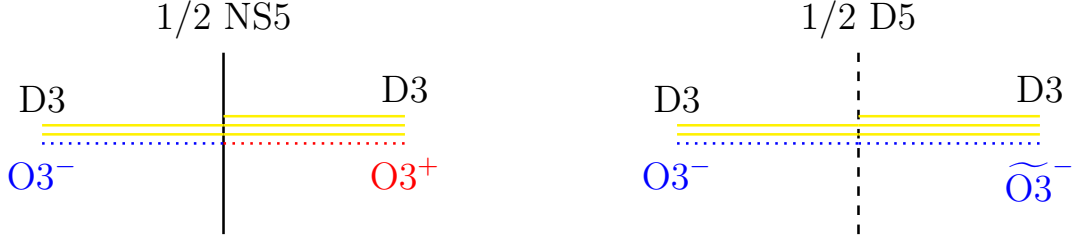
\begin{figure}
\centering
\scalebox{1.2}{
\begin{tikzpicture}
\draw[thick,solid,black] (-3.5,1) -- (-3.5,3) node[above] {\textrm{1/2 NS5}};
\draw[thick,dashed,black] (3.5,1) -- (3.5,3) node[above] {\textrm{1/2 D5}};
\draw[thick,dotted,blue] (-5.5,2) node[below] {$\textrm{O3}^-$} -- (-3.5,2);
\draw[thick,dotted,red] (-3.5,2) -- (-1.5,2) node[below] {$\textrm{O3}^+$};
\draw[thick,dotted,blue] (1.5,2) node[below] {$\textrm{O3}^-$} -- (3.5,2);
\draw[thick,dotted,blue] (3.5,2) -- (5.5,2) node[below] {$\widetilde{\textrm{O3}}^-$};
\draw[thick,solid,yellow] (-5.5,2.2) node[above] {\textcolor{black}{\textrm{D3}}} -- (-3.5,2.2);
\draw[thick,solid,yellow] (-5.5,2.1)  -- (-3.5,2.1);
\draw[thick,solid,yellow] (-3.5,2.3)  -- (-1.5,2.3) node[above] {\textcolor{black}{\textrm{D3}}};
\draw[thick,solid,yellow] (-3.5,2.2)  -- (-1.5,2.2);
\draw[thick,solid,yellow] (-3.5,2.1)  -- (-1.5,2.1);
\draw[thick,solid,yellow] (1.5,2.2) node[above] {\textcolor{black}{\textrm{D3}}} -- (3.5,2.2);
\draw[thick,solid,yellow] (1.5,2.1)  -- (3.5,2.1);
\draw[thick,solid,yellow] (3.5,2.3)  -- (5.5,2.3) node[above] {\textcolor{black}{\textrm{D3}}};
\draw[thick,solid,yellow] (3.5,2.2)  -- (5.5,2.2);
\draw[thick,solid,yellow] (3.5,2.1)  -- (5.5,2.1);
\end{tikzpicture}
}
\caption{\textrm{
The S-dual brane configurations for the NS5-type $\mathfrak{so}(2N)|\mathfrak{usp}(2M)$ interface and D5-type $\mathfrak{so}(2N)|\mathfrak{so}(2M+1)$ interface.  
}}
\label{fig_o2N|usp2M}
\end{figure}

%%%%%%%%%%%%%%%%%%%%%%%%%%%%%%%%%%%%%%%%%%%%%%%%%
%%%%%%%%%%%%%%%%%%%%%%%%%%%%%%%%%%%%%%%%%%%%%%%%%
Let us examine the NS5-type interface between 4d $\mathcal{N}=4$ gauge theories with gauge algebras $\mathfrak{so}(2N)$ and $\mathfrak{usp}(2M)$. 
At the interface there is a 3d half-hypermultiplet transforming in the bifundamental representation under a pair of orthogonal and symplectic gauge groups. 
It can be constructed in Type IIB string theory by introducing a half NS5-brane at $x^6=0$ on which $N$ D3-branes in the background of an O3$^-$-plane for $x^6<0$ 
and $M$ D3-branes in the background of an O3$^+$-plane for $x^6>0$ end. 
The brane configuration is illustrated in Figure \ref{fig_o2N|usp2M}. 
%
%
%
%
%
%%%%%%%%%%%%%%%%%%%%%%%%%%%%%%%%%%%%%%%%%%%%%%%%%
\subsection{Wilson lines}
\label{sec_DCNS5W}
%%%%%%%%%%%%%%%%%%%%%%%%%%%%%%%%%%%%%%%%%%%%%%%%%
The Wilson line at the NS5-type $\mathfrak{so}(2N)|\mathfrak{usp}(2M)$ interface can be labeled by a pair $(\mathcal{R}_1,\mathcal{R}_2)$ of representations 
under the gauge groups associated with the Lie algebras $\mathfrak{so}(2N)$ and $\mathfrak{usp}(2M)$. 

%half-index
The line defect half-index of the Wilson line operator $W_{\mathcal{R}_1;\mathcal{R}_2}$ at the NS5-type $SO(2N)|USp(2M)$ interface 
can be evaluated from the matrix integral of the form
\begin{align}
\label{h_so2N|usp2M_W}
&
\langle W_{\mathcal{R}_1;\mathcal{R}_2}\rangle_{\mathcal{N}}^{\mathfrak{so}(2N)|\mathfrak{usp}(2M)}(t;q)
\nonumber\\
&=\frac{1}{2^{N-1}N!} \frac{(q)_{\infty}^{N}}{(q^{\frac12}t^{-2};q)_{\infty}^N}
\oint \prod_{i=1}^N \frac{ds_i^{(1)}}{2\pi is_i^{(1)}}
\prod_{i<j}
\frac{(s_i^{(1)\pm}s_j^{(1)\pm};q)_{\infty}}
{(q^{\frac12}t^{-2}s_i^{(1)\pm}s_j^{(1)\pm};q)_{\infty}}
\nonumber\\
&\times 
\frac{1}{2^M M!} \frac{(q)_{\infty}^M}{(q^{\frac12}t^{-2};q)_{\infty}^M}
\oint \prod_{i=1}^M \frac{ds_i^{(2)}}{2\pi is_i^{(2)}}
\frac{(s_i^{(2)\pm2};q)_{\infty}}
{(q^{\frac12}t^{-2}s_i^{(2)\pm2};q)_{\infty}}
\prod_{i<j}
\frac{(s_i^{(2)\pm}s_j^{(2)\pm};q)_{\infty}}
{(q^{\frac12}t^{-2}s_i^{(2)\pm}s_j^{(2)\pm};q)_{\infty}}
\nonumber\\
&\times 
\prod_{i=1}^N\prod_{j=1}^M
\frac{(q^{\frac34}t^{-1}s_i^{(1)\pm}s_j^{(2)\pm};q)_{\infty}}
{(q^{\frac14}ts_i^{(1)\pm}s_j^{(2)\pm};q)_{\infty}}
\chi_{\mathcal{R}_1}^{\mathfrak{so}(2N)}(s^{(1)})
\chi_{\mathcal{R}_2}^{\mathfrak{usp}(2M)}(s^{(2)}). 
\end{align}
Here $\chi_{\mathcal{R}_1}^{\mathfrak{so}(2N)}(s^{(1)})$ and $\chi_{\mathcal{R}_2}^{\mathfrak{usp}(2M)}(s^{(2)})$ are the character of the representation $\mathcal{R}_1$ of $\mathfrak{so}(2N)$ and the character of the representation $\mathcal{R}_2$ of $\mathfrak{usp}(2M)$ respectively. 

%O(2N)
The $SO(2N)$ gauge group can be replaced by the disconnected gauge group $O(2N)$, 
which has two connected components distinguished by the determinant $\det g$ $=$ $+1$ or $-1$ where $g\in O(2N)$. 
The connected component with $\det g=+1$ is $SO(2N)$. 
The other connected component with $\det g=-1$ is the twisted sector, which we denote by $SO(2N)^-$. 
However, it is not a group as the product of two elements has determinant $+1$ and the identity is not included in it. 
For $O(2N)^+$ the twisted sectors contribute with positive weight, while for $O(2N)^-$ they contribute with a relative minus sign. 
For the NS5-type $O(2N)^{\pm}|USp(2M)$ interface, the line defect half-index is obtained by
\begin{align}
\label{h_o2N|usp2M_W}
&
\langle W_{\mathcal{R}_1;\mathcal{R}_2}\rangle_{\mathcal{N}}^{\mathfrak{o}(2N)^{\pm}|\mathfrak{usp}(2M)}(t;q)
\nonumber\\
&=\frac12
\left(
\langle W_{\mathcal{R}_1;\mathcal{R}_2}\rangle_{\mathcal{N}}^{\mathfrak{so}(2N)|\mathfrak{usp}(2M)}(t;q)
\pm \langle W_{\mathcal{R}_1;\mathcal{R}_2}\rangle_{\mathcal{N}}^{\mathfrak{so}(2N)^-|\mathfrak{usp}(2M)}(t;q)
\right), 
\end{align}
where the contribution $\langle W_{\mathcal{R}_1;\mathcal{R}_2}\rangle_{\mathcal{N}}^{\mathfrak{so}(2N)^-|\mathfrak{usp}(2M)}$ 
from the disconnected component $SO(2N)^-$ is evaluated as
\begin{align}
\label{h_so2N-|usp2M_W}
&
\langle W_{\mathcal{R}_1;\mathcal{R}_2}\rangle_{\mathcal{N}}^{\mathfrak{so}(2N)^{-}|\mathfrak{usp}(2M)}(t;q)
\nonumber\\
&=\frac{1}{2^{N-1}(N-1)!}
\frac{(q)_{\infty}^{N-1}(-q;q)_{\infty}}
{(q^{\frac12}t^{-2};q)_{\infty}^{N-1}(-q^{\frac12}t^{-2};q)_{\infty}}
\oint \prod_{i=1}^{N-1}
\frac{ds_i^{(1)}}{2\pi is_i^{(1)}}
\nonumber\\
&\times 
\prod_{i=1}^{N-1}
\frac{(s_i^{(1)\pm};q)_{\infty}(-s_i^{(1)\pm};q)_{\infty}}
{(q^{\frac12}t^{-2}s_i^{(1)\pm};q)_{\infty} (-q^{\frac12}t^{-2}s_i^{(1)\pm};q)_{\infty}}
\prod_{i<j}
\frac{(s_i^{(1)\pm}s_j^{(1)\pm};q)_{\infty}}
{(q^{\frac12}t^{-2}s_i^{(1)\pm}s_j^{(1)\pm};q)_{\infty}}
\nonumber\\
&\times 
\frac{1}{2^M M!}\frac{(q)_{\infty}^M}{(q^{\frac12}t^{-2};q)_{\infty}^M}
\oint \prod_{i=1}^M \frac{ds_i^{(2)}}{2\pi is_i^{(2)}}
\frac{(s_i^{(2)\pm2};q)_{\infty}}{(q^{\frac12}t^{-2}s_i^{(2)\pm2};q)_{\infty}}
\prod_{i<j}\frac{
(s_i^{(2)\pm}s_j^{(2)\pm};q)_{\infty}
}
{
(q^{\frac12}t^{-2}s_i^{(2)\pm}s_j^{(2)\pm};q)_{\infty}
}
\nonumber\\
&\times 
\prod_{i=1}^{N-1}\prod_{j=1}^{M}
\frac{
(q^{\frac34}t^{-1}s_i^{(1)\pm}s_j^{(2)\pm};q)_{\infty}
}
{
(q^{\frac14}ts_i^{(1)\pm}s_j^{(2)\pm};q)_{\infty}
}
\prod_{j=1}^M
\frac{
(q^{\frac34}t^{-1}s_j^{(2)\pm};q)_{\infty}
(-q^{\frac34}t^{-1}s_j^{(2)\pm};q)_{\infty}
}
{
(q^{\frac14}ts_j^{(2)\pm};q)_{\infty}
(-q^{\frac14}ts_j^{(2)\pm};q)_{\infty}
}
\nonumber\\
&\times 
\chi_{\mathcal{R}_1}^{\mathfrak{usp}(2N-2)}(s^{(1)})
\chi_{\mathcal{R}_2}^{\mathfrak{usp}(2M)}(s^{(2)}). 
\end{align}
Here $\chi_{\mathcal{R}_1}^{\mathfrak{usp}(2N-2)}(s^{(1)})$ and $\chi_{\mathcal{R}_2}^{\mathfrak{usp}(2M)}(s^{(2)})$ are 
the character of the representation $\mathcal{R}_1$ of $\mathfrak{usp}(2N-2)$ 
and the character of the representation $\mathcal{R}_2$ of $\mathfrak{usp}(2M)$ respectively. 

%so(2N)
The Wilson lines transforming under the right symplectic gauge group associated with $\mathfrak{usp}(2M)$ 
behave similarly to those in the NS5-type $\mathfrak{so}(2N+1)|\mathfrak{usp}(2M)'$ interface discussed in section \ref{sec_BC1}. 
Here we discuss representations of the left orthogonal gauge group associated with $\mathfrak{so}(2N)$.
For the gauge algebra $\mathfrak{so}(2N)$ we can take the $Spin(2N)$ gauge group, which has the center symmetry 
\begin{align}
Z&=\begin{cases}
\mathbb{Z}_2\times \mathbb{Z}_2&\textrm{for even $N$}\cr 
\mathbb{Z}_4&\textrm{for odd $N$}\cr 
\end{cases}. 
\end{align}
The possible choices of the gauge groups depend on whether $N$ is even or odd 
by gauging different subgroups of the center. 

%even N
For even $N$ the chiral spinor representation $S_+$ and the conjugate chiral spinor representation $S_-$ are inequivalent self-conjugate representations, which are real for $N\equiv 0$ mod $4$ and pseudoreal for $N\equiv 2$ mod $4$. 
The $\mathbb{Z}_2\times \mathbb{Z}_2$ center contains three distinct subgroups generated by $-1$, $\omega$ and $-\omega$
\begin{align}
\mathbb{Z}_2^{V}&=\langle-1\rangle,& 
\mathbb{Z}_2^{S}&=\langle \omega \rangle, & 
\mathbb{Z}_2^{C}&=\langle-\omega\rangle. 
\end{align}
These subgroups allow for five distinct global gauge groups via standard discrete quotients as follows \cite{Aharony:2013hda}: 
\begin{align}
&Spin(2N), \\
SO(2N)=&Spin(2N)/\mathbb{Z}_2^V, \\
Ss(2N)=&Spin(2N)/\mathbb{Z}_2^S, \\
Sc(2N)=&Spin(2N)/\mathbb{Z}_2^C, \\
SO(2N)/\mathbb{Z}_2=&Spin(2N)/(\mathbb{Z}_2^S\times \mathbb{Z}_2^C). 
\end{align}

%odd N
For odd $N$ the chiral spinor representation $S_+$ and the conjugate chiral spinor representation $S_-$ are complex conjugates, i.e. $S_-=\overline{S_+}$. 
There are three kinds of gauge groups corresponding to 
gauging a non-trivial $\mathbb{Z}_2$ subgroup generated by $\omega^2$ and $\mathbb{Z}_4$ \cite{Aharony:2013hda}
\begin{align}
&Spin(2N), \\
SO(2N)=&Spin(2N)/\mathbb{Z}_2, \\
SO(2N)/\mathbb{Z}_2=&Spin(2N)/\mathbb{Z}_4.
\end{align}

For $Spin(2N)$ gauge theory we have the spinor Wilson lines. 
%spinor character
The character of the chiral spinor representation is given by
\begin{align}
\chi_{\textrm{sp}}^{\mathfrak{so}(2N)}(s)
&=\frac12 \left[
\prod_{i=1}^N (s_i^{\frac12}+s_i^{-\frac12})
+ \prod_{i=1}^N (s_i^{\frac12}-s_i^{-\frac12})
\right], 
\end{align}
and that of the conjugate chiral spinor representation is
\begin{align}
\chi_{\overline{\textrm{sp}}}^{\mathfrak{so}(2N)}(s)
&=\frac12 \left[
\prod_{i=1}^N (s_i^{\frac12}+s_i^{-\frac12})
-\prod_{i=1}^N (s_i^{\frac12}-s_i^{-\frac12})
\right]. 
\end{align}
For $Spin(2N)$, the global structure of the center symmetry and its representation characters depend on the parity of $N$, as discussed above. 
Nevertheless, we obtain a parallel vanishing mechanism for an isolated spinor Wilson line defect half-index based on the failure of discrete charge neutralization at the interface.
The local operators appearing in the integrand of the half-index 
without a spinor Wilson line are completely invariant under the spinor inverting elements of the center subgroup. 
On the other hand, the chiral spinor representations $S_\pm$ carry non-trivial center charges 
acting as $-1$ under the respective chiral inversion generators for even $N$, and as $\pm i$ under the cyclic $\mathbb{Z}_4$ generator for odd $N$. 
Consequently, the gauge invariant singlet at the interface which must carry a trivial total center charge 
cannot be formed by inserting an odd number of spinor lines. 
Hence we have a selection rule according to the total center charge constraint:
\begin{align}
\langle W_{\textrm{sp};\mathcal{R}_2}\rangle_{\mathcal{N}}^{\mathfrak{so}(2N)|\mathfrak{usp}(2M)}(t;q)&=0, \\
\langle W_{\overline{\textrm{sp}};\mathcal{R}_2}\rangle_{\mathcal{N}}^{\mathfrak{so}(2N)|\mathfrak{usp}(2M)}(t;q)&=0. 
\end{align}
Here $\lambda=\textrm{sp}=(1/2,\dots,1/2,1/2)$ and $\lambda=\overline{\textrm{sp}}=(1/2,\dots,1/2,-1/2)$ 
denote the spinor representations carrying half-integer weights in the Young diagram notation.

%general character
More generally, the character of the irreducible representation of $\mathfrak{so}(2N)$ with highest weight 
corresponding to the Young diagram $\lambda=(\lambda_1,\dots, \lambda_N)$ 
with $\lambda_1 \geq \dots \geq \lambda_{N-1} \geq |\lambda_N| \geq 0$ is given by \cite{MR1153249}
\begin{align}
\label{ch_so2N_irrep}
\chi_{\lambda}^{\mathfrak{so}(2N)}
&=\frac{\det (s_j^{\lambda_i+N-i}+s_j^{-\lambda_i-N+i})+\det(s_j^{\lambda_i+N-i}-s_j^{-\lambda_i-N+i})
}{\det(s_j^{N-i}+s_j^{-N+i})}. 
\end{align}

%%%%%%%%%%%%%%%%%%%%%%%%%%%%%%%%%%%%%%%%%%%%%%%%%
\subsubsection{Coulomb limit}
%%%%%%%%%%%%%%%%%%%%%%%%%%%%%%%%%%%%%%%%%%%%%%%%%
%Coulomb limit
As in the NS5-type $\mathfrak{so}(2N+1)|\mathfrak{usp}(2M)'$ interface, 
the Coulomb limit factorizes into a product of the half-BPS line defect indices of the left $SO(2N)$ gauge theory and the right $USp(2M)$ gauge theory.
It follows that the matrix integrals (\ref{h_so2N|usp2M_W}) and (\ref{h_so2N-|usp2M_W}) are factorized as
\begin{align}
\label{hC_so2N|usp2M_W}
{\langle W_{\mathcal{R}_1;\mathcal{R}_2}\rangle_{\mathcal{N}}^{\mathfrak{so}(2N)|\mathfrak{usp}(2M)}}^{(C)}(\mathfrak{q})
&=\langle W_{\mathcal{R}_1}\rangle_{\textrm{$\frac12$BPS}}^{\mathfrak{so}(2N)}(\mathfrak{q})
\langle W_{\mathcal{R}_2}\rangle_{\textrm{$\frac12$BPS}}^{\mathfrak{usp}(2M)}(\mathfrak{q}), \\
\label{hC_so2N-|usp2M_W}
{\langle W_{\mathcal{R}_1;\mathcal{R}_2}\rangle_{\mathcal{N}}^{\mathfrak{so}(2N)^{-}|\mathfrak{usp}(2M)}}^{(C)}(\mathfrak{q})
&=\langle W_{\mathcal{R}_1}\rangle_{\textrm{$\frac12$BPS}}^{\mathfrak{so}(2N)^{-}}(\mathfrak{q})
\langle W_{\mathcal{R}_2}\rangle_{\textrm{$\frac12$BPS}}^{\mathfrak{usp}(2M)}(\mathfrak{q}), 
\end{align}
in the Coulomb limit. 
When no line operators are inserted, they are given by \cite{Hatsuda:2024lcc}
\begin{align}
\label{hC_so2N|usp2M}
{\mathbb{II}_{\mathcal{N}}^{\mathfrak{so}(2N)|\mathfrak{usp}(2M)}}^{(C)}(\mathfrak{q})
&=\frac{1}{1-\mathfrak{q}^{2N}}
\prod_{n=1}^{N-1}\frac{1}{1-\mathfrak{q}^{4n}}
\prod_{n=1}^{M}\frac{1}{1-\mathfrak{q}^{4n}}, \\
\label{hC_so2N-|usp2M}
{\mathbb{II}_{\mathcal{N}}^{\mathfrak{so}(2N)^{-}|\mathfrak{usp}(2M)}}^{(C)}(\mathfrak{q})
&=\frac{1}{1+\mathfrak{q}^{2N}}
\prod_{n=1}^{N-1}\frac{1}{1-\mathfrak{q}^{4n}}
\prod_{n=1}^{M}\frac{1}{1-\mathfrak{q}^{4n}}. 
\end{align}
Accordingly, the normalized Wilson line defect half-indices also factorize into the normalized half-BPS line defect indices in an analogous fashion
\begin{align}
{\langle \mathcal{W}_{\mathcal{R}_1;\mathcal{R}_2}\rangle_{\mathcal{N}}^{\mathfrak{so}(2N)|\mathfrak{usp}(2M)}}^{(C)}(\mathfrak{q})
&=\langle \mathcal{W}_{\mathcal{R}_1}\rangle_{\textrm{$\frac12$BPS}}^{\mathfrak{so}(2N)}(\mathfrak{q})
\langle \mathcal{W}_{\mathcal{R}_2}\rangle_{\textrm{$\frac12$BPS}}^{\mathfrak{usp}(2M)}(\mathfrak{q}), \\
\label{NhC_so2N-|usp2M_W}
{\langle \mathcal{W}_{\mathcal{R}_1;\mathcal{R}_2}\rangle_{\mathcal{N}}^{\mathfrak{so}(2N)^{-}|\mathfrak{usp}(2M)}}^{(C)}(\mathfrak{q})
&=\langle \mathcal{W}_{\mathcal{R}_1}\rangle_{\textrm{$\frac12$BPS}}^{\mathfrak{so}(2N)^{-}}(\mathfrak{q})
\langle \mathcal{W}_{\mathcal{R}_2}\rangle_{\textrm{$\frac12$BPS}}^{\mathfrak{usp}(2M)}(\mathfrak{q}). 
\end{align}
As in (\ref{WI_so2N+1|usp2M_C}), 
the line defect Witten index for the $\mathfrak{so}(2N)|\mathfrak{usp}(2M)$ interface is given by the integral over the maximal torus $T^N\times T^M$ 
\begin{align}
\label{WI_so2N|usp2M_C}
{\langle \mathcal{W}_{\mathcal{R}_1;\mathcal{R}_2}\rangle_{\mathcal{N}}^{\mathfrak{so}(2N)|\mathfrak{usp}(2M)}}^{(C)}(1)
&=\oint \prod_{i=1}^{N}\frac{ds_i^{(1)}}{2\pi is_i^{(1)}}\prod_{j=1}^{M}\frac{ds_j^{(2)}}{2\pi is_j^{(2)}}
\chi_{\mathcal{R}_1}^{\mathfrak{so}(2N)}(s^{(1)})\chi_{\mathcal{R}_2}^{\mathfrak{usp}(2M)}(s^{(2)})
\nonumber\\
&=\dim \mathcal{R}_1[0]\cdot \dim \mathcal{R}_2[0]. 
\end{align}
In particular, it vanishes whenever $\mathcal{R}_1$ or $\mathcal{R}_2$ is the fundamental representation, 
since the fundamental representations of $\mathfrak{so}(2N)$ and $\mathfrak{usp}(2M)$ have no zero weight. 

%%%%%%%%%%%%%%%%%%%%%%%%%%%%%%%%%%%%%%%%%%%%%%%%%
\subsubsection{Higgs limit}
%%%%%%%%%%%%%%%%%%%%%%%%%%%%%%%%%%%%%%%%%%%%%%%%%
%Higgs limit
In the Higgs limit the matrix integral (\ref{h_so2N|usp2M_W}) reduces to
\begin{align}
\label{hH_so2N|usp2M_W}
{\langle W_{\mathcal{R}_1;\mathcal{R}_2}\rangle_{\mathcal{N}}^{\mathfrak{so}(2N)|\mathfrak{usp}(2M)}}^{(H)}(\mathfrak{q})
&=\frac{1}{2^{N-1}N!}\oint \prod_{i=1}^{N}\frac{ds^{(1)}_i}{2\pi is^{(1)}_i}
\prod_{i<j}(1-s_i^{(1)\pm}s_{j}^{(1)\pm})
\nonumber\\
&\times \frac{1}{2^{M}M!}\oint \prod_{i=1}^{M}\frac{ds^{(2)}_i}{2\pi is^{(2)}_i}
(1-s_i^{(2)\pm 2})
\prod_{i<j}(1-s_i^{(2)\pm}s_{j}^{(2)\pm})
\nonumber\\
&\times 
\prod_{i=1}^N\prod_{j=1}^M
\frac{1}{1-\mathfrak{q}s_i^{(1)\pm}s_j^{(2)\pm}}
\chi_{\mathcal{R}_1}^{\mathfrak{so}(2N)}(s^{(1)})
\chi_{\mathcal{R}_2}^{\mathfrak{usp}(2M)}(s^{(2)})
\end{align}
and (\ref{h_so2N-|usp2M_W}) becomes 
\begin{align}
\label{hH_so2N-|usp2M_W}
&
{\langle W_{\mathcal{R}_1;\mathcal{R}_2}\rangle_{\mathcal{N}}^{\mathfrak{so}(2N)^-|\mathfrak{usp}(2M)}}^{(H)}(\mathfrak{q})
\nonumber\\
&=\frac{1}{2^{N-1}(N-1)!}
\oint \prod_{i=1}^{N-1}\frac{ds_i^{(1)}}{2\pi is_i^{(1)}}(1-s^{(1)\pm}_i)(1+s^{(1)\pm}_i)
\prod_{i<j}(1-s^{(1)\pm}_i s^{(1)\pm}_j)
\nonumber\\
&\times 
\frac{1}{2^{M}M!}\oint \prod_{i=1}^{M}\frac{ds^{(2)}_i}{2\pi is^{(2)}_i}
(1-s_i^{(2)\pm 2})
\prod_{i<j}(1-s_i^{(2)\pm}s_{j}^{(2)\pm})
\nonumber\\
&\times 
\prod_{i=1}^{N-1}\prod_{j=1}^M
\frac{1}{1-\mathfrak{q}s_i^{(1)\pm}s_j^{(2)\pm}}
\prod_{j=1}^M
\frac{1}{(1-\mathfrak{q}s^{(2)\pm}_j)(1+\mathfrak{q}s^{(2)\pm}_j)}
\chi_{\mathcal{R}_1}^{\mathfrak{usp}(2N-2)}(s^{(1)})
\chi_{\mathcal{R}_2}^{\mathfrak{usp}(2M)}(s^{(2)}). 
\end{align}

The line defect Witten indices are determined by the residual gauge group 
as in the $\mathfrak{so}(2N+1)|\mathfrak{usp}(2M)'$ interface (see Appendix \ref{app_residual}). 
In this case the action of the gauge group is stable for any $N$ and $M$. 
The residual gauge group is $U(1)^N\times USp(2(M-N))$ for $N\le M$ and $U(1)^M\times SO(2(N-M))$ for $N>M$, 
corresponding to the NS5-type $\mathfrak{so}(0)|\mathfrak{usp}(2(M-N))$ and $\mathfrak{so}(2(N-M))|\mathfrak{usp}(0)$ interfaces, 
which carry no hypermultiplet, in the brane picture. 
We obtain 
\begin{align}
\label{WI_so2N|usp2M_H}
{\langle \mathcal{W}_{\mathcal{R}_1;\mathcal{R}_2}\rangle_{\mathcal{N}}^{\mathfrak{so}(2N)|\mathfrak{usp}(2M)}}^{(H)}(1)
&=
\begin{cases}
\dim (\mathcal{R}_1\otimes \mathcal{R}_2)^{U(1)^N\times USp(2(M-N))}&\textrm{for $N\le M$}\cr
\dim (\mathcal{R}_1\otimes \mathcal{R}_2)^{U(1)^M\times SO(2(N-M))}&\textrm{for $N>M$}\cr
\end{cases}, 
\end{align}
where $U(1)^{\min(N,M)}$ is embedded diagonally as $s_i^{(1)}=s_i^{(2)}=u_i$ 
and the residual $USp(2(M-N))$ or $SO(2(N-M))$ acts on the remaining entries 
of $s^{(2)}$ or $s^{(1)}$ respectively. 
Since the element $(-1,-1)$ of $SO(2N)\times USp(2M)$ acts trivially on the bifundamental half-hypermultiplet, 
it belongs to the residual gauge group, 
so that the Witten index vanishes unless the total center charge is trivial, 
e.g. for $\mathcal{W}_{\tiny \yng(1);0}$ and $\mathcal{W}_{0;\tiny \yng(1)}$. 
For the Wilson line in the bifundamental representation we get 
\begin{align}
{\langle \mathcal{W}_{\tiny \yng(1);\yng(1)}\rangle_{\mathcal{N}}^{\mathfrak{so}(2N)|\mathfrak{usp}(2M)}}^{(H)}(1)
&=2\min(N,M), 
\end{align}
which agrees with the Coulomb limit Witten index of the S-dual minuscule vortex line at the D5-type $\mathfrak{so}(2N)|\mathfrak{so}(2M+1)$ interface 
obtained in section \ref{sec_DC2}. 
For the NS5-type $\mathfrak{so}(2)|\mathfrak{usp}(2M)$ interface we find 
\begin{align}
{\langle \mathcal{W}_{n;(n)}\rangle_{\mathcal{N}}^{\mathfrak{so}(2)|\mathfrak{usp}(2M)}}^{(H)}(\mathfrak{q})
&=\mathfrak{q}^n
\end{align}
for $1\le n\le 3$ and $M\le 3$, whose Witten index $1$ is again reproduced by (\ref{WI_so2N|usp2M_H}).

%%%%%%%%%%%%%%%%%%%%%%%%%%%%%%%%%%%%%%%%%%%%%%%%%
\subsection{Vortex lines}
\label{sec_DCNS5V}
%%%%%%%%%%%%%%%%%%%%%%%%%%%%%%%%%%%%%%%%%%%%%%%%%
Next consider the magnetic lines at the NS5-type $\mathfrak{so}(2N)|\mathfrak{usp}(2M)$ interface. 
They can be characterized by a pair of magnetic charges $(k^{(1)},k^{(2)})$ for the two gauge group factors 
associated with the gauge algebras $\mathfrak{so}(2N)$ and $\mathfrak{usp}(2M)$. 

%so(2N)
For 4d $\mathcal{N}=4$ $Spin(2N)$ gauge theory 
there is no monopole bubbling for the spinor magnetic line of half-integral magnetic charge $(1/2,\cdots,1/2)$. 
For $SO(2N)$ $=$ $Spin(2N)/\mathbb{Z}_2$ gauge theory the allowed magnetic charges are restricted to be the integral values, 
for which only the vector magnetic line with magnetic charge $(1,0^{N-1})$ is minuscule and free from the monopole bubbling \cite{Kapustin:2006pk}. 

%half-index
We expect that 
the line defect half-index of the vortex line with $k^{(1)}$ $=$ $(1,0^{N-1})$ and $k^{(2)}=0$ 
at the NS5-type $\mathfrak{so}(2N)|\mathfrak{usp}(2M)$ interface for $N\le M$ 
can be evaluated by taking into account the shifted spins due to the singular profile of the 4d and 3d fields 
which are charged under the $SO(2N)$ gauge group. 
We propose that it is given by the following matrix integral: 
\begin{align}
\label{h_so2N|usp2M_V}
&
\langle V_{(1,0^{N-1});0}\rangle_{\mathcal{N}}^{\mathfrak{so}(2N)|\mathfrak{usp}(2M)}(t;q)
\nonumber\\
&=q^{\frac12+\frac{M-N}{2}}t^{-2-2(M-N)}
\frac{1}{2^{N-2}(N-1)!} \frac{(q)_{\infty}^N}{(q^{\frac12}t^{-2};q)_{\infty}^N}
\oint \prod_{i=1}^N \frac{ds_i^{(1)}}{2\pi is_i^{(1)}}
\nonumber\\
&\times 
\prod_{i=2}^N
\frac{(q^{\frac12}s_1^{(1)\pm}s_i^{(1)\pm};q)_{\infty}}
{(qt^{-2}s_1^{(1)\pm}s_i^{(1)\pm};q)_{\infty}}
\prod_{2\le i<j\le N}
\frac{(s_i^{(1)\pm}s_j^{(1)\pm};q)_{\infty}}
{(q^{\frac12}t^{-2}s_i^{(1)\pm}s_j^{(1)\pm};q)_{\infty}}
\nonumber\\
&\times 
\frac{1}{2^M M!} \frac{(q)_{\infty}^M}{(q^{\frac12}t^{-2};q)_{\infty}^M}
\oint \prod_{i=1}^M \frac{ds_i^{(2)}}{2\pi is_i^{(2)}}
\frac{(s_i^{(2)\pm2};q)_{\infty}}
{(q^{\frac12}t^{-2}s_i^{(2)\pm2};q)_{\infty}}
\prod_{i<j}
\frac{(s_i^{(2)\pm}s_j^{(2)\pm};q)_{\infty}}
{(q^{\frac12}t^{-2}s_i^{(2)\pm}s_j^{(2)\pm};q)_{\infty}}
\nonumber\\
&\times 
\prod_{j=1}^M
\frac{(q^{\frac54}t^{-1}s_1^{(1)\pm}s_j^{(2)\pm};q)_{\infty}}
{(q^{\frac34}ts_1^{(1)\pm}s_j^{(2)\pm};q)_{\infty}}
\prod_{i=2}^N
\prod_{j=1}^M
\frac{(q^{\frac34}t^{-1}s_i^{(1)\pm}s_j^{(2)\pm};q)_{\infty}}
{(q^{\frac14}ts_i^{(1)\pm}s_j^{(2)\pm};q)_{\infty}}, 
\end{align}
where the overall factor $q^{\frac12+\frac{M-N}{2}}t^{-2-2(M-N)}$ is a contribution from the background bare monopole operator. 

%%%%%%%%%%%%%%%%%%%%%%%%%%%%%%%%%%%%%%%%%%%%%%%%%
%ADDED
The structure of the matrix integral (\ref{h_so2N|usp2M_V}) is the same as that of (\ref{h_so2N+1|usp2M_V}). 
The fugacity of a field of weight $\rho$ is shifted by $q^{\frac{|\rho(k)|}{2}}$ 
and the background bare monopole operator of the $SU(2)_C$ R-charge (\ref{bare_monopole_D}) 
gives rise to the overall factor $q^{\frac{\Delta(k)}{4}}t^{-\Delta(k)}$. 
For $k^{(1)}=(1,0^{N-1})$ and $k^{(2)}=0$ the weights $\pm e_1\pm f_j$ 
of the 3d bifundamental half-hypermultiplet and the positive roots $e_1\pm e_i$ with $i\neq 1$ of $\mathfrak{so}(2N)$ 
are the only ones which are charged under $k$, so that 
\begin{align}
\label{Delta_so2N|usp2M}
\Delta\bigl((1,0^{N-1});0\bigr)&=\frac12\cdot 4M-2(N-1)=2+2(M-N), 
\end{align}
which precisely accounts for the non-trivial prefactor in (\ref{h_so2N|usp2M_V}), 
in agreement with the leading power $\mathfrak{q}^{2+2(M-N)}$ of the Coulomb limit (\ref{hC_so2N|usp2M_V}). 
Unlike the $\mathfrak{so}(2N+1)$ case (\ref{Delta_so2N+1|usp2M}), 
the short roots $\pm e_1$ are absent for $\mathfrak{so}(2N)$, 
which is why $\Delta$ is larger by one unit. 

The expression (\ref{h_so2N|usp2M_V}) is written for $N\ge 2$. 
For $N=1$ the gauge algebra $\mathfrak{so}(2)$ is Abelian and has no root. 
Hence the residual gauge symmetry in the vortex line background is $U(1)$ itself, whose Weyl group is trivial, 
so that the factor $\frac{1}{2^{N-2}(N-1)!}$ in (\ref{h_so2N|usp2M_V}), 
which is the inverse of the order $2^{N-2}(N-1)!$ of the Weyl group of the residual $U(1)\times SO(2N-2)$ gauge symmetry, 
is replaced by $1$. 
Furthermore, the vortex lines of the Abelian gauge group are free from the monopole bubbling effects, 
so that the above prescription can be generalized to an arbitrary magnetic charge $n\in \mathbb{Z}$ 
in a straightforward manner. 
Since $\mathfrak{so}(2)$ has no root, the 4d vector multiplet factor is not modified, 
while the 3d bifundamental half-hypermultiplet, which carries the $\mathfrak{so}(2)$ charge $\pm1$, 
is shifted by $q^{\frac{|n|}{2}}$. 
All the $4M$ weights of the bifundamental half-hypermultiplet obey $|\rho(n)|=|n|$ 
so that (\ref{bare_monopole_D}) gives $\Delta(n;0)=2M|n|$. 
Consequently, we propose that the line defect half-index of the vortex line of magnetic charge $n$ 
at the NS5-type $\mathfrak{so}(2)|\mathfrak{usp}(2M)$ interface is given by
\begin{align}
\label{h_so2|usp2M_V}
&
\langle V_{n;0}\rangle_{\mathcal{N}}^{\mathfrak{so}(2)|\mathfrak{usp}(2M)}(t;q)
\nonumber\\
&=q^{\frac{M|n|}{2}}t^{-2M|n|}
\frac{(q)_{\infty}}{(q^{\frac12}t^{-2};q)_{\infty}}
\oint \frac{ds^{(1)}}{2\pi is^{(1)}}
\nonumber\\
&\times 
\frac{1}{2^M M!} \frac{(q)_{\infty}^M}{(q^{\frac12}t^{-2};q)_{\infty}^M}
\oint \prod_{i=1}^M \frac{ds_i^{(2)}}{2\pi is_i^{(2)}}
\frac{(s_i^{(2)\pm2};q)_{\infty}}
{(q^{\frac12}t^{-2}s_i^{(2)\pm2};q)_{\infty}}
\prod_{i<j}
\frac{(s_i^{(2)\pm}s_j^{(2)\pm};q)_{\infty}}
{(q^{\frac12}t^{-2}s_i^{(2)\pm}s_j^{(2)\pm};q)_{\infty}}
\nonumber\\
&\times 
\prod_{j=1}^M
\frac{(q^{\frac34+\frac{|n|}{2}}t^{-1}s^{(1)\pm}s_j^{(2)\pm};q)_{\infty}}
{(q^{\frac14+\frac{|n|}{2}}ts^{(1)\pm}s_j^{(2)\pm};q)_{\infty}}. 
\end{align}
It depends on $n$ only through $|n|$, as it should be, 
because the vortex lines of charge $n$ and $-n$ are exchanged by the charge conjugation. 
For $n=\pm 1$ it reduces to (\ref{h_so2N|usp2M_V}) with $N=1$ and the prefactor $\frac{1}{2^{N-2}(N-1)!}$ replaced by $1$. 
In section \ref{sec_DCD5W_Sduality} we prove analytically, for arbitrary $M\ge 1$ and $n\in\mathbb{Z}$, that (\ref{h_so2|usp2M_V}) agrees with the Wilson line half-index of charge $n$ at the dual D5-type $\mathfrak{so}(2)|\mathfrak{so}(2M+1)$ interface, i.e. $\langle W_{n}\rangle^{\mathfrak{so}(2)|\mathfrak{so}(2M+1)}_{\mathcal{D}}(t^{-1};q)$ obtained from (\ref{h_so2N|so2M+1_Wa}) with $N=1$. 
The explicit $q$-series in Appendix \ref{app_expansion} provide independent low-rank checks of this exact identity. 
%%%%%%%%%%%%%%%%%%%%%%%%%%%%%%%%%%%%%%%%%%%%%%%%%

%O(2N)
Furthermore, for the NS5-type $\mathfrak{o}(2N)^{\pm}|\mathfrak{usp}(2M)$ interface 
we propose that 
\begin{align}
\label{h_o2N|usp2M_V}
&
\langle V_{(1,0^{N-1});0}\rangle_{\mathcal{N}}^{\mathfrak{o}(2N)^{\pm}|\mathfrak{usp}(2M)}(t;q)
\nonumber\\
&=\frac12
\left(
\langle V_{(1,0^{N-1});0}\rangle_{\mathcal{N}}^{\mathfrak{so}(2N)|\mathfrak{usp}(2M)}(t;q)
\pm 
\langle V_{(1,0^{N-2});0}\rangle_{\mathcal{N}}^{\mathfrak{so}(2N)^{-}|\mathfrak{usp}(2M)}(t;q)
\right), 
\end{align}
where the contribution from the twisted sector is given by
\begin{align}
\label{h_so2N-|usp2M_V}
&
\langle V_{(1,0^{N-2});0}\rangle_{\mathcal{N}}^{\mathfrak{so}(2N)^{-}|\mathfrak{usp}(2M)}(t;q)
\nonumber\\
&=q^{\frac12+\frac{M-N}{2}}t^{-2-2(M-N)}
\frac{1}{2^{N-2}(N-2)!}
\frac{(q)_{\infty}^{N-1}(-q;q)_{\infty}}
{(q^{\frac12}t^{-2};q)_{\infty}^{N-1}(-q^{\frac12}t^{-2};q)_{\infty}}
\oint \prod_{i=1}^{N-1}
\frac{ds_i^{(1)}}{2\pi is_i^{(1)}}
\nonumber\\
&\times 
\frac{(q^{\frac12}s_1^{(1)\pm};q)_{\infty}(-q^{\frac12}s_1^{(1)\pm};q)_{\infty}}
{(qt^{-2}s_1^{(1)\pm};q)_{\infty} (-qt^{-2}s_1^{(1)\pm};q)_{\infty}}
\prod_{i=2}^{N-1}
\frac{(s_i^{(1)\pm};q)_{\infty}(-s_i^{(1)\pm};q)_{\infty}}
{(q^{\frac12}t^{-2}s_i^{(1)\pm};q)_{\infty} (-q^{\frac12}t^{-2}s_i^{(1)\pm};q)_{\infty}}
\nonumber\\
&\times 
\prod_{i=2}^{N-1}
\frac{(q^{\frac12}s_1^{(1)\pm}s_i^{(1)\pm};q)_{\infty}}
{(qt^{-2}s_1^{(1)\pm}s_i^{(1)\pm};q)_{\infty}}
\prod_{2\le i<j\le N-1}
\frac{(s_i^{(1)\pm}s_j^{(1)\pm};q)_{\infty}}
{(q^{\frac12}t^{-2}s_i^{(1)\pm}s_j^{(1)\pm};q)_{\infty}}
\nonumber\\
&\times 
\frac{1}{2^M M!}\frac{(q)_{\infty}^M}{(q^{\frac12}t^{-2};q)_{\infty}^M}
\oint \prod_{i=1}^M \frac{ds_i^{(2)}}{2\pi is_i^{(2)}}
\frac{(s_i^{(2)\pm2};q)_{\infty}}{(q^{\frac12}t^{-2}s_i^{(2)\pm2};q)_{\infty}}
\prod_{i<j}\frac{
(s_i^{(2)\pm}s_j^{(2)\pm};q)_{\infty}
}
{
(q^{\frac12}t^{-2}s_i^{(2)\pm}s_j^{(2)\pm};q)_{\infty}
}
\nonumber\\
&\times 
\prod_{j=1}^M
\frac{
(q^{\frac54}t^{-1}s_1^{(1)\pm}s_j^{(2)\pm};q)_{\infty}
}
{
(q^{\frac34}ts_1^{(1)\pm}s_j^{(2)\pm};q)_{\infty}
}
\prod_{i=2}^{N-1}\prod_{j=1}^{M}
\frac{
(q^{\frac34}t^{-1}s_i^{(1)\pm}s_j^{(2)\pm};q)_{\infty}
}
{
(q^{\frac14}ts_i^{(1)\pm}s_j^{(2)\pm};q)_{\infty}
}
\nonumber\\
&\times 
\prod_{j=1}^M
\frac{
(q^{\frac34}t^{-1}s_j^{(2)\pm};q)_{\infty}
(-q^{\frac34}t^{-1}s_j^{(2)\pm};q)_{\infty}
}
{
(q^{\frac14}ts_j^{(2)\pm};q)_{\infty}
(-q^{\frac14}ts_j^{(2)\pm};q)_{\infty}
}. 
\end{align}

%%%%%%%%%%%%%%%%%%%%%%%%%%%%%%%%%%%%%%%%%%%%%%%%%
\subsubsection{Coulomb limit}
%%%%%%%%%%%%%%%%%%%%%%%%%%%%%%%%%%%%%%%%%%%%%%%%%
%Coulomb limit
The Coulomb limit of the line defect half-index (\ref{h_so2N|usp2M_V}) is exactly computed as
\begin{align}
\label{hC_so2N|usp2M_V}
{\langle V_{(1,0^{N-1});0}\rangle_{\mathcal{N}}^{\mathfrak{so}(2N)|\mathfrak{usp}(2M)}}^{(C)}(\mathfrak{q})
&=\frac{\mathfrak{q}^{2+2(M-N)}}{1-\mathfrak{q}^2}
\mathcal{I}_{\textrm{$\frac12$BPS}}^{\mathfrak{so}(2N-2)}
\mathcal{I}_{\textrm{$\frac12$BPS}}^{\mathfrak{usp}(2M)}
\nonumber\\
&=\frac{\mathfrak{q}^{2+2(M-N)}}{(1-\mathfrak{q}^2)(1-\mathfrak{q}^{2(N-1)})}
\prod_{n=1}^{N-2}
\frac{1}{1-\mathfrak{q}^{4n}}
\prod_{n=1}^{M}
\frac{1}{1-\mathfrak{q}^{4n}}. 
\end{align}
Here $\mathcal{I}_{\textrm{$\frac12$BPS}}^{\mathfrak{so}(2N-2)}(\mathfrak{q})$ and 
$\mathcal{I}_{\textrm{$\frac12$BPS}}^{\mathfrak{usp}(2M)}(\mathfrak{q})$ are the half-BPS indices 
of 4d $\mathcal{N}=4$ $SO(2N-2)$ and $USp(2M)$ gauge theories \cite{Caputa:2013vla,Hatsuda:2024lcc}. 
Similarly, the Coulomb limit of the expression (\ref{h_so2N-|usp2M_V}) is 
\begin{align}
{\langle V_{(1,0^{N-1});0}\rangle_{\mathcal{N}}^{\mathfrak{so}(2N)^-|\mathfrak{usp}(2M)}}^{(C)}(\mathfrak{q})
&=\frac{\mathfrak{q}^{2+2(M-N)}}{1-\mathfrak{q}^2}
\mathcal{I}_{\textrm{$\frac12$BPS}}^{\mathfrak{so}(2N-2)^-}
\mathcal{I}_{\textrm{$\frac12$BPS}}^{\mathfrak{usp}(2M)}
\nonumber\\
&=\frac{\mathfrak{q}^{2+2(M-N)}}{(1-\mathfrak{q}^2)(1+\mathfrak{q}^{2(N-1)})}
\prod_{n=1}^{N-2}
\frac{1}{1-\mathfrak{q}^{4n}}
\prod_{n=1}^{M}
\frac{1}{1-\mathfrak{q}^{4n}}. 
\end{align}
Taking into account that the interface half-index without line insertions is given by (\ref{hC_so2N|usp2M}) and (\ref{hC_so2N-|usp2M}), 
the normalized line defect half-indices are computed as follows:
\begin{align}
{\langle \mathcal{V}_{(1,0^{N-1});0}\rangle_{\mathcal{N}}^{\mathfrak{so}(2N)|\mathfrak{usp}(2M)}}^{(C)}(\mathfrak{q})
&=\mathfrak{q}^{2+2(M-N)}(1+\mathfrak{q}^{2(N-1)})(1+\mathfrak{q}^2+\cdots+\mathfrak{q}^{2(N-1)}), \\
{\langle \mathcal{V}_{(1,0^{N-1});0}\rangle_{\mathcal{N}}^{\mathfrak{o}(2N)^+|\mathfrak{usp}(2M)}}^{(C)}(\mathfrak{q})
&=\mathfrak{q}^{2+2(M-N)}(1+\mathfrak{q}^2+\cdots+\mathfrak{q}^{2(2N-1)}), \\
{\langle \mathcal{V}_{(1,0^{N-1});0}\rangle_{\mathcal{N}}^{\mathfrak{o}(2N)^-|\mathfrak{usp}(2M)}}^{(C)}(\mathfrak{q})
&=\mathfrak{q}^{2(M-N)}(1+\mathfrak{q}^2+\cdots+\mathfrak{q}^{2(2N-1)}). 
\end{align}

%%%%%%%%%%%%%%%%%%%%%%%%%%%%%%%%%%%%%%%%%%%%%%%%%
\subsubsection{Higgs limit}
%%%%%%%%%%%%%%%%%%%%%%%%%%%%%%%%%%%%%%%%%%%%%%%%%
%Higgs limit
In the Higgs limit the line defect half-indices (\ref{h_so2N|usp2M_V}) and (\ref{h_o2N|usp2M_V}) vanish
\begin{align}
\label{hH_so2N|usp2M_V}
{\langle \mathcal{V}_{(1,0^{N-1});0}\rangle_{\mathcal{N}}^{\mathfrak{so}(2N)|\mathfrak{usp}(2M)}}^{(H)}(\mathfrak{q})
&=0, \\
\label{hH_o2N|usp2M_V}
{\langle \mathcal{V}_{(1,0^{N-1});0}\rangle_{\mathcal{N}}^{\mathfrak{o}(2N)^{\pm}|\mathfrak{usp}(2M)}}^{(H)}(\mathfrak{q})
&=0. 
\end{align}
This implies that the corresponding vortex line cannot combine with the interface bifundamental half-hypermultiplet 
to form a gauge-invariant operator.

%%%%%%%%%%%%%%%%%%%%%%%%%%%%%%%%%%%%%%%%%%%%%%%%%
%%%%%%%%%%%%%%%%%%%%%%%%%%%%%%%%%%%%%%%%%%%%%%%%%
\section{$\mathfrak{so}(2N)|\mathfrak{so}(2M+1)$}
\label{sec_DC2}
%%%%%%%%%%%%%%%%%%%%%%%%%%%%%%%%%%%%%%%%%%%%%%%%%
%%%%%%%%%%%%%%%%%%%%%%%%%%%%%%%%%%%%%%%%%%%%%%%%%
The S-dual of the NS5-type $\mathfrak{so}(2N)|\mathfrak{usp}(2M)$ interface is the D5-type $\mathfrak{so}(2N)|\mathfrak{so}(2M+1)$ interface. 
For $N=M$ the $\mathfrak{so}(2N)\oplus \mathfrak{so}(2N+1)$ gauge algebra is broken down to $\mathfrak{so}(2N)$ 
and involves the 3d hypermultiplet transforming as the fundamental representation. 
For $N<M$ the interface again keeps the $\mathfrak{so}(2N)$ gauge algebra but it contains the $\mathfrak{so}(2(M-N)+1)$ Nahm pole.   
For $N>M$ the gauge algebra is broken down to $\mathfrak{so}(2M+1)$ and the $\mathfrak{so}(2(N-M))$ Nahm pole appears. 
When $N\neq M$ there are additional 3d matter fields transforming as the fundamental representation of the preserved gauge group \cite{Gaiotto:2019jvo,Hatsuda:2024lcc}. 
In the brane configuration we introduce a half D5-brane at $x^6=0$, $N$ D3-branes along with an O3$^{-}$-plane in $x^6<0$ 
and $M$ D3-branes along with an $\widetilde{\textrm{O3}}^-$-plane in $x^6>0$. 

%%%%%%%%%%%%%%%%%%%%%%%%%%%%%%%%%%%%%%%%%%%%%%%%%
\subsection{Wilson lines}
\label{sec_DCD5W}
%%%%%%%%%%%%%%%%%%%%%%%%%%%%%%%%%%%%%%%%%%%%%%%%%

%half-index
%N\le M
For $N\le M$ the line defect half-index of the Wilson lines $W_{\mathcal{R}_j}$ transforming in the representation $\mathcal{R}_j$ 
with $j=1,\cdots,k$ of $\mathfrak{so}(2N)$ at the D5-type $\mathfrak{so}(2N)|\mathfrak{so}(2M+1)$ interface is given by
\begin{align}
\label{h_so2N|so2M+1_Wa}
&
\langle W_{\mathcal{R}_1}\cdots W_{\mathcal{R}_k}\rangle_{\mathcal{D}}^{\mathfrak{so}(2N)|\mathfrak{so}(2M+1)}(t;q)
\nonumber\\
&=\frac{1}{2^{N-1} N!} \frac{(q)_{\infty}^{2N}}{(q^{\frac12}t^{\pm2};q)_{\infty}^N}
\oint \prod_{i=1}^{N} \frac{ds_i}{2\pi is_i}
\prod_{i<j}
\frac{(s_i^{\pm}s_j^{\pm};q)_{\infty}(qs_i^{\pm}s_j^{\pm};q)_{\infty}}
{(q^{\frac12}t^{2}s_i^{\pm}s_j^{\pm};q)_{\infty}(q^{\frac12}t^{-2}s_i^{\pm}s_j^{\pm};q)_{\infty}}
\nonumber\\
&\times 
\prod_{\ell=1}^{M-N}
\frac{(q^{\frac12+\ell}t^{4\ell-2};q)_{\infty}}
{(q^{\ell} t^{4\ell};q)_{\infty}}
\prod_{i=1}^{N}
\frac{(q^{1+\frac{M-N}{2}} t^{2(M-N)} s_i^{\pm};q)_{\infty}}{(q^{\frac12+\frac{M-N}{2}}t^{2+2(M-N)} s_i^{\pm};q)_{\infty}}
\prod_{j=1}^k \chi^{\mathfrak{so}(2N)}_{\mathcal{R}_j}(s). 
\end{align}

Analogously, the Wilson line half-index for the D5-type $\mathfrak{o}(2N)^{\pm}|\mathfrak{so}(2M+1)$ interface is given by
\begin{align}
\label{h_o2N|so2M+1_W}
&
\langle W_{\mathcal{R}_1}\cdots W_{\mathcal{R}_k}\rangle_{\mathcal{D}}^{\mathfrak{o}(2N)^{\pm}|\mathfrak{so}(2M+1)}(t;q)
\nonumber\\
&=\frac12
\left(
\langle W_{\mathcal{R}_1}\cdots W_{\mathcal{R}_k} \rangle_{\mathcal{D}}^{\mathfrak{so}(2N)|\mathfrak{so}(2M+1)}(t;q)
\pm 
\langle W_{\mathcal{R}_1}\cdots W_{\mathcal{R}_k} \rangle_{\mathcal{D}}^{\mathfrak{so}(2N)^{-}|\mathfrak{so}(2M+1)}(t;q)
\right), 
\end{align}
where the contribution from the twisted sector is given by
\begin{align}
\label{h_so2N|so2M+1_Wa-}
&
\langle W_{\mathcal{R}_1}\cdots W_{\mathcal{R}_k} \rangle^{\mathfrak{so}(2N)^-|\mathfrak{so}(2M+1)}_{\mathcal{D}}(t;q)
\nonumber\\
&=\frac{1}{2^{N-1} (N-1)!} 
\frac{(q)_{\infty}^{2N-2} (-q;q)_{\infty}^2}
{(q^{\frac12}t^{\pm2};q)_{\infty}^{N-1}(-q^{\frac12}t^{\pm2};q)_{\infty}}
\oint \prod_{i=1}^{N-1} \frac{ds_i}{2\pi is_i}
\nonumber\\
&\times
\prod_{i=1}^{N-1} 
\frac{(s_i^{\pm};q)_{\infty}(-s_i^{\pm};q)_{\infty} (qs_i^{\pm};q)_{\infty}(-qs_i^{\pm};q)_{\infty}}
{(q^{\frac12}t^{2}s_i^{\pm};q)_{\infty}(-q^{\frac12}t^{2}s_i^{\pm};q)_{\infty}(q^{\frac12}t^{-2}s_i^{\pm};q)_{\infty}(-q^{\frac12}t^{-2}s_i^{\pm};q)_{\infty}}
\nonumber\\
&\times \prod_{i<j}
\frac{(s_i^{\pm}s_j^{\pm};q)_{\infty}(qs_i^{\pm}s_j^{\pm};q)_{\infty}}
{(q^{\frac12}t^{2}s_i^{\pm}s_j^{\pm};q)_{\infty}(q^{\frac12}t^{-2}s_i^{\pm}s_j^{\pm};q)_{\infty}}
\prod_{\ell=1}^{M-N}
\frac{(q^{\frac12+\ell}t^{4\ell-2};q)_{\infty}}
{(q^{\ell} t^{4\ell};q)_{\infty}}
\nonumber\\
&\times 
\prod_{i=1}^{N-1}
\frac{(q^{1+\frac{M-N}{2}} t^{2(M-N)} s_i^{\pm};q)_{\infty}}{(q^{\frac12+\frac{M-N}{2}}t^{2+2(M-N)} s_i^{\pm};q)_{\infty}}
\prod_{\epsilon=\pm1}
\frac{(\epsilon q^{1+\frac{M-N}{2}} t^{2(M-N)};q)_{\infty}}{(\epsilon q^{\frac12+\frac{M-N}{2}}t^{2+2(M-N)};q)_{\infty}}
\prod_{j=1}^k \chi_{\mathcal{R}_j}^{\mathfrak{usp}(2N-2)}(s). 
\end{align}
Here the last one-body factor, which is a product over the two signs $\epsilon=\pm1$, 
is the contribution of the two eigenvalues $\pm1$ 
of the holonomy in the disconnected component of $O(2N)$. 

%N>M
When $N>M$ the interface keeps the $\mathfrak{so}(2M+1)$ gauge algebra. 
The Wilson line defect half-index at the D5-type $\mathfrak{so}(2N)|\mathfrak{so}(2M+1)$ interface reads
\begin{align}
\label{h_so2N|so2M+1_Wb}
&
\langle W_{\mathcal{R}_1}\cdots W_{\mathcal{R}_k} \rangle_{\mathcal{D}}^{\mathfrak{so}(2N)|\mathfrak{so}(2M+1)}(t;q)
\nonumber\\
&=\frac{1}{2^M M!} \frac{(q)_{\infty}^{2M}}{(q^{\frac12}t^{\pm2};q)_{\infty}^M}
\oint \prod_{i=1}^{M} \frac{ds_i}{2\pi is_i}
\frac{(s_i^{\pm};q)_{\infty} (qs_i^{\pm};q)_{\infty}}{(q^{\frac12}t^2 s_i^{\pm};q)_{\infty}(q^{\frac12}t^{-2}s_i^{\pm};q)_{\infty}}
\nonumber\\
&\times \prod_{i<j}
\frac{(s_i^{\pm}s_j^{\pm};q)_{\infty}(qs_i^{\pm}s_j^{\pm};q)_{\infty}}
{(q^{\frac12}t^{2}s_i^{\pm}s_j^{\pm};q)_{\infty}(q^{\frac12}t^{-2}s_i^{\pm}s_j^{\pm};q)_{\infty}}
\frac{(q^{\frac12+\frac{N-M}{2}} t^{2(N-M)-2};q)_{\infty}}
{(q^{\frac{N-M}{2}}t^{2(N-M)};q)_{\infty}}
\prod_{\ell=1}^{N-M-1}
\frac{(q^{\frac12+\ell}t^{4\ell-2};q)_{\infty}}
{(q^{\ell} t^{4\ell};q)_{\infty}}
\nonumber\\
&\times 
\prod_{i=1}^{M}
\frac{(q^{1+\frac{N-M-1}{2}} t^{2(N-M-1)} s_i^{\pm};q)_{\infty}}{(q^{\frac12+\frac{N-M-1}{2}}t^{2+2(N-M-1)} s_i^{\pm};q)_{\infty}}
\prod_{j=1}^k \chi_{\mathcal{R}_j}^{\mathfrak{so}(2M+1)}(s). 
\end{align}

Also the Wilson line half-index for the D5-type $\mathfrak{o}(2N)^{\pm}|\mathfrak{so}(2M+1)$ interface 
can be obtained from (\ref{h_o2N|so2M+1_W}) by taking into account the following twisted sector contribution for $N>M$: 
\begin{align}
\label{h_so2N|so2M+1_Wb-}
&
\langle W_{\mathcal{R}_1}\cdots W_{\mathcal{R}_k} \rangle_{\mathcal{D}}^{\mathfrak{so}(2N)^-|\mathfrak{so}(2M+1)}(t;q)
\nonumber\\
&=\frac{1}{2^M M!} \frac{(q)_{\infty}^{2M}}{(q^{\frac12}t^{\pm2};q)_{\infty}^M}
\oint \prod_{i=1}^{M} \frac{ds_i}{2\pi is_i}
\frac{(s_i^{\pm};q)_{\infty} (qs_i^{\pm};q)_{\infty}}{(q^{\frac12}t^2 s_i^{\pm};q)_{\infty}(q^{\frac12}t^{-2}s_i^{\pm};q)_{\infty}}
\nonumber\\
&\times \prod_{i<j}
\frac{(s_i^{\pm}s_j^{\pm};q)_{\infty}(qs_i^{\pm}s_j^{\pm};q)_{\infty}}
{(q^{\frac12}t^{2}s_i^{\pm}s_j^{\pm};q)_{\infty}(q^{\frac12}t^{-2}s_i^{\pm}s_j^{\pm};q)_{\infty}}
\frac{(-q^{\frac12+\frac{N-M}{2}} t^{2(N-M)-2};q)_{\infty}}
{(-q^{\frac{N-M}{2}}t^{2(N-M)};q)_{\infty}}
\prod_{\ell=1}^{N-M-1}
\frac{(q^{\frac12+\ell}t^{4\ell-2};q)_{\infty}}
{(q^{\ell} t^{4\ell};q)_{\infty}}
\nonumber\\
&\times 
\prod_{i=1}^{M}
\frac{(-q^{1+\frac{N-M-1}{2}} t^{2(N-M-1)} s_i^{\pm};q)_{\infty}}{(-q^{\frac12+\frac{N-M-1}{2}}t^{2+2(N-M-1)} s_i^{\pm};q)_{\infty}}
\prod_{j=1}^k \chi_{\mathcal{R}_j}^{\mathfrak{so}(2M+1)}(s). 
\end{align}

%%%%%%%%%%%%%%%%%%%%%%%%%%%%%%%%%%%%%%%%%%%%%%%%%
\subsubsection{S-duality}
\label{sec_DCD5W_Sduality}
%%%%%%%%%%%%%%%%%%%%%%%%%%%%%%%%%%%%%%%%%%%%%%%%%
We next examine the action of S-duality on interface line operators. 
In the bulk, S-duality maps the fundamental Wilson line of 4d $\mathcal{N}=4$ SYM theory of gauge algebra $\mathfrak{so}(2N)$ 
to the minuscule magnetic 't Hooft line characterized by magnetic charge $k=(1,0^{N-1})$ in the same theory. 
This correspondence admits a natural extension to the interfaces. 

The case with $N=1$ is particularly simple 
because the preserved $\mathfrak{so}(2)$ gauge algebra is Abelian. 
The Wilson lines are then classified by an integer electric charge $n\in \mathbb{Z}$, 
whereas the vortex lines are classified by an integer magnetic charge. 
Furthermore, the absence of monopole bubbling implies that the magnetic charge uniquely labels the vortex line defect. 
In fact, we find the following non-trivial matching of the associated line defect half-indices:
\begin{align}
\langle V_{n;0}\rangle_{\mathcal{N}}^{\mathfrak{so}(2)|\mathfrak{usp}(2M)}(t;q)
&=\langle W_{n}\rangle^{\mathfrak{so}(2)|\mathfrak{so}(2M+1)}_{\mathcal{D}}(t^{-1};q). 
\label{eq:V_{n;0}=W_{n}}
\end{align}
Here the left-hand side is the matrix integral (\ref{h_so2|usp2M_V}) for the vortex line of magnetic charge $n$ at the NS5-type $\mathfrak{so}(2)|\mathfrak{usp}(2M)$ interface, while the right-hand side is given by (\ref{h_so2N|so2M+1_Wa}) with $N=1$ and $\chi_{\mathcal{R}}^{\mathfrak{so}(2)}(s)=s^n$. 
In this Abelian case the equality (\ref{eq:V_{n;0}=W_{n}}) can be established analytically for arbitrary $M\ge 1$ and $n\in\mathbb{Z}$. 
The $q$-series in Appendix \ref{app_expansion} provide independent low-rank checks of the exact identity.

We use the variables $\tau$ and $u$ introduced in (\ref{app_tau_u}) and set
\begin{align}
\label{abelian_variables}
\sigma&:=\frac{q}{\tau}=u^2,
\qquad m:=|n|.
\end{align}
For the NS5-type matrix integral (\ref{h_so2|usp2M_V}), let
\begin{align}
X_{m}&=\left(q^{\frac{m}{2}}s,q^{\frac{m}{2}}s^{-1}\right),
&
X_C&=\left(s_1^{(2)\pm},\ldots,s_M^{(2)\pm}\right).
\end{align}
In the present case the bifundamental contribution in (\ref{h_so2|usp2M_V}) is therefore the type-$A$ Macdonald Cauchy kernel
\begin{align}
\prod_{x\in X_{m}}\prod_{y\in X_C}
\frac{(\tau uxy;q)_\infty}{(uxy;q)_\infty}
&=
\sum_{\ell(\lambda)\le 2}
u^{|\lambda|}b_\lambda(q,\tau)
P_\lambda(X_{m};q,\tau)P_\lambda(X_C;q,\tau).
\label{abelian_cauchy_kernel}
\end{align}
This is the specialization of (\ref{app_cauchy_kernel}) to the two-letter alphabet $X_{m}$. 
Since $X_{m}$ contains only two letters, only partitions with $\ell(\lambda)\le 2$ occur. 
The $C_M$ Macdonald vanishing integral (\ref{app_CM_vanishing_rule}) then implies that only
\begin{align}
\lambda&=(r,r),\qquad r=0,1,2,\ldots
\end{align}
contributes.  For these partitions,
\begin{align}
 b_{(r,r)}(q,\tau)
 &=\frac{(\tau;q)_r(\tau^2;q)_r}{(q;q)_r(q\tau;q)_r},
 &
 P_{(r,r)}(X_{m};q,\tau)&=q^{mr}.
\label{abelian_rr_data}
\end{align}
Using (\ref{app_CM_rr}), the $C_M$ variables can therefore be integrated out exactly. 
Since $P_{(r,r)}(X_{m};q,\tau)$ is independent of $s$, the remaining $U(1)$ contour integral is trivial. 
We obtain
\begin{align}
\label{abelian_NS5_2phi1}
\langle V_{n;0}\rangle_{\mathcal{N}}^{\mathfrak{so}(2)|\mathfrak{usp}(2M)}(t;q)
&=\tau^{Mm}\frac{(q;q)_\infty}{(\tau;q)_\infty}\mathcal{Z}_M(\tau)
\,{}_2\phi_1\!\left[
\begin{matrix}
\tau,\tau^{2M}\\
q\tau^{2M-1}
\end{matrix};q,q^{m}\sigma
\right],
\end{align}
where $\mathcal{Z}_M(\tau)$ is the normalization of the $C_M$ Macdonald measure,
\begin{align*}
\mathcal{Z}_M(\tau)
&=\prod_{j=1}^{M}
\frac{(q\tau^{2j-1};q)_\infty}{(\tau^{2j};q)_\infty},
\end{align*}
as defined in (\ref{app_CM_partition}), and ${}_2\phi_1$ is the basic hypergeometric, or $q$-hypergeometric, series
\begin{align}
\label{abelian_2phi1_def}
{}_2\phi_1\!\left[
\begin{matrix}a,b\\c\end{matrix};q,z
\right]
&:=\sum_{r=0}^{\infty}\frac{(a;q)_r(b;q)_r}{(q;q)_r(c;q)_r}z^r.
\end{align}

We next evaluate the D5-type Wilson line half-index.  For $n\ge 0$, setting $N=1$ in (\ref{h_so2N|so2M+1_Wa}) and replacing $t\to t^{-1}$ gives
\begin{align}
\label{abelian_D5_contour}
\langle W_{n}\rangle_{\mathcal{D}}^{\mathfrak{so}(2)|\mathfrak{so}(2M+1)}(t^{-1};q)
&=\mathcal{Z}_{M-1}(\tau)
\frac{(q;q)_\infty^2}{(\tau;q)_\infty(\sigma;q)_\infty}
\oint\frac{ds}{2\pi i s}\,s^{n}
\frac{(q\tau^{M-1}s^{\pm};q)_\infty}{(\tau^M s^{\pm};q)_\infty}.
\end{align}
Applying the $q$-binomial theorem \cite{MR2128719} to the two factors in the integrand yields
\begin{align}
\label{abelian_D5_2phi1}
\langle W_{n}\rangle_{\mathcal{D}}^{\mathfrak{so}(2)|\mathfrak{so}(2M+1)}(t^{-1};q)
&=\tau^{Mm}\mathcal{Z}_{M-1}(\tau)
\frac{(q;q)_\infty^2}{(\tau;q)_\infty(\sigma;q)_\infty}
\frac{(\sigma;q)_{m}}{(q;q)_{m}}
\nonumber\\
&\quad\times{}_2\phi_1\!\left[
\begin{matrix}
\sigma,q^{m}\sigma\\
q^{m+1}
\end{matrix};q,\tau^{2M}
\right].
\end{align}
The result for charge $-n$ is identical by the change of variable $s\to s^{-1}$, so that the right-hand side depends on $n$ only through $m=|n|$.

Finally, Heine's transformation \cite{MR2128719}
\begin{align}
\label{abelian_Heine}
{}_2\phi_1\!\left[
\begin{matrix}a,b\\c\end{matrix};q,z
\right]
&=\frac{(b,az;q)_\infty}{(c,z;q)_\infty}
{}_2\phi_1\!\left[
\begin{matrix}c/b,z\\az\end{matrix};q,b
\right]
\end{align}
with
\begin{align}
a&=\tau,
&b&=\tau^{2M},
&c&=q\tau^{2M-1},
&z&=q^{m}\sigma
\end{align}
gives $c/b=\sigma$ and $az=q^{m+1}$.  Together with
\begin{align}
\label{abelian_Z_recursion}
\mathcal{Z}_M(\tau)
&=\mathcal{Z}_{M-1}(\tau)
\frac{(q\tau^{2M-1};q)_\infty}{(\tau^{2M};q)_\infty},\\
\label{abelian_pochhammer_ratio}
\frac{(q^{m+1};q)_\infty}{(q^{m}\sigma;q)_\infty}
&=\frac{(q;q)_\infty}{(\sigma;q)_\infty}
\frac{(\sigma;q)_{m}}{(q;q)_{m}},
\end{align}
this maps (\ref{abelian_NS5_2phi1}) precisely to (\ref{abelian_D5_2phi1}).  This proves (\ref{eq:V_{n;0}=W_{n}}) analytically for arbitrary $M\ge 1$ and $n\in\mathbb{Z}$.

For generic $N$ and $M$, the $\mathfrak{so}(2N)$ gauge algebra remains unbroken only for $N\le M$ 
in the D5-type $\mathfrak{so}(2N)|\mathfrak{so}(2M+1)$ interface. 
We therefore propose that, in this regime, 
the fundamental Wilson line at the D5-type interface is exchanged under S-duality with the minuscule vortex line 
with $k^{(1)}=(1,0^{N-1})$ and $k^{(2)}=0$ supported on the dual NS5-type $\mathfrak{so}(2N)|\mathfrak{usp}(2M)$ interface. 
A quantitative consequence of this proposal is the following identities of the line defect half-indices 
\begin{align}
\langle V_{(1,0^{N-1});0}\rangle_{\mathcal{N}}^{\mathfrak{so}(2N)|\mathfrak{usp}(2M)}(t;q)
&=\langle W_{\tiny\yng(1)}\rangle^{\mathfrak{so}(2N)|\mathfrak{so}(2M+1)}_{\mathcal{D}}(t^{-1};q), \\
\langle V_{(1,0^{N-1});0}\rangle_{\mathcal{N}}^{\mathfrak{o}(2N)^{\pm}|\mathfrak{usp}(2M)}(t;q)
&=\langle W_{\tiny\yng(1)}\rangle^{\mathfrak{o}(2N)^{\pm}|\mathfrak{so}(2M+1)}_{\mathcal{D}}(t^{-1};q) 
\end{align}
for $N\le M$. 
Here the two sides are represented by the matrix integrals 
in equations (\ref{h_so2N|usp2M_V}), (\ref{h_o2N|usp2M_V}), (\ref{h_so2N|so2M+1_Wa}) and (\ref{h_o2N|so2M+1_W}). 
We have verified these identities in several low-rank examples. 
The corresponding $q$-series expansions are listed in Appendix \ref{app_expansion}.

%%%%%%%%%%%%%%%%%%%%%%%%%%%%%%%%%%%%%%%%%%%%%%%%%
\subsubsection{Coulomb limit}
%%%%%%%%%%%%%%%%%%%%%%%%%%%%%%%%%%%%%%%%%%%%%%%%%
%Coulomb limit
In the Coulomb limit the line defect half-indices (\ref{h_so2N|so2M+1_Wa}), (\ref{h_so2N|so2M+1_Wa-}), (\ref{h_o2N|so2M+1_W}), (\ref{h_so2N|so2M+1_Wb}) and (\ref{h_so2N|so2M+1_Wb-})
coincide with the half-BPS line defect indices \cite{Hatsuda:2025jze}
\begin{align}
\label{hC_so2N|so2M+1_Wa}
{\langle W_{\mathcal{R}_1}\cdots W_{\mathcal{R}_k}\rangle_{\mathcal{D}}^{\mathfrak{so}(2N)|\mathfrak{so}(2M+1)}}^{(C)}(\mathfrak{q})
&=
\begin{cases}
\langle  W_{\mathcal{R}_1}\cdots W_{\mathcal{R}_k} \rangle_{\textrm{$\frac12$BPS}}^{\mathfrak{so}(2N)}&\textrm{for $N\le M$}\cr
\langle  W_{\mathcal{R}_1}\cdots W_{\mathcal{R}_k} \rangle_{\textrm{$\frac12$BPS}}^{\mathfrak{so}(2M+1)}&\textrm{for $N>M$}\cr
\end{cases}
, \\
\label{hC_o2N|so2M+1_W}
{\langle W_{\mathcal{R}_1}\cdots W_{\mathcal{R}_k}\rangle_{\mathcal{D}}^{\mathfrak{o}(2N)^{\pm}|\mathfrak{so}(2M+1)}}^{(C)}(\mathfrak{q})
&=
\begin{cases}
\langle W_{\mathcal{R}_1}\cdots W_{\mathcal{R}_k}\rangle_{\textrm{$\frac12$BPS}}^{\mathfrak{o}(2N)^{\pm}}&\textrm{for $N\le M$}\cr
\langle  W_{\mathcal{R}_1}\cdots W_{\mathcal{R}_k}\rangle_{\textrm{$\frac12$BPS}}^{\mathfrak{so}(2M+1)}&\textrm{for $N>M$}\cr
\end{cases}
\end{align}
The half-indices without line defect insertions are computed as \cite{Hatsuda:2024lcc}
\begin{align}
\label{hC_so2N|so2M+1}
{\mathbb{II}^{\mathfrak{so}(2N)|\mathfrak{so}(2M+1)}_{\mathcal{D}}}^{(C)}(\mathfrak{q})
&=
\begin{cases}
\frac{1}{1-\mathfrak{q}^{2N}}\prod_{n=1}^{N-1}\frac{1}{1-\mathfrak{q}^{4n}}&\textrm{for $N\le M$}\cr
\prod_{n=1}^{M}\frac{1}{1-\mathfrak{q}^{4n}}&\textrm{for $N>M$}\cr
\end{cases}, \\
\label{hC_o2N+|so2M+1}
{\mathbb{II}^{\mathfrak{o}(2N)^+|\mathfrak{so}(2M+1)}_{\mathcal{D}}}^{(C)}(\mathfrak{q})
&=
\begin{cases}
\prod_{n=1}^{N}\frac{1}{1-\mathfrak{q}^{4n}}&\textrm{for $N\le M$}\cr
\prod_{n=1}^{M}\frac{1}{1-\mathfrak{q}^{4n}}&\textrm{for $N>M$}\cr
\end{cases}, \\
\label{hC_o2N-|so2M+1}
{\mathbb{II}^{\mathfrak{o}(2N)^-|\mathfrak{so}(2M+1)}_{\mathcal{D}}}^{(C)}(\mathfrak{q})
&=
\begin{cases}
\mathfrak{q}^{2N}\prod_{n=1}^{N}\frac{1}{1-\mathfrak{q}^{4n}}&\textrm{for $N\le M$}\cr
\prod_{n=1}^{M}\frac{1}{1-\mathfrak{q}^{4n}}&\textrm{for $N>M$}\cr
\end{cases}. 
\end{align}
It follows from the half-BPS line defect indices \cite{Hatsuda:2025jze} 
that for a single Wilson line labeled by an irreducible representation $\mathcal{R}$ without zero weight, the line defect half-index vanishes. 
This is the case for the fundamental representation of $\mathfrak{so}(2N)$, 
whereas the fundamental representation of $\mathfrak{so}(2M+1)$ relevant for $N>M$ contains the zero weight 
and yields $\mathfrak{q}^{2M}$ according to (\ref{halfBPS_K}). 
When $N\le M$, this agrees with the Higgs limit behaviors of the proposed S-dual minuscule vortex line defect half-indices (\ref{hH_so2N|usp2M_V}) and (\ref{hH_o2N|usp2M_V}) 
for the NS5-type $\mathfrak{so}(2N)|\mathfrak{usp}(2M)$ interface, providing a non-trivial consistency check of the duality proposal. 

%Witten ind
The line defect Witten index is naturally captured by the $\mathfrak{q}\rightarrow 1$ limit of the corresponding half-BPS line defect indices. 
We conjecture that it is given by the constant term of the product of the characters
\begin{align}
{\langle W_{\mathcal{R}_1}\cdots W_{\mathcal{R}_k}\rangle_{\mathcal{D}}^{\mathfrak{so}(2N)|\mathfrak{so}(2M+1)}}^{(C)}(1)
&=
\begin{cases}
\oint \prod_{i=1}^{N}\frac{ds_i}{2\pi is_i}
\prod_{j=1}^k \chi_{\mathcal{R}_j}^{\mathfrak{so}(2N)}(s)&\textrm{for $N\le M$}\cr
\oint \prod_{i=1}^{M}\frac{ds_i}{2\pi is_i}
\prod_{j=1}^k \chi_{\mathcal{R}_j}^{\mathfrak{so}(2M+1)}(s)&\textrm{for $N>M$}\cr
\end{cases}. 
\end{align}
As for the $\mathfrak{usp}(2N)|\mathfrak{usp}(2M)'$ interface, 
this follows from (\ref{halfBPS_K}) and Kostant's theorem $K_{\lambda,0}(1)=\dim \mathcal{R}[0]$. 

%%%%%%%%%%%%%%%%%%%%%%%%%%%%%%%%%%%%%%%%%%%%%%%%%
\subsubsection{Higgs limit}
%%%%%%%%%%%%%%%%%%%%%%%%%%%%%%%%%%%%%%%%%%%%%%%%%
%Higgs limit
In the Higgs limit the line defect half-indices reduce to matrix integrals, 
which cannot be simply expressed as the half-BPS line defect indices.  
For $N\le M$ we get
\begin{align}
\label{hH_so2N|so2M+1_Wa}
&
{\langle W_{\mathcal{R}_1}\cdots W_{\mathcal{R}_k}\rangle_{\mathcal{D}}^{\mathfrak{so}(2N)|\mathfrak{so}(2M+1)}}^{(H)}(\mathfrak{q})
\nonumber\\
&=\frac{1}{2^{N-1}N!}\frac{1}{(1-\mathfrak{q}^2)^N}
\oint \prod_{i=1}^N \frac{ds_i}{2\pi is_i}
\prod_{i<j}\frac{(1-s_i^{\pm}s_j^{\pm})}
{(1-\mathfrak{q}^2s_i^{\pm}s_j^{\pm})}
\nonumber\\
&\times 
\prod_{\ell=1}^{M-N}
\frac{1}{1-\mathfrak{q}^{4\ell}}
\prod_{i=1}^N \frac{1}{1-\mathfrak{q}^{2+2(M-N)}s_i^{\pm}}
\prod_{j=1}^k \chi_{\mathcal{R}_j}^{\mathfrak{so}(2N)}(s), \\
&
\label{hH_so2N|so2M+1_Wa-}
{\langle W_{\mathcal{R}_1}\cdots W_{\mathcal{R}_k}\rangle_{\mathcal{D}}^{\mathfrak{so}(2N)^-|\mathfrak{so}(2M+1)}}^{(H)}(\mathfrak{q})
\nonumber\\
&=\frac{1}{2^{N-1}(N-1)!}
\frac{1}{(1-\mathfrak{q}^2)^{N-1}(1+\mathfrak{q}^2)}
\oint \prod_{i=1}^{N-1}
\frac{ds_i}{2\pi is_i}\frac{(1-s_i^{\pm})(1+s_i^{\pm})}{(1-\mathfrak{q}^2s_i^{\pm})(1+\mathfrak{q}^2s_i^{\pm})}
\prod_{i<j}\frac{(1-s_i^{\pm}s_j^{\pm})}{(1-\mathfrak{q}^2s_i^{\pm}s_j^{\pm})}
\nonumber\\
&\times 
\prod_{\ell=1}^{M-N}
\frac{1}{1-\mathfrak{q}^{4\ell}}
\prod_{i=1}^{N-1}
\frac{1}{1-\mathfrak{q}^{2+2(M-N)}s_i^{\pm}}
\frac{1}{(1- \mathfrak{q}^{2+2(M-N)})(1+ \mathfrak{q}^{2+2(M-N)})}
\prod_{j=1}^k \chi_{\mathcal{R}_j}^{\mathfrak{usp}(2N-2)}(s). 
\end{align}
When $N>M$ we obtain
\begin{align}
\label{hH_so2N|so2M+1_Wb}
&
{\langle W_{\mathcal{R}_1}\cdots W_{\mathcal{R}_k}\rangle_{\mathcal{D}}^{\mathfrak{so}(2N)|\mathfrak{so}(2M+1)}}^{(H)}(\mathfrak{q})
\nonumber\\
&=\frac{1}{2^M M!}
\frac{1}{(1-\mathfrak{q}^{2})^M}
\oint \prod_{i=1}^M \frac{ds_i}{2\pi is_i}
\frac{(1-s_i^{\pm})}{(1-\mathfrak{q}^2s_i^{\pm})}
\prod_{i<j}\frac{(1-s_i^{\pm}s_j^{\pm})}{(1-\mathfrak{q}^2s_i^{\pm}s_j^{\pm})}
\nonumber\\
&\times 
\frac{1}{1-\mathfrak{q}^{2(N-M)}}
\prod_{\ell=1}^{N-M-1}
\frac{1}{1-\mathfrak{q}^{4\ell}}
\prod_{i=1}^M \frac{1}{1-\mathfrak{q}^{2+2(N-M-1)}s_i^{\pm}}
\prod_{j=1}^k \chi_{\mathcal{R}_j}^{\mathfrak{so}(2M+1)}(s), \\
&
\label{hH_so2N|so2M+1_Wb-}
{\langle W_{\mathcal{R}_1}\cdots W_{\mathcal{R}_k}\rangle_{\mathcal{D}}^{\mathfrak{so}(2N)^-|\mathfrak{so}(2M+1)}}^{(H)}(\mathfrak{q})
\nonumber\\
&=\frac{1}{2^M M!}
\frac{1}{(1-\mathfrak{q}^2)^M}
\oint \prod_{i=1}^M \frac{ds_i}{2\pi is_i}\frac{(1-s_i^{\pm})}{(1-\mathfrak{q}^2s_i^{\pm})}
\prod_{i<j}\frac{(1-s_i^{\pm}s_j^{\pm})}{(1-\mathfrak{q}^2s_i^{\pm}s_j^{\pm})}
\nonumber\\
&\times 
\frac{1}{1+\mathfrak{q}^{2(N-M)}}
\prod_{\ell=1}^{N-M-1}
\frac{1}{1-\mathfrak{q}^{4\ell}}
\prod_{i=1}^M \frac{1}{(1+\mathfrak{q}^{2+2(N-M-1)}s_i^{\pm})}
\prod_{j=1}^k \chi_{\mathcal{R}_j}^{\mathfrak{so}(2M+1)}(s). 
\end{align}
The Higgs half-indices without line defect insertions are computed as \cite{Hatsuda:2024lcc}
\begin{align}
\label{hH_so2N|so2M+1}
{\mathbb{II}^{\mathfrak{so}(2N)|\mathfrak{so}(2M+1)}_{\mathcal{D}}}^{(H)}(\mathfrak{q})
&=\frac{1}{1-\mathfrak{q}^{2N}}\prod_{n=1}^{N-1}\frac{1}{1-\mathfrak{q}^{4n}} \prod_{n=1}^{M}\frac{1}{1-\mathfrak{q}^{4n}}, \\
{\mathbb{II}^{\mathfrak{o}(2N)^+|\mathfrak{so}(2M+1)}_{\mathcal{D}}}^{(H)}(\mathfrak{q})
&= \prod_{n=1}^{N}\frac{1}{1-\mathfrak{q}^{4n}} \prod_{n=1}^{M}\frac{1}{1-\mathfrak{q}^{4n}}, \\
{\mathbb{II}^{\mathfrak{o}(2N)^-|\mathfrak{so}(2M+1)}_{\mathcal{D}}}^{(H)}(\mathfrak{q})
&=\mathfrak{q}^{2N} \prod_{n=1}^{N}\frac{1}{1-\mathfrak{q}^{4n}} \prod_{n=1}^{M}\frac{1}{1-\mathfrak{q}^{4n}}. 
\end{align}

We now consider several explicit examples of the interfaces with lower-rank gauge groups. 
For example, in the case with $N\le M\le 3$, 
the line defect half-indices associated with the fundamental Wilson line can be computed as follows:
\begin{align}
{\langle \mathcal{W}_{\tiny \yng(1)}\rangle_{\mathcal{D}}^{\mathfrak{so}(2)|\mathfrak{so}(3)}}^{(H)}(\mathfrak{q})
&=2\mathfrak{q}^2, \\
{\langle \mathcal{W}_{\tiny \yng(1)}\rangle_{\mathcal{D}}^{\mathfrak{so}(2)|\mathfrak{so}(5)}}^{(H)}(\mathfrak{q})
&=2\mathfrak{q}^4, \\
{\langle \mathcal{W}_{\tiny \yng(1)}\rangle_{\mathcal{D}}^{\mathfrak{so}(2)|\mathfrak{so}(7)}}^{(H)}(\mathfrak{q})
&=2\mathfrak{q}^6, \\
{\langle \mathcal{W}_{\tiny \yng(1)}\rangle_{\mathcal{D}}^{\mathfrak{so}(4)|\mathfrak{so}(5)}}^{(H)}(\mathfrak{q})
&=\mathfrak{q}^2+2\mathfrak{q}^4+\mathfrak{q}^6, \\
{\langle \mathcal{W}_{\tiny \yng(1)}\rangle_{\mathcal{D}}^{\mathfrak{so}(4)|\mathfrak{so}(7)}}^{(H)}(\mathfrak{q})
&=\mathfrak{q}^4+2\mathfrak{q}^6+\mathfrak{q}^8, \\
{\langle \mathcal{W}_{\tiny \yng(1)}\rangle_{\mathcal{D}}^{\mathfrak{so}(6)|\mathfrak{so}(7)}}^{(H)}(\mathfrak{q})
&=\mathfrak{q}^2+\mathfrak{q}^4+2\mathfrak{q}^6+\mathfrak{q}^8+\mathfrak{q}^{10}. 
\end{align}
For $M<N\le 4$ the line defect half-indices associated with the fundamental Wilson line are evaluated as
\begin{align}
{\langle \mathcal{W}_{\tiny \yng(1)}\rangle_{\mathcal{D}}^{\mathfrak{so}(4)|\mathfrak{so}(3)}}^{(H)}(\mathfrak{q})
&=\mathfrak{q}^2(2+\mathfrak{q}^2), \\
{\langle \mathcal{W}_{\tiny \yng(1)}\rangle_{\mathcal{D}}^{\mathfrak{so}(6)|\mathfrak{so}(3)}}^{(H)}(\mathfrak{q})
&=\mathfrak{q}^2+\mathfrak{q}^4+\mathfrak{q}^6, \\
{\langle \mathcal{W}_{\tiny \yng(1)}\rangle_{\mathcal{D}}^{\mathfrak{so}(8)|\mathfrak{so}(3)}}^{(H)}(\mathfrak{q})
&=\mathfrak{q}^2+\mathfrak{q}^6+\mathfrak{q}^8, \\
{\langle \mathcal{W}_{\tiny \yng(1)}\rangle_{\mathcal{D}}^{\mathfrak{so}(6)|\mathfrak{so}(5)}}^{(H)}(\mathfrak{q})
&=\mathfrak{q}^2+2\mathfrak{q}^4+\mathfrak{q}^6+\mathfrak{q}^8, \\
{\langle \mathcal{W}_{\tiny \yng(1)}\rangle_{\mathcal{D}}^{\mathfrak{so}(8)|\mathfrak{so}(5)}}^{(H)}(\mathfrak{q})
&=2\mathfrak{q}^4+\mathfrak{q}^6+\mathfrak{q}^8+\mathfrak{q}^{10}, \\
{\langle \mathcal{W}_{\tiny \yng(1)}\rangle_{\mathcal{D}}^{\mathfrak{so}(8)|\mathfrak{so}(7)}}^{(H)}(\mathfrak{q})
&=\mathfrak{q}^2+\mathfrak{q}^4+2\mathfrak{q}^6+\mathfrak{q}^8+\mathfrak{q}^{10}+\mathfrak{q}^{12}. 
\end{align}

%Witten ind
The exact computations suggest that 
the corresponding line defect Witten index, extracted from the $\mathfrak{q}\rightarrow 1$ limit, is given by the following formula:
\begin{align}
{\langle \mathcal{W}_{\mathcal{R}_1} \cdots \mathcal{W}_{\mathcal{R}_{k}} \rangle_{\mathcal{D}}^{\mathfrak{so}(2N)|\mathfrak{so}(2M+1)}}^{(H)}(1)
&=
\begin{cases}
\prod_{j=1}^k \chi_{\mathcal{R}_j}^{\mathfrak{so}(2N)}(1)&\textrm{for $N\le M$}\cr
\prod_{j=1}^k \chi_{\mathcal{R}_j}^{\mathfrak{so}(2M+1)}(1)&\textrm{for $N> M$}\cr
\end{cases}, 
\end{align}
where 
\begin{align}
\chi_{\mathcal{R}}^{\mathfrak{so}(2n)}(1)
&=\prod_{1\le i<j\le n}
\frac{(\lambda_i-\lambda_j+j-i) (\lambda_i+\lambda_j+2n-i-j)}{(j-i) (2n-i-j)}
\end{align}
and 
\begin{align}
\chi_{\mathcal{R}}^{\mathfrak{so}(2n+1)}(1)
&=\prod_{1\le i<j\le n}
\frac{\lambda_i-\lambda_j+j-i}{j-i}
\prod_{1\le i\le j\le n}
\frac{\lambda_i+\lambda_j+2n+1-i-j}{2n+1-i-j}
\end{align}
are the dimensions of the $SO(2n)$ and $SO(2n+1)$ irreducible representations 
labeled by the Young diagram $\lambda$ respectively. 
As for the $\mathfrak{usp}(2N)|\mathfrak{usp}(2M)'$ interface, 
this follows from the triviality of the residual gauge group at a generic point of the Higgs branch (see Appendix \ref{app_residual}). 
We have also confirmed it for non-minuscule representations, e.g. 
\begin{align}
{\langle \mathcal{W}_{\tiny \yng(2)}\rangle_{\mathcal{D}}^{\mathfrak{so}(4)|\mathfrak{so}(5)}}^{(H)}(\mathfrak{q})
&=2\mathfrak{q}^4+2\mathfrak{q}^6+3\mathfrak{q}^8+2\mathfrak{q}^{10}, \\
{\langle \mathcal{W}_{\tiny \yng(1,1)}\rangle_{\mathcal{D}}^{\mathfrak{so}(6)|\mathfrak{so}(5)}}^{(H)}(\mathfrak{q})
&=\mathfrak{q}^2+\mathfrak{q}^4+3\mathfrak{q}^6+2\mathfrak{q}^8+2\mathfrak{q}^{10}+\mathfrak{q}^{12}, 
\end{align}
whose values at $\mathfrak{q}=1$ are $9$ and $10$. 

%%%%%%%%%%%%%%%%%%%%%%%%%%%%%%%%%%%%%%%%%%%%%%%%%
\subsection{Vortex lines}
\label{sec_DCD5V}
%%%%%%%%%%%%%%%%%%%%%%%%%%%%%%%%%%%%%%%%%%%%%%%%%
%vortex
Let us consider the vortex lines at the D5-type $\mathfrak{so}(2N)|\mathfrak{so}(2M+1)$ interface. 

%N\le M
When $N\le M$, they are labeled by a magnetic charge $k$ for the preserved gauge algebra $\mathfrak{so}(2N)$ at the interface. 
For the $SO(2N)$ gauge group, the magnetic lines can carry integral magnetic charges. 
The magnetic lines of $k=(1,0^{N-1})$ associated with the fundamental coweight are minuscule 
and therefore free from the monopole bubbling effect. 

%half-index
Here we propose that
the line defect half-index of the minuscule vortex line with $k=(1,0^{N-1})$ 
at the D5-type $\mathfrak{so}(2N)|\mathfrak{so}(2M+1)$ interface with $N\le M$ is given by
\begin{align}
\label{h_so2N|so2M+1_Va}
&
\langle V_{(1,0^{N-1})}\rangle^{\mathfrak{so}(2N)|\mathfrak{so}(2M+1)}_{\mathcal{D}}(t;q)
\nonumber\\
&=q^{\frac14}t^{-1}
\frac{1}{2^{N-2}(N-1)!}
\frac{(q)_{\infty}^{2N}}{(q^{\frac12}t^{\pm2};q)_{\infty}^N}
\oint \prod_{i=1}^N \frac{ds_i}{2\pi is_i}
\nonumber\\
&\times 
\prod_{i=2}^N
\frac{(q^{\frac12}s_1^{\pm}s_i^{\pm};q)_{\infty}
(q^{\frac32}s_1^{\pm}s_i^{\pm};q)_{\infty}}
{(qt^2s_1^{\pm}s_i^{\pm};q)_{\infty}(qt^{-2}s_1^{\pm}s_i^{\pm};q)_{\infty}}
\prod_{2\le i<j\le N}
\frac{(s_i^{\pm}s_j^{\pm};q)_{\infty}
(qs_i^{\pm}s_j^{\pm};q)_{\infty}}
{(q^{\frac12}t^2s_i^{\pm}s_j^{\pm};q)_{\infty}(q^{\frac12}t^{-2}s_i^{\pm}s_j^{\pm};q)_{\infty}}
\nonumber\\
&\times 
\prod_{\ell=1}^{M-N}
\frac{(q^{\frac12+\ell}t^{4\ell-2};q)_{\infty}}
{(q^{\ell} t^{4\ell};q)_{\infty}}
\frac{(q^{1+\frac{M-N}{2}+\frac12}t^{2(M-N)}s_1^{\pm};q)_{\infty}}
{(q^{\frac12+\frac{M-N}{2}+\frac12}t^{2+2(M-N)}s_1^{\pm};q)_{\infty}}
\prod_{i=2}^{N}
\frac{(q^{1+\frac{M-N}{2}}t^{2(M-N)}s_i^{\pm};q)_{\infty}}
{(q^{\frac12+\frac{M-N}{2}}t^{2+2(M-N)}s_i^{\pm};q)_{\infty}}. 
\end{align}

%%%%%%%%%%%%%%%%%%%%%%%%%%%%%%%%%%%%%%%%%%%%%%%%%
%ADDED
As in the NS5-type case, the expression (\ref{h_so2N|so2M+1_Va}) is obtained 
by shifting the fugacity of a field of weight $\rho$ by $q^{\frac{|\rho(k)|}{2}}$ 
and by including the overall factor $q^{\frac{\Delta(k)}{4}}t^{-\Delta(k)}$ 
of the background bare monopole operator. 
For $k=(1,0^{N-1})$ the only fields charged under $k$ are 
the 3d fundamental hypermultiplet of weights $\pm e_1$ 
and the components of the 4d vector multiplet associated with the roots $\pm e_1\pm e_i$ with $i\neq 1$, 
whose fugacities are accordingly multiplied by $q^{\frac12}$. 
A new feature of the D5-type interface is that the root contribution to (\ref{bare_monopole_D}) is absent. 
This is because the 4d vector multiplet obeying the D5-type boundary condition 
contributes to the integrand of (\ref{h_so2N|so2M+1_Wa}) the balanced factor 
$(s^{\alpha};q)_{\infty}(qs^{\alpha};q)_{\infty}/[(q^{\frac12}t^2s^{\alpha};q)_{\infty}(q^{\frac12}t^{-2}s^{\alpha};q)_{\infty}]$, 
in which the numerator and the denominator involve the same number of $q$-Pochhammer factors, 
so that the shift $q^{\frac{|\alpha(k)|}{2}}$ cancels out in the associated R-charge, 
in contrast to the Neumann-type factor 
$(s^{\alpha};q)_{\infty}/(q^{\frac12}t^{-2}s^{\alpha};q)_{\infty}$ of the NS5-type interface. 
Hence only the 3d fundamental hypermultiplet contributes and 
\begin{align}
\label{bare_monopole_N}
\Delta(k)&=\frac12\sum_{\rho}|\rho(k)|
=\frac12\cdot 2=1, 
\end{align}
which reproduces the prefactor $q^{\frac14}t^{-1}$ in (\ref{h_so2N|so2M+1_Va}) 
as well as the leading power $\mathfrak{q}$ of its Coulomb limit. 

Again the case with $N=1$ deserves a separate treatment. 
The expression (\ref{h_so2N|so2M+1_Va}) is written for $N\ge 2$ 
and for $N=1$ the factor $\frac{1}{2^{N-2}(N-1)!}$, 
which is the inverse of the order of the Weyl group of the residual $U(1)\times SO(2N-2)$ gauge symmetry, 
is replaced by $1$ since the Weyl group of the preserved $U(1)$ gauge symmetry is trivial. 
As the Abelian gauge algebra $\mathfrak{so}(2)$ has no root, 
the vortex lines are free from the monopole bubbling effects 
and they are completely characterized by an arbitrary magnetic charge $n\in \mathbb{Z}$. 
The 4d vector multiplet factor is intact, 
while the 3d fundamental hypermultiplet of the $\mathfrak{so}(2)$ charges $\pm1$ is shifted by $q^{\frac{|n|}{2}}$ 
and it gives rise to $\Delta(n)=\frac12\cdot 2|n|=|n|$ in (\ref{bare_monopole_N}). 
Hence we propose that the line defect half-index of the vortex line of magnetic charge $n$ 
at the D5-type $\mathfrak{so}(2)|\mathfrak{so}(2M+1)$ interface is given by
\begin{align}
\label{h_so2|so2M+1_V}
&
\langle V_{n}\rangle^{\mathfrak{so}(2)|\mathfrak{so}(2M+1)}_{\mathcal{D}}(t;q)
\nonumber\\
&=q^{\frac{|n|}{4}}t^{-|n|}
\frac{(q)_{\infty}^{2}}{(q^{\frac12}t^{\pm2};q)_{\infty}}
\oint \frac{ds}{2\pi is}
\prod_{\ell=1}^{M-1}
\frac{(q^{\frac12+\ell}t^{4\ell-2};q)_{\infty}}
{(q^{\ell} t^{4\ell};q)_{\infty}}
\frac{(q^{1+\frac{M-1}{2}+\frac{|n|}{2}}t^{2(M-1)}s^{\pm};q)_{\infty}}
{(q^{\frac12+\frac{M-1}{2}+\frac{|n|}{2}}t^{2+2(M-1)}s^{\pm};q)_{\infty}}. 
\end{align}
It depends on $n$ only through $|n|$ as the charge conjugation exchanges the vortex lines of charge $n$ and $-n$, 
and it reduces to (\ref{h_so2N|so2M+1_Va}) with $N=1$ for $n=\pm1$. 
We have checked that the $q$-expansion of (\ref{h_so2|so2M+1_V}) coincides with that of 
$\langle W_{n;(n)}\rangle_{\mathcal{N}}^{\mathfrak{so}(2)|\mathfrak{usp}(2M)}(t^{-1};q)$ 
obtained from (\ref{h_so2N|usp2M_W}) with $N=1$ for $M=1,2$ and $n=1,2,3$. 
In fact this agreement is proved analytically for arbitrary $M\ge 1$ and $n\in\mathbb{Z}$ in section \ref{sec_DCD5V_Sduality}. 
The resulting $q$-series are presented in Appendix \ref{app_expansion}. 
%%%%%%%%%%%%%%%%%%%%%%%%%%%%%%%%%%%%%%%%%%%%%%%%%

%O(2N)
For the D5-type $\mathfrak{o}(2N)^{\pm}|\mathfrak{so}(2M+1)$ interface with $N\le M$ 
there is a contribution from the twisted sector associated with the disconnected component of $\mathfrak{o}(2N)$ with the form 
\begin{align}
\label{h_so2N|so2M+1_Va-}
&
\langle V_{(1,0^{N-2})}\rangle^{\mathfrak{so}(2N)^-|\mathfrak{so}(2M+1)}_{\mathcal{D}}(t;q)
\nonumber\\
&=
q^{\frac14}t^{-1}
\frac{1}{2^{N-2} (N-2)!} 
\frac{(q)_{\infty}^{2N-2} (-q;q)_{\infty}^2}
{(q^{\frac12}t^{\pm2};q)_{\infty}^{N-1}(-q^{\frac12}t^{\pm2};q)_{\infty}}
\nonumber\\
&\times 
\oint \prod_{i=1}^{N-1} \frac{ds_i}{2\pi is_i}
\frac{(q^{\frac12}s_1^{\pm};q)_{\infty}(-q^{\frac12}s_1^{\pm};q)_{\infty} (q^{\frac32}s_1^{\pm};q)_{\infty}(-q^{\frac32}s_1^{\pm};q)_{\infty}}
{(qt^{2}s_1^{\pm};q)_{\infty}(-qt^{2}s_1^{\pm};q)_{\infty}(qt^{-2}s_1^{\pm};q)_{\infty}(-qt^{-2}s_1^{\pm};q)_{\infty}}
\nonumber\\
&\times
\prod_{i=2}^{N-1} 
\frac{(s_i^{\pm};q)_{\infty}(-s_i^{\pm};q)_{\infty} (qs_i^{\pm};q)_{\infty}(-qs_i^{\pm};q)_{\infty}}
{(q^{\frac12}t^{2}s_i^{\pm};q)_{\infty}(-q^{\frac12}t^{2}s_i^{\pm};q)_{\infty}(q^{\frac12}t^{-2}s_i^{\pm};q)_{\infty}(-q^{\frac12}t^{-2}s_i^{\pm};q)_{\infty}}
\nonumber\\
&\times 
\prod_{j=2}^{N-1}
\frac{(q^{\frac12}s_1^{\pm}s_j^{\pm};q)_{\infty}(q^{\frac32}s_1^{\pm}s_j^{\pm};q)_{\infty}}
{(qt^{2}s_1^{\pm}s_j^{\pm};q)_{\infty}(qt^{-2}s_1^{\pm}s_j^{\pm};q)_{\infty}}
\prod_{2\le i<j\le N-1}
\frac{(s_i^{\pm}s_j^{\pm};q)_{\infty}(qs_i^{\pm}s_j^{\pm};q)_{\infty}}
{(q^{\frac12}t^{2}s_i^{\pm}s_j^{\pm};q)_{\infty}(q^{\frac12}t^{-2}s_i^{\pm}s_j^{\pm};q)_{\infty}}
\nonumber\\
&\times 
\prod_{\ell=1}^{M-N}
\frac{(q^{\frac12+\ell}t^{4\ell-2};q)_{\infty}}
{(q^{\ell} t^{4\ell};q)_{\infty}}
\frac{(q^{\frac32+\frac{M-N}{2}} t^{2(M-N)} s_1^{\pm};q)_{\infty}}
{(q^{1+\frac{M-N}{2}}t^{2+2(M-N)} s_1^{\pm};q)_{\infty}}
\nonumber\\
&\times 
\prod_{i=2}^{N-1}
\frac{(q^{1+\frac{M-N}{2}} t^{2(M-N)} s_i^{\pm};q)_{\infty}}{(q^{\frac12+\frac{M-N}{2}}t^{2+2(M-N)} s_i^{\pm};q)_{\infty}}
\prod_{\epsilon=\pm1}
\frac{(\epsilon q^{1+\frac{M-N}{2}} t^{2(M-N)};q)_{\infty}}{(\epsilon q^{\frac12+\frac{M-N}{2}}t^{2+2(M-N)};q)_{\infty}}. 
\end{align}
So we have
\begin{align}
\label{h_o2N|so2M+1_Va}
&
\langle V_{(1,0^{N-1})}\rangle^{\mathfrak{o}(2N)^{\pm}|\mathfrak{so}(2M+1)}_{\mathcal{D}}(t;q)
\nonumber\\
&=\frac12
\left(
\langle V_{(1,0^{N-1})}\rangle^{\mathfrak{so}(2N)|\mathfrak{so}(2M+1)}_{\mathcal{D}}(t;q)
\pm 
\langle V_{(1,0^{N-2})}\rangle^{\mathfrak{so}(2N)^-|\mathfrak{so}(2M+1)}_{\mathcal{D}}(t;q)
\right). 
\end{align}

%N>M 
When $N>M$, the preserved gauge algebra of the D5-type $\mathfrak{so}(2N)|\mathfrak{so}(2M+1)$ interface is $\mathfrak{so}(2M+1)$, 
and the relevant vortex line operators are classified by magnetic charges of the preserved algebra. 
Since magnetic charges $k$ $=$ $(1,0^{M-1})$ corresponding to the fundamental coweights are minuscule, 
the associated vortex lines do not receive monopole bubbling effects. 
This leads us to propose that the corresponding vortex line defect half-index is given by the following spin shift prescription:
\begin{align}
\label{h_so2N|so2M+1_Vb}
&
\langle V_{(1,0^{M-1})}\rangle^{\mathfrak{so}(2N)|\mathfrak{so}(2M+1)}_{\mathcal{D}}(t;q)
\nonumber\\
&=q^{\frac14}t^{-1}
\frac{1}{2^{M-1}(M-1)!}
\frac{(q)_{\infty}^{2M}}{(q^{\frac12}t^{\pm2};q)_{\infty}^{M}}
\oint \prod_{i=1}^M 
\frac{ds_i}{2\pi is_i}
\frac{(q^{\frac12\delta_{i,1}}s_i^{\pm};q)_{\infty} (q^{1+\frac12\delta_{i,1}}s_i^{\pm};q)_{\infty}}
{(q^{\frac{1+\delta_{i,1}}{2}}t^2s_i^{\pm};q)_{\infty} (q^{\frac{1+\delta_{i,1}}{2}}t^{-2}s_i^{\pm};q)_{\infty}}
\nonumber\\
&\times \prod_{i<j}
\frac{(q^{\frac12 \delta_{i,1}}s_i^{\pm}s_j^{\pm};q)_{\infty}(q^{1+\frac12\delta_{i,1}}s_i^{\pm}s_j^{\pm};q)_{\infty}}
{(q^{\frac{1+\delta_{i,1}}{2}}t^{2}s_i^{\pm}s_j^{\pm};q)_{\infty}(q^{\frac{1+\delta_{i,1}}{2}}t^{-2}s_i^{\pm}s_j^{\pm};q)_{\infty}}
\frac{(q^{\frac12+\frac{N-M}{2}} t^{2(N-M)-2};q)_{\infty}}
{(q^{\frac{N-M}{2}}t^{2(N-M)};q)_{\infty}}
\prod_{\ell=1}^{N-M-1}
\frac{(q^{\frac12+\ell}t^{4\ell-2};q)_{\infty}}
{(q^{\ell} t^{4\ell};q)_{\infty}}
\nonumber\\
&\times 
\prod_{i=1}^{M}
\frac{(q^{1+\frac{N-M-1+\delta_{i,1}}{2}} t^{2(N-M-1)} s_i^{\pm};q)_{\infty}}{(q^{\frac12+\frac{N-M-1+\delta_{i,1}}{2}}t^{2+2(N-M-1)} s_i^{\pm};q)_{\infty}}. 
\end{align}
Similarly, upon incorporating the spin shift, 
we have the following expression for the line defect half-index in the twisted sector:
\begin{align}
\label{h_so2N|so2M+1_Vb-}
&
\langle V_{(1,0^{M-1})}\rangle^{\mathfrak{so}(2N)^{-}|\mathfrak{so}(2M+1)}_{\mathcal{D}}(t;q)
\nonumber\\
&=q^{\frac14}t^{-1}
\frac{1}{2^{M-1} (M-1)!} \frac{(q)_{\infty}^{2M}}{(q^{\frac12}t^{\pm2};q)_{\infty}^M}
\oint \prod_{i=1}^{M} \frac{ds_i}{2\pi is_i}
\frac{(q^{\frac12 \delta_{i,1}}s_i^{\pm};q)_{\infty} (q^{1+\frac12\delta_{i,1}}s_i^{\pm};q)_{\infty}}
{(q^{\frac{1+\delta_{i,1}}{2}}t^2 s_i^{\pm};q)_{\infty}(q^{\frac{1+\delta_{i,1}}{2}}t^{-2}s_i^{\pm};q)_{\infty}}
\nonumber\\
&\times \prod_{i<j}
\frac{(q^{\frac12\delta_{i,1}}s_i^{\pm}s_j^{\pm};q)_{\infty}(q^{1+\frac12\delta_{i,1}}s_i^{\pm}s_j^{\pm};q)_{\infty}}
{(q^{\frac{1+\delta_{i,1}}{2}}t^{2}s_i^{\pm}s_j^{\pm};q)_{\infty}(q^{\frac{1+\delta_{i,1}}{2}}t^{-2}s_i^{\pm}s_j^{\pm};q)_{\infty}}
\frac{(-q^{\frac12+\frac{N-M}{2}} t^{2(N-M)-2};q)_{\infty}}
{(-q^{\frac{N-M}{2}}t^{2(N-M)};q)_{\infty}}
\prod_{\ell=1}^{N-M-1}
\frac{(q^{\frac12+\ell}t^{4\ell-2};q)_{\infty}}
{(q^{\ell} t^{4\ell};q)_{\infty}}
\nonumber\\
&\times 
\prod_{i=1}^{M}
\frac{(-q^{1+\frac{N-M-1+\delta_{i,1}}{2}} t^{2(N-M-1)} s_i^{\pm};q)_{\infty}}{(-q^{\frac12+\frac{N-M-1+\delta_{i,1}}{2}}t^{2+2(N-M-1)} s_i^{\pm};q)_{\infty}}. 
\end{align}
Combining (\ref{h_so2N|so2M+1_Vb}) and (\ref{h_so2N|so2M+1_Vb-}), 
we obtain the vortex line defect half-index for the D5-type $\mathfrak{o}(2N)^{\pm}|\mathfrak{so}(2M+1)$ interface for $N>M$
\begin{align}
\label{h_o2N|so2M+1_Vb}
&
\langle V_{(1,0^{M-1})}\rangle^{\mathfrak{o}(2N)^{\pm}|\mathfrak{so}(2M+1)}_{\mathcal{D}}(t;q)
\nonumber\\
&=\frac12
\left(
\langle V_{(1,0^{M-1})}\rangle^{\mathfrak{so}(2N)|\mathfrak{so}(2M+1)}_{\mathcal{D}}(t;q)
\pm 
\langle V_{(1,0^{M-1})}\rangle^{\mathfrak{so}(2N)^-|\mathfrak{so}(2M+1)}_{\mathcal{D}}(t;q)
\right). 
\end{align}

%%%%%%%%%%%%%%%%%%%%%%%%%%%%%%%%%%%%%%%%%%%%%%%%%
\subsubsection{S-duality}
\label{sec_DCD5V_Sduality}
%%%%%%%%%%%%%%%%%%%%%%%%%%%%%%%%%%%%%%%%%%%%%%%%%
We are now in a position to discuss the action of S-duality on the interface line operators. 
The D5-type $\mathfrak{so}(2N)|\mathfrak{so}(2M+1)$ interface preserves an orthogonal gauge algebra. 
For gauge theories with orthogonal gauge groups, line operators carrying magnetic charges 
associated with the fundamental weights are minuscule and do not receive monopole bubbling corrections. 
As a consequence, the action of S-duality can be realized more transparently than for the $\mathfrak{usp}(2N)|\mathfrak{usp}(2M)'$ interface. 

%so(2)|USp(2M)
For $N=1$ the gauge group associated with $\mathfrak{so}(2N)$ is Abelian and isomorphic to $U(1)$. 
Hence the irreducible representations of $\mathfrak{so}(2)$ are one-dimensional 
and the Wilson line at the NS5-type interface is classified by an integer charge $n\in \mathbb{Z}$ 
and the irreducible representation of $\mathfrak{usp}(2M)$. 
On the other hand, the magnetic vortex lines are labeled by coweights of the Abelian gauge group and therefore carry a well-defined magnetic charge. 
In the Abelian setting, monopole bubbling is expected to be absent, so that the magnetic charge provides a complete characterization of the defect. 
In particular, we conjecture that the Wilson line $W_{n;(n)}$ on the NS5-type $\mathfrak{so}(2)|\mathfrak{usp}(2M)$ interface, 
carrying electric charge $n$ and transforming in the rank-$n$ symmetric representation, 
is mapped to the vortex line $V_{n}$ of magnetic charge $n$ on the D5-type $\mathfrak{so}(2)|\mathfrak{so}(2M+1)$ interface. 
Correspondingly, we find the following agreement of the line defect half-indices:
\begin{align}
\langle W_{n;(n)}\rangle_{\mathcal{N}}^{\mathfrak{so}(2)|\mathfrak{usp}(2M)}(t;q)
&=\langle V_{n}\rangle^{\mathfrak{so}(2)|\mathfrak{so}(2M+1)}_{\mathcal{D}}(t^{-1};q). 
\label{eq:W_{n;(n)}=V_{n}}
\end{align}
Here the left-hand side is given by (\ref{h_so2N|usp2M_W}) with $N=1$, 
where $\chi_{\mathcal{R}_1}^{\mathfrak{so}(2)}(s^{(1)})=(s^{(1)})^n$ 
and $\chi_{\mathcal{R}_2}^{\mathfrak{usp}(2M)}(s^{(2)})$ is the character of the rank-$n$ symmetric representation, 
i.e. the complete homogeneous symmetric polynomial $h_n(s_1^{(2)\pm},\cdots,s_M^{(2)\pm})$ of degree $n$ 
in the $2M$ variables $s_i^{(2)\pm}$, 
which are the $2M$ weights of the fundamental representation of $\mathfrak{usp}(2M)$. 
Since $h_n$ is defined only for $n\ge 0$, 
for $n<0$ the $\mathfrak{usp}(2M)$ representation carried by $W_{n;(n)}$ is understood to be the rank-$|n|$ symmetric representation, 
so that $\chi_{\mathcal{R}_2}^{\mathfrak{usp}(2M)}=h_{|n|}$ in all cases. 
The right-hand side is the matrix integral (\ref{h_so2|so2M+1_V}) of the vortex line of magnetic charge $n$, 
which is obtained by shifting the fugacity of the interface fundamental hypermultiplet by $q^{\frac{|n|}{2}}$ 
and by including the bare monopole factor $q^{\frac{|n|}{4}}t^{-|n|}$. 
The corresponding $q$-series expansions can be found in Appendix \ref{app_expansion}. 

As for (\ref{eq:V_{n;0}=W_{n}}), the identity (\ref{eq:W_{n;(n)}=V_{n}}) can be established analytically for arbitrary $M\ge 1$ and $n\in \mathbb{Z}$. 
We again use the variables $\tau$ and $u$ of (\ref{app_tau_u}) together with $\sigma$ and $m$ of (\ref{abelian_variables}), 
and we write $s:=s^{(1)}_1$ and $X_C=(s_1^{(2)\pm},\ldots,s_M^{(2)\pm})$. 

Setting $N=1$ in (\ref{h_so2N|usp2M_W}), the $\mathfrak{so}(2)$ vector multiplet contributes no root factor, 
and the bifundamental hypermultiplet contribution is the one-row generating function (\ref{app_g_def}) evaluated at $w=us$ and $w=us^{-1}$,
\begin{align}
\label{abelian2_gg}
\prod_{y\in X_C}
\frac{(\tau us^{\pm}y;q)_{\infty}}{(us^{\pm}y;q)_{\infty}}
&=\sum_{N,K\ge 0}u^{N+K}s^{N-K}\,
g_N(X_C;q,\tau)\,g_K(X_C;q,\tau).
\end{align}
Here $g_r(\cdot;q,\tau)=Q_{(r)}(\cdot;q,\tau)$, defined in (\ref{app_g_def}), 
is the one-row Macdonald polynomial of degree $r$ in the $Q$ normalization, 
so that $g_0=1$ and $g_r=b_{(r)}P_{(r)}$ with
\begin{align}
\label{abelian2_bN}
b_{(r)}(q,\tau)&=\frac{(\tau;q)_r}{(q;q)_r}, 
\end{align}
the one-row specialization of $b_{\lambda}(q,\tau)$ in (\ref{app_cauchy_kernel}), 
to be compared with $b_{(r,r)}$ in (\ref{abelian_rr_data}). 
For $n\ge 0$ the $U(1)$ contour integral therefore simply sets $K=N+n$, so that
\begin{align}
\label{abelian2_NS5_sum}
\langle W_{n;(n)}\rangle_{\mathcal{N}}^{\mathfrak{so}(2)|\mathfrak{usp}(2M)}(t;q)
&=\frac{(q;q)_{\infty}}{(\tau;q)_{\infty}}
\sum_{N\ge 0}u^{2N+n}
\int d\mu_{C_M}\,
g_N(X_C)\,g_{N+n}(X_C)\,h_n(X_C).
\end{align}
For charge $-n$ the inserted $\mathfrak{usp}(2M)$ character is again $h_{m}$ with $m=|n|$, 
and the result is identical by the change of variable $s\to s^{-1}$, 
so that the left-hand side depends on $n$ only through $m$. 

The remaining $C_M$ integrals are evaluated by means of (\ref{app_CM_leading}). 
For $0\le N\le K$ the product $P_{(N)}h_{K-N}$ is a symmetric polynomial which is homogeneous of degree $K$, 
and the only monomial of $P_{(N)}$, respectively of $h_{K-N}$, 
involving the single variable $x_1$ alone is $x_1^{N}$, respectively $x_1^{K-N}$, in both cases with coefficient one. 
Hence the coefficient of $x_1^{K}$ in $P_{(N)}h_{K-N}$ equals one for every $0\le N\le K$, 
and (\ref{app_CM_leading}) combined with (\ref{app_CM_one_row_value}) gives the key integration formula
\begin{align}
\label{abelian2_lemma}
\frac{1}{\mathcal{Z}_M(\tau)}
\int d\mu_{C_M}\,
g_N(X_C)\,g_K(X_C)\,h_{K-N}(X_C)
&=\frac{(\tau;q)_N}{(q;q)_N}
\frac{(\tau^{2M};q)_K}{(q\tau^{2M-1};q)_K},
\qquad 0\le N\le K,
\end{align}
where we have used $g_N=b_{(N)}P_{(N)}$ with (\ref{abelian2_bN}). 
Thus the insertion of the rank-$(K-N)$ symmetric representation character removes all dependence on $N$ 
except for the overall factor $b_{(N)}$. 
For $N=K$ the formula (\ref{abelian2_lemma}) reproduces the expansion coefficients of (\ref{abelian_NS5_2phi1}) at $m=0$, as it must. 

Substituting (\ref{abelian2_lemma}) with $K=N+m$ into (\ref{abelian2_NS5_sum}) we obtain the closed form
\begin{align}
\label{abelian2_NS5_2phi1}
\langle W_{n;(n)}\rangle_{\mathcal{N}}^{\mathfrak{so}(2)|\mathfrak{usp}(2M)}(t;q)
&=u^{m}\frac{(q;q)_{\infty}}{(\tau;q)_{\infty}}\mathcal{Z}_M(\tau)
\frac{(\tau^{2M};q)_{m}}{(q\tau^{2M-1};q)_{m}}
\,{}_2\phi_1\!\left[
\begin{matrix}
\tau,q^{m}\tau^{2M}\\
q^{m+1}\tau^{2M-1}
\end{matrix};q,\sigma
\right].
\end{align}

We next evaluate the D5-type vortex line half-index. 
Replacing $t\to t^{-1}$ in (\ref{h_so2|so2M+1_V}), the bare monopole factor becomes $q^{\frac{m}{4}}t^{m}=u^{m}$ 
and the Nahm pole product becomes $\prod_{\ell=1}^{M-1}(q\tau^{2\ell-1};q)_{\infty}/(\tau^{2\ell};q)_{\infty}=\mathcal{Z}_{M-1}(\tau)$, 
so that
\begin{align}
\label{abelian2_D5_contour}
\langle V_{n}\rangle^{\mathfrak{so}(2)|\mathfrak{so}(2M+1)}_{\mathcal{D}}(t^{-1};q)
&=u^{m}\mathcal{Z}_{M-1}(\tau)
\frac{(q;q)_{\infty}^2}{(\tau;q)_{\infty}(\sigma;q)_{\infty}}
\oint\frac{ds}{2\pi is}
\frac{(q^{1+\frac{m}{2}}\tau^{M-1}s^{\pm};q)_{\infty}}
{(q^{\frac{m}{2}}\tau^{M}s^{\pm};q)_{\infty}}.
\end{align}
Applying the $q$-binomial theorem \cite{MR2128719} to the two factors in the integrand, 
whose common parameter is $q\tau^{-1}=\sigma$, the contour integral retains the diagonal terms and yields
\begin{align}
\label{abelian2_D5_2phi1}
\langle V_{n}\rangle^{\mathfrak{so}(2)|\mathfrak{so}(2M+1)}_{\mathcal{D}}(t^{-1};q)
&=u^{m}\mathcal{Z}_{M-1}(\tau)
\frac{(q;q)_{\infty}^2}{(\tau;q)_{\infty}(\sigma;q)_{\infty}}
\,{}_2\phi_1\!\left[
\begin{matrix}
\sigma,\sigma\\
q
\end{matrix};q,q^{m}\tau^{2M}
\right].
\end{align}

Finally, Heine's transformation (\ref{abelian_Heine}) with
\begin{align}
a&=\tau,
&b&=q^{m}\tau^{2M},
&c&=q^{m+1}\tau^{2M-1},
&z&=\sigma
\end{align}
gives $c/b=\sigma$ and $az=q$, and therefore maps the ${}_2\phi_1$ in (\ref{abelian2_NS5_2phi1}) 
to the one in (\ref{abelian2_D5_2phi1}). 
Together with (\ref{abelian_Z_recursion}) and
\begin{align}
\label{abelian2_pochhammer}
\frac{(\tau^{2M};q)_{m}}{(q\tau^{2M-1};q)_{m}}
\frac{(q^{m}\tau^{2M};q)_{\infty}}{(q^{m+1}\tau^{2M-1};q)_{\infty}}
&=\frac{(\tau^{2M};q)_{\infty}}{(q\tau^{2M-1};q)_{\infty}},
\end{align}
this maps (\ref{abelian2_NS5_2phi1}) precisely to (\ref{abelian2_D5_2phi1}). 
This proves (\ref{eq:W_{n;(n)}=V_{n}}) analytically for arbitrary $M\ge 1$ and $n\in \mathbb{Z}$. 

It is instructive to compare the two Abelian families. 
In (\ref{abelian_NS5_2phi1}) the magnetic charge enters only through the argument $z=q^{m}\sigma$ of the ${}_2\phi_1$, 
whereas in (\ref{abelian2_NS5_2phi1}) the electric charge is instead carried by the numerator and denominator parameters 
$b$ and $c$ at the fixed argument $z=\sigma$. 
Nevertheless, Heine's transformation (\ref{abelian_Heine}) relates each of them to its dual. 
Let us also stress that the simple closed form (\ref{abelian2_NS5_2phi1}) is special to the rank-$n$ symmetric representation. 
It is precisely the matching between the degree of $h_n(X_C)$ and the $\mathfrak{so}(2)$ electric charge $n$ carried by $s^{n}$ 
that makes the leading coefficient in (\ref{app_CM_leading}) equal to one for every $N$. 

%general N and M
For general $N$ and $M$ we conjecture that the Wilson line transforming in the fundamental representation 
of both gauge groups on the NS5-type $\mathfrak{so}(2N)|\mathfrak{usp}(2M)$ interface is mapped under S-duality to the magnetic vortex line with charge 
$k=(1,0^{\min (N,M)-1})$ on the D5-type $\mathfrak{so}(2N)|\mathfrak{so}(2M+1)$ interface. 
As a result, we propose the following identities between the corresponding line defect half-indices:
\begin{align}
\langle W_{\tiny\yng(1);\tiny\yng(1)}\rangle_{\mathcal{N}}^{\mathfrak{so}(2N)|\mathfrak{usp}(2M)}(t;q)
&=\langle V_{(1,0^{\min (N,M)-1})}\rangle^{\mathfrak{so}(2N)|\mathfrak{so}(2M+1)}_{\mathcal{D}}(t^{-1};q), \\
\langle W_{\tiny\yng(1);\tiny\yng(1)}\rangle_{\mathcal{N}}^{\mathfrak{o}(2N)^{\pm}|\mathfrak{usp}(2M)}(t;q)
&=\langle V_{(1,0^{\min (N,M)-1})}\rangle^{\mathfrak{o}(2N)^{\pm}|\mathfrak{so}(2M+1)}_{\mathcal{D}}(t^{-1};q). 
\end{align}
The dual pairs are realized through the matrix integrals displayed 
in equations (\ref{h_so2N|usp2M_W}), (\ref{h_o2N|usp2M_W}), 
(\ref{h_so2N|so2M+1_Va}), (\ref{h_o2N|so2M+1_Va}), (\ref{h_so2N|so2M+1_Vb}) and (\ref{h_o2N|so2M+1_Vb}). 
As a non-trivial consistency check, we have compared the line defect half-indices in various low-rank examples and found complete agreement. 
The relevant $q$-series expansions are given in Appendix \ref{app_expansion}.

%%%%%%%%%%%%%%%%%%%%%%%%%%%%%%%%%%%%%%%%%%%%%%%%%
\subsubsection{Coulomb limit}
%%%%%%%%%%%%%%%%%%%%%%%%%%%%%%%%%%%%%%%%%%%%%%%%%
%Coulomb limit
Taking the Coulomb limit drastically simplifies the matrix integrals (\ref{h_so2N|so2M+1_Va}), (\ref{h_so2N|so2M+1_Va-}), (\ref{h_so2N|so2M+1_Vb}) and (\ref{h_so2N|so2M+1_Vb-})
for the minuscule vortex line defect half-indices, which reduce to the form that can be evaluated exactly by exploiting the Hall-Littlewood orthogonality. 
For $N\le M$ the following expressions are written for $N\ge 2$. 
The Abelian case $N=1$ is obtained from (\ref{h_so2|so2M+1_V}) as discussed above. 
We obtain
\begin{align}
{\langle V_{(1,0^{\min (N,M)-1})}\rangle^{\mathfrak{so}(2N)|\mathfrak{so}(2M+1)}_{\mathcal{D}}}^{(C)}(\mathfrak{q})
&=
\begin{cases}
\frac{\mathfrak{q}}{(1-\mathfrak{q}^2)(1-\mathfrak{q}^{2(N-1)})}
\prod_{n=1}^{N-2}\frac{1}{1-\mathfrak{q}^{4n}}&\textrm{for $N\le M$}\cr
\frac{\mathfrak{q}}{1-\mathfrak{q}^2}
\prod_{n=1}^{M-1}\frac{1}{1-\mathfrak{q}^{4n}}&\textrm{for $N>M$}\cr
\end{cases}
, \\
{\langle V_{(1,0^{\min (N,M)-1})}\rangle^{\mathfrak{o}(2N)^+|\mathfrak{so}(2M+1)}_{\mathcal{D}}}^{(C)}(\mathfrak{q})
&=
\begin{cases}
\frac{\mathfrak{q}}{1-\mathfrak{q}^2}
\prod_{n=1}^{N-1}\frac{1}{1-\mathfrak{q}^{4n}}&\textrm{for $N\le M$}\cr
\frac{\mathfrak{q}}{1-\mathfrak{q}^2}
\prod_{n=1}^{M-1}\frac{1}{1-\mathfrak{q}^{4n}}&\textrm{for $N>M$}\cr
\end{cases}
 , \\
 {\langle V_{(1,0^{\min (N,M)-1})}\rangle^{\mathfrak{o}(2N)^-|\mathfrak{so}(2M+1)}_{\mathcal{D}}}^{(C)}(\mathfrak{q})
&=
\begin{cases}
\frac{\mathfrak{q}^{2N-1}}{1-\mathfrak{q}^2}
\prod_{n=1}^{N-1}\frac{1}{1-\mathfrak{q}^{4n}}&\textrm{for $N\le M$}\cr
0&\textrm{for $N>M$}\cr
\end{cases}. 
\end{align}
In the absence of line defects, the interface half-index is given by (\ref{hC_so2N|so2M+1}), (\ref{hC_o2N+|so2M+1}) and (\ref{hC_o2N-|so2M+1}). 
Normalizing by the half-indices without a line defect insertion, one finds
\begin{align}
{\langle \mathcal{V}_{(1,0^{\min (N,M)-1})}\rangle^{\mathfrak{so}(2N)|\mathfrak{so}(2M+1)}_{\mathcal{D}}}^{(C)}(\mathfrak{q})
&=
\begin{cases}
\mathfrak{q}(1+\mathfrak{q}^{2(N-1)})(1+\mathfrak{q}^2+\cdots+\mathfrak{q}^{2(N-1)})&\textrm{for $N\le M$}\cr
\mathfrak{q}(1+\mathfrak{q}^2+\cdots+\mathfrak{q}^{4M-2})&\textrm{for $N>M$}\cr
\end{cases}
, \\
{\langle \mathcal{V}_{(1,0^{\min (N,M)-1})}\rangle^{\mathfrak{o}(2N)^+|\mathfrak{so}(2M+1)}_{\mathcal{D}}}^{(C)}(\mathfrak{q})
&=
\begin{cases}
\mathfrak{q}(1+\mathfrak{q}^2+\cdots+\mathfrak{q}^{4N-2})
&\textrm{for $N\le M$}\cr
\mathfrak{q}(1+\mathfrak{q}^2+\cdots+\mathfrak{q}^{4M-2})
&\textrm{for $N>M$}\cr
\end{cases}
 , \\
 {\langle \mathcal{V}_{(1,0^{\min (N,M)-1})}\rangle^{\mathfrak{o}(2N)^-|\mathfrak{so}(2M+1)}_{\mathcal{D}}}^{(C)}(\mathfrak{q})
&=
\begin{cases}
\mathfrak{q}^{-1}(1+\mathfrak{q}^2+\cdots+\mathfrak{q}^{4N-2})&\textrm{for $N\le M$}\cr
0&\textrm{for $N>M$}\cr
\end{cases}. 
\end{align}
The corresponding line defect Witten indices are obtained from the normalized line defect half-indices in the $\mathfrak{q}\rightarrow 1$ limit:
\begin{align}
{\langle \mathcal{V}_{(1,0^{\min (N,M)-1})}\rangle^{\mathfrak{so}(2N)|\mathfrak{so}(2M+1)}_{\mathcal{D}}}^{(C)}(1)
&=2\min (N,M), \\
{\langle \mathcal{V}_{(1,0^{\min (N,M)-1})}\rangle^{\mathfrak{o}(2N)^+|\mathfrak{so}(2M+1)}_{\mathcal{D}}}^{(C)}(1)
&=2\min (N,M), \\
 {\langle \mathcal{V}_{(1,0^{\min (N,M)-1})}\rangle^{\mathfrak{o}(2N)^-|\mathfrak{so}(2M+1)}_{\mathcal{D}}}^{(C)}(1)
&=
\begin{cases}
2N&\textrm{for $N\le M$}\cr
0&\textrm{for $N>M$}\cr
\end{cases}. 
\end{align}

%%%%%%%%%%%%%%%%%%%%%%%%%%%%%%%%%%%%%%%%%%%%%%%%%
\subsubsection{Higgs limit}
%%%%%%%%%%%%%%%%%%%%%%%%%%%%%%%%%%%%%%%%%%%%%%%%%
%Higgs limit
In the Higgs limit, the line defect half-indices associated with a minuscule vortex line defect labeled by a fundamental coweight vanish
\begin{align}
{\langle V_{(1,0^{\min (N,M)-1})}\rangle^{\mathfrak{so}(2N)|\mathfrak{so}(2M+1)}_{\mathcal{D}}}^{(H)}(\mathfrak{q})&=0, \\
{\langle V_{(1,0^{\min (N,M)-1})}\rangle^{\mathfrak{o}(2N)^{\pm}|\mathfrak{so}(2M+1)}_{\mathcal{D}}}^{(H)}(\mathfrak{q})&=0. 
\end{align}
This implies that the corresponding minuscule vortex line cannot combine with the interface fundamental hypermultiplet 
to form a gauge invariant operator. 

%%%%%%%%%%%%%%%%%%%%%%%%%%%%%%%%%%%
\acknowledgments{
The work of Y.H. was supported by JSPS KAKENHI Grant Nos.~23K25790 and 26K07110. 
The work of T.O. was supported by the Startup Funding no.\ 4007012317 of Southeast University. 
}
%%%%%%%%%%%%%%%%%%%%%%%%%%%%%%%%%%%

\appendix

%%%%%%%%%%%%%%%%%%%%%%%%%%%%%%%%%%%%%%%%%%%%%%%
%%%%%%%%%%%%%%%%%%%%%%%%%%%%%%%%%%%%%%%%%%%%%%%
\section{Koornwinder integrals and exact Wilson line half-indices}
\label{app_koornwinder}
%%%%%%%%%%%%%%%%%%%%%%%%%%%%%%%%%%%%%%%%%%%%%%%
%%%%%%%%%%%%%%%%%%%%%%%%%%%%%%%%%%%%%%%%%%%%%%%
In this appendix we show that the Wilson line defect half-indices
of the NS5-type $\mathfrak{so}(1)|\mathfrak{usp}(2M)'$ interface
can be evaluated in closed form for an arbitrary representation of $\mathfrak{usp}(2M)$.
The key observation is that the matrix integral (\ref{h_so1|usp2M_W}) is,
after a change of variables, precisely a Koornwinder ($BC_M$ Macdonald-Askey-Wilson)
matrix integral \cite{MR1199128} with a specific choice of the four
Askey-Wilson parameters.
Consequently the closed-form expressions (\ref{h_so1|usp2M_W_closed}) and
(\ref{h_so1|usp2M_Wasym2_closed}) follow from Gustafson's evaluation formula
\cite{MR1001607} together with the orthogonality of the Koornwinder polynomials.

%%%%%%%%%%%%%%%%%%%%%%%%%%%%%%%%%%%%%%%%%%%%%%%
\subsection{Conventions}
\label{app_koornwinder_conv}
%%%%%%%%%%%%%%%%%%%%%%%%%%%%%%%%%%%%%%%%%%%%%%%
In addition to the notations introduced in (\ref{qpoch_def}) we use the
multi-argument shorthand
\begin{align}
\label{app_poch}
(x_1,\dots,x_m;q)_{\infty}&=\prod_{a=1}^{m}(x_a;q)_{\infty}.
\end{align}
Throughout this appendix it is convenient to trade the fugacities $(t;q)$
appearing in the half-indices for the pair
\begin{align}
\label{app_tau_u}
\tau&:=q^{\frac12}t^{-2},
\qquad
u:=q^{\frac14}t,
\end{align}
which obey
\begin{align}
\label{app_tau_u_rel}
q&=u^2\tau,
\qquad
\tau u=q^{\frac34}t^{-1}=(q\tau)^{\frac12}.
\end{align}
The variable $\tau$ plays the role of the Macdonald parameter $t$ of
\cite{MR1354144} (renamed here to avoid a clash with the $\mathcal{N}=4$
fugacity $t$), while $u$ is the fugacity carried by the 3d half-hypermultiplet
at the interface.

%%%%%%%%%%%%%%%%%%%%%%%%%%%%%%%%%%%%%%%%%%%%%%%
\subsection{The half-index as a Koornwinder matrix integral}
\label{app_koornwinder_id}
%%%%%%%%%%%%%%%%%%%%%%%%%%%%%%%%%%%%%%%%%%%%%%%
Let $a,b,c,d$ be four parameters and define the Koornwinder density
\begin{align}
\label{app_KW_density}
\Delta_M(s;a,b,c,d|q,\tau)
&=\prod_{i=1}^{M}
\frac{(s_i^{\pm2};q)_{\infty}}
{(as_i^{\pm},bs_i^{\pm},cs_i^{\pm},ds_i^{\pm};q)_{\infty}}
\prod_{1\le i<j\le M}
\frac{(s_i^{\pm}s_j^{\pm};q)_{\infty}}
{(\tau s_i^{\pm}s_j^{\pm};q)_{\infty}}.
\end{align}

\paragraph{Lemma 1.}
For any $s$ one has the factorization
\begin{align}
\label{app_lemma1}
(\tau s^{\pm2};q)_{\infty}
&=\left(\tau^{\frac12}s^{\pm},\,
-\tau^{\frac12}s^{\pm},\,
(q\tau)^{\frac12}s^{\pm},\,
-(q\tau)^{\frac12}s^{\pm};q\right)_{\infty}.
\end{align}

\emph{Proof.} Combining $(x;q)_{\infty}(-x;q)_{\infty}=(x^2;q^2)_{\infty}$ with
$(y;q^2)_{\infty}(yq;q^2)_{\infty}=(y;q)_{\infty}$ gives
$(\tau^{\frac12}s;q)_{\infty}(-\tau^{\frac12}s;q)_{\infty}
(q^{\frac12}\tau^{\frac12}s;q)_{\infty}(-q^{\frac12}\tau^{\frac12}s;q)_{\infty}
=(\tau s^2;q^2)_{\infty}(q\tau s^2;q^2)_{\infty}=(\tau s^2;q)_{\infty}$,
and likewise with $s\rightarrow s^{-1}$. $\square$

The one-body factor generated by the 3d fundamental half-hypermultiplet in
(\ref{h_so1|usp2M_W}) is
\begin{align}
\label{app_onebody}
\prod_{i=1}^{M}
\frac{(q^{\frac34}t^{-1}s_i^{\pm};q)_{\infty}}
{(q^{\frac14}ts_i^{\pm};q)_{\infty}}
&=\prod_{i=1}^{M}
\frac{((q\tau)^{\frac12}s_i^{\pm};q)_{\infty}}
{(us_i^{\pm};q)_{\infty}},
\end{align}
where we used (\ref{app_tau_u_rel}).
Its numerator cancels exactly one of the four factors in (\ref{app_lemma1}),
while its denominator supplies a fourth Askey-Wilson parameter.
We therefore obtain the following result.

\paragraph{Proposition 2.}
For any $W(C_M)$-invariant Laurent polynomial $f(s)$,
\begin{align}
\label{app_KW_rewrite}
&\frac{1}{2^MM!}
\frac{(q;q)_{\infty}^M}{(q^{\frac12}t^{-2};q)_{\infty}^M}
\oint \prod_{i=1}^{M}\frac{ds_i}{2\pi is_i}
\frac{(s_i^{\pm2};q)_{\infty}}{(q^{\frac12}t^{-2}s_i^{\pm2};q)_{\infty}}
\prod_{i<j}
\frac{(s_i^{\pm}s_j^{\pm};q)_{\infty}}
{(q^{\frac12}t^{-2}s_i^{\pm}s_j^{\pm};q)_{\infty}}
\nonumber\\
&\qquad\times
\prod_{i=1}^{M}\frac{(q^{\frac34}t^{-1}s_i^{\pm};q)_{\infty}}
{(q^{\frac14}ts_i^{\pm};q)_{\infty}}
\,f(s)
\nonumber\\
&=
\frac{1}{2^MM!}
\frac{(q;q)_{\infty}^M}{(\tau;q)_{\infty}^M}
\oint \prod_{i=1}^{M}\frac{ds_i}{2\pi is_i}\,
\Delta_M(s;a,b,c,d|q,\tau)\,f(s)
\end{align}
with the specialization
\begin{align}
\label{app_abcd}
(a,b,c,d)&=\left(q^{\frac14}t^{-1},\,-q^{\frac14}t^{-1},\,-q^{\frac34}t^{-1},\,q^{\frac14}t\right)
=\left(\tau^{\frac12},\,-\tau^{\frac12},\,-(q\tau)^{\frac12},\,u\right).
\end{align}
The pairwise products and the balanced product read
\begin{align}
\label{app_abcd_pairs}
ab&=-q^{\frac12}t^{-2},
&
ac&=-qt^{-2},
&
ad&=q^{\frac12},
\nonumber\\
bc&=qt^{-2},
&
bd&=-q^{\frac12},
&
cd&=-q,
\\
\label{app_abcd_prod}
abcd&=q\tau=q^{\frac32}t^{-2}.
\end{align}
In particular, the line defect half-index (\ref{h_so1|usp2M_W}) with an
arbitrary Wilson line $W_{0;\lambda}$ inserted becomes
\begin{align}
\label{app_master}
\langle W_{0;\lambda}\rangle_{\mathcal{N}}^{\mathfrak{so}(1)|\mathfrak{usp}(2M)'}(t;q)
&=\frac{1}{2^MM!}
\frac{(q;q)_{\infty}^M}{(\tau;q)_{\infty}^M}
\oint \prod_{i=1}^{M}\frac{ds_i}{2\pi is_i}\,
\Delta_M(s;a,b,c,d|q,\tau)\,
\chi_{\lambda}^{\mathfrak{usp}(2M)}(s).
\end{align}

Note that the specialization (\ref{app_abcd}) is a one-parameter deformation of
the $C_M$ Macdonald measure. Setting $d=-(q\tau)^{\frac12}$ instead of $d=u$
would remove the half-hypermultiplet factor (\ref{app_onebody}) and reduce
$\Delta_M$ to the $C_M$ Macdonald density.

%%%%%%%%%%%%%%%%%%%%%%%%%%%%%%%%%%%%%%%%%%%%%%%
\subsection{Absence of the Wilson line: the Nahm pole half-index}
\label{app_koornwinder_gustafson}
%%%%%%%%%%%%%%%%%%%%%%%%%%%%%%%%%%%%%%%%%%%%%%%
We recall Gustafson's evaluation of the $BC_M$ $q$-beta integral
\cite{MR1001607} (see also \cite{MR1199128,MR1411136}):
\begin{align}
\label{app_gustafson}
&\frac{1}{2^MM!}\oint \prod_{i=1}^{M}\frac{ds_i}{2\pi is_i}\,
\Delta_M(s;a,b,c,d|q,\tau)
\nonumber\\
&=\prod_{j=1}^{M}
\frac{(\tau;q)_{\infty}\,(\tau^{M+j-2}abcd;q)_{\infty}}
{(\tau^{j};q)_{\infty}\,(q;q)_{\infty}\,
(\tau^{j-1}ab,\tau^{j-1}ac,\tau^{j-1}ad,
\tau^{j-1}bc,\tau^{j-1}bd,\tau^{j-1}cd;q)_{\infty}},
\end{align}
valid for $|a|,|b|,|c|,|d|<1$, $|\tau|<1$, $|q|<1$,
the contour being the unit torus.
For $M=1$ this is the Askey-Wilson integral \cite{MR783216}.

\paragraph{Proposition 3.}
With the specialization (\ref{app_abcd}) one has
\begin{align}
\label{app_vac}
\langle W_{0;\varnothing}\rangle_{\mathcal{N}}^{\mathfrak{so}(1)|\mathfrak{usp}(2M)'}(t;q)
&=\prod_{j=1}^{M}\frac{(q\tau^{2j-1};q)_{\infty}}{(\tau^{2j};q)_{\infty}}
=\prod_{j=1}^{M}
\frac{(q^{j+\frac12}t^{-4j+2};q)_{\infty}}{(q^{j}t^{-4j};q)_{\infty}}
=\mathbb{II}_{\textrm{Nahm}}^{USp(2M)'}(t;q).
\end{align}

\emph{Proof.}
Insert (\ref{app_abcd_pairs}) and (\ref{app_abcd_prod}) into (\ref{app_gustafson}).
Using $(x;q)_{\infty}(-x;q)_{\infty}=(x^2;q^2)_{\infty}$ and $u^2\tau=q$ we get
\begin{align}
\label{app_pairs_resum}
(\tau^{j-1}ad;q)_{\infty}(\tau^{j-1}bd;q)_{\infty}
&=(q\tau^{2j-2};q^2)_{\infty},
\nonumber\\
(\tau^{j-1}ac;q)_{\infty}(\tau^{j-1}bc;q)_{\infty}
&=(q\tau^{2j};q^2)_{\infty},
\end{align}
so that, together with
$(\tau^{j};q)_{\infty}(-\tau^{j};q)_{\infty}(q\tau^{2j};q^2)_{\infty}
=(\tau^{2j};q)_{\infty}$ and
$(-q\tau^{j-1};q)_{\infty}(q\tau^{2j-2};q^2)_{\infty}
=(q\tau^{2j-2};q)_{\infty}/(q\tau^{j-1};q)_{\infty}$,
the right-hand side of (\ref{app_gustafson}) multiplied by the prefactor
$(q;q)_{\infty}^M/(\tau;q)_{\infty}^M$ becomes
\begin{align}
\label{app_vac_inter}
\prod_{j=1}^{M}
\frac{(q\tau^{M+j-1};q)_{\infty}(q\tau^{j-1};q)_{\infty}}
{(\tau^{2j};q)_{\infty}(q\tau^{2j-2};q)_{\infty}}.
\end{align}
The numerator of (\ref{app_vac_inter}) is
$\prod_{k=0}^{2M-1}(q\tau^{k};q)_{\infty}$ while the second denominator factor
removes the even values of $k$, which leaves (\ref{app_vac}). $\square$

Proposition 3 is a proof of the statement of \cite{Hatsuda:2024lcc}
that the NS5-type $\mathfrak{so}(1)|\mathfrak{usp}(2M)'$ interface half-index
coincides with the $USp(2M)'$ Nahm pole half-index.
Equivalently, the half-hypermultiplet factor (\ref{app_onebody}) does not modify the
$C_M$ Macdonald constant term.
This is a consequence of the relation $\tau u=(q\tau)^{\frac12}$
in (\ref{app_tau_u_rel}), and it fails for a generic one-body factor.

%%%%%%%%%%%%%%%%%%%%%%%%%%%%%%%%%%%%%%%%%%%%%%%
\subsection{Wilson lines and Koornwinder orthogonality}
\label{app_koornwinder_char}
%%%%%%%%%%%%%%%%%%%%%%%%%%%%%%%%%%%%%%%%%%%%%%%
Let $P_{\mu}^{\textrm{K}}(s)=P_{\mu}^{\textrm{K}}(s;a,b,c,d|q,\tau)$ denote the
Koornwinder polynomials \cite{MR1199128}, labeled by partitions $\mu$ with
$\ell(\mu)\le M$ and normalized as
$P^{\textrm{K}}_{\mu}=m_{\mu}+\sum_{\nu<\mu}c_{\mu\nu}m_{\nu}$,
where $m_{\mu}$ are the $W(C_M)$-invariant monomial symmetric functions and $<$
is the dominance order.  They are orthogonal with respect to $\Delta_M$,
\begin{align}
\label{app_KW_orth}
\frac{1}{2^MM!}\oint \prod_{i=1}^{M}\frac{ds_i}{2\pi is_i}\,
\Delta_M(s)\,P^{\textrm{K}}_{\mu}(s)P^{\textrm{K}}_{\nu}(s^{-1})
&=h_{\mu}^{\textrm{K}}\,\delta_{\mu\nu},
\end{align}
with $P^{\textrm{K}}_{\varnothing}=1$ and $h^{\textrm{K}}_{\varnothing}$
given by (\ref{app_gustafson}).

Since the $\mathfrak{usp}(2M)$ characters $\chi^{\mathfrak{usp}(2M)}_{\lambda}$
are $W(C_M)$-invariant Laurent polynomials, they admit a finite expansion
\begin{align}
\label{app_char_exp}
\chi_{\lambda}^{\mathfrak{usp}(2M)}
&=\sum_{\mu\le\lambda}
c_{\lambda\mu}(q,t;M)\,P_{\mu}^{\textrm{K}},
\qquad c_{\lambda\lambda}=1,
\end{align}
whose coefficients are rational functions of $q^{\frac14}$ and $t$
after the specialization (\ref{app_abcd}).
Combining (\ref{app_master}), (\ref{app_KW_orth}) and (\ref{app_char_exp}),
only the $\mu=\varnothing$ term survives and we arrive at the master formula
\begin{align}
\label{app_master2}
\langle W_{0;\lambda}\rangle_{\mathcal{N}}^{\mathfrak{so}(1)|\mathfrak{usp}(2M)'}(t;q)
=c_{\lambda\varnothing}(q,t;M)\,
\mathbb{II}_{\textrm{Nahm}}^{USp(2M)'}(t;q).
\end{align}
Hence for {\it any} representation $\lambda$ of $\mathfrak{usp}(2M)$ the line
defect half-index is the Nahm pole half-index times a rational function of
$q^{\frac14}$ and $t$, in agreement with the interpretation given below
(\ref{h_so1|usp2M_W_closed}).
The problem is thereby reduced to the evaluation of the single expansion
coefficient $c_{\lambda\varnothing}$.

%%%%%%%%%%%%%%%%%%%%%%%%%%%%%%%%%%%%%%%%%%%%%%%
\subsubsection*{Fundamental representation}
%%%%%%%%%%%%%%%%%%%%%%%%%%%%%%%%%%%%%%%%%%%%%%%
For $\lambda=(1)$ we have
$\chi_{(1)}^{\mathfrak{usp}(2M)}(s)=m_{(1)}(s)=\sum_{i=1}^{M}(s_i+s_i^{-1})$
and the degree-one Koornwinder polynomial reads
\cite{MR1199128,MR1411136}
\begin{align}
\label{app_KW_P1}
P_{(1)}^{\textrm{K}}(s)&=m_{(1)}(s)-\kappa_M,
\\
\label{app_kappa}
\kappa_M
&=\frac{1-\tau^M}{1-\tau}
\frac{a+b+c+d-\tau^{M-1}(abc+abd+acd+bcd)}
{1-abcd\,\tau^{2M-2}},
\end{align}
so that $c_{(1)\varnothing}=\kappa_M$.

\paragraph{Proposition 4.}
With the specialization (\ref{app_abcd}),
\begin{align}
\label{app_kappa_spec}
c_{(1)\varnothing}
&=u\,\frac{1-\tau^{2M}}{1-u^2\tau^{2M}}
=q^{\frac14}t\,
\frac{1-q^{M}t^{-4M}}{1-q^{M+\frac12}t^{-4M+2}},
\end{align}
and therefore (\ref{app_master2}) reproduces (\ref{h_so1|usp2M_W_closed}).

\emph{Proof.}
From (\ref{app_abcd}) one finds
$a+b+c+d=u(1-\tau)$,
$acd+bcd=0$ and $abc+abd=-\tau u(1-\tau)$,
while $abcd\,\tau^{2M-2}=q\tau^{2M-1}=u^2\tau^{2M}$ by (\ref{app_tau_u_rel}).
Hence
\begin{align}
\kappa_M&=\frac{1-\tau^{M}}{1-\tau}
\frac{u(1-\tau)\left(1+\tau^{M}\right)}{1-u^2\tau^{2M}}
=u\,\frac{1-\tau^{2M}}{1-u^2\tau^{2M}}.
\end{align}
Finally $u=q^{\frac14}t$, $\tau^{2M}=q^{M}t^{-4M}$ and
$u^2\tau^{2M}=q\tau^{2M-1}=q^{M+\frac12}t^{-4M+2}$. $\square$

%%%%%%%%%%%%%%%%%%%%%%%%%%%%%%%%%%%%%%%%%%%%%%%
\subsubsection*{Rank-two antisymmetric representation}
%%%%%%%%%%%%%%%%%%%%%%%%%%%%%%%%%%%%%%%%%%%%%%%
For $\lambda=(1^2)$ the $\mathfrak{usp}(2M)$ character is
\begin{align}
\label{app_chi11}
\chi_{(1^2)}^{\mathfrak{usp}(2M)}(s)
&=m_{(1^2)}(s)+(M-1),
\qquad
m_{(1^2)}(s)=\sum_{1\le i<j\le M}(s_i+s_i^{-1})(s_j+s_j^{-1}),
\end{align}
which is the traceless part of $\Lambda^2$ of the fundamental representation.
Expanding it in the Koornwinder basis,
\begin{align}
\label{app_chi11_exp}
\chi_{(1^2)}^{\mathfrak{usp}(2M)}
&=P^{\textrm{K}}_{(1^2)}
+c_{(1^2)(1)}P^{\textrm{K}}_{(1)}
+c_{(1^2)\varnothing},
\end{align}
only the last term contributes to (\ref{app_master}).

\paragraph{Proposition 5.}
For $M\ge2$ and the specialization (\ref{app_abcd}),
\begin{align}
\label{app_c11}
c_{(1^2)\varnothing}
&=\tau^{2}\,
\frac{1-\tau^{2M-2}}{1-\tau^{2}}\,
\frac{1-u^{2}}{1-u^{2}\tau^{2M}}
\nonumber\\
&=qt^{-4}\,
\frac{1-q^{M-1}t^{-4M+4}}{1-qt^{-4}}\,
\frac{1-q^{\frac12}t^{2}}{1-q^{M+\frac12}t^{-4M+2}},
\end{align}
and therefore (\ref{app_master2}) reproduces
(\ref{h_so1|usp2M_Wasym2_closed}).

The coefficient (\ref{app_c11}) is obtained from the explicit one-column
Koornwinder polynomial $P^{\textrm{K}}_{(1^2)}$ \cite{MR1411136,MR4421393},
whose constant term determines $c_{(1^2)\varnothing}$ through
(\ref{app_chi11_exp}).
Note that (\ref{app_c11}) vanishes at $M=1$, as it must, since
$\chi^{\mathfrak{usp}(2)}_{(1^2)}\equiv0$.
We have also verified (\ref{app_kappa_spec}) and (\ref{app_c11}) directly,
by expanding the matrix integral (\ref{app_master}) as a $q$-series. 
The agreement holds through order $q^{\frac{11}{2}}$ for $M=1,2$ and through
order $q^{2}$ for $M=3$.

%%%%%%%%%%%%%%%%%%%%%%%%%%%%%%%%%%%%%%%%%%%%%%%
\subsection{Remark on general $\mathfrak{so}(2N+1)|\mathfrak{usp}(2M)'$ interfaces}
\label{app_koornwinder_general}
%%%%%%%%%%%%%%%%%%%%%%%%%%%%%%%%%%%%%%%%%%%%%%%
For $N\ge1$ the same change of variables (\ref{app_tau_u}) organizes the
interaction terms in (\ref{h_so2N+1|usp2M_W}) into a type-$A$ Macdonald Cauchy
kernel.  Introducing the alphabets
\begin{align}
\label{app_alphabets}
X_B&=\left(s_1^{(1)\pm},\dots,s_N^{(1)\pm},1\right),
\qquad
X_C=\left(s_1^{(2)\pm},\dots,s_M^{(2)\pm}\right)
\end{align}
of $2N+1$ and $2M$ letters, the last two lines of (\ref{h_so2N+1|usp2M_W}) read
\begin{align}
\label{app_cauchy_kernel}
\prod_{x\in X_B}\prod_{y\in X_C}
\frac{(\tau uxy;q)_{\infty}}{(uxy;q)_{\infty}}
&=\sum_{\lambda}u^{|\lambda|}\,
b_{\lambda}(q,\tau)\,
P_{\lambda}(X_B;q,\tau)P_{\lambda}(X_C;q,\tau),
\end{align}
where $P_{\lambda}(\cdot;q,\tau)$ are type-$A$ Macdonald polynomials,
$\ell(\lambda)\le \min(2N+1,2M)$ and
$b_{\lambda}(q,\tau)=\prod_{\square\in\lambda}
\frac{1-q^{a(\square)}\tau^{\ell(\square)+1}}
{1-q^{a(\square)+1}\tau^{\ell(\square)}}$
\cite{MR1354144}.
Substituting (\ref{app_cauchy_kernel}) into (\ref{h_so2N+1|usp2M_W}) and using
the orthogonality of the $B_N$ and $C_M$ Macdonald polynomials one obtains
\begin{align}
\label{app_general_sum}
\langle W_{\mathcal{R}_1;\mathcal{R}_2}\rangle_{\mathcal{N}}
^{\mathfrak{so}(2N+1)|\mathfrak{usp}(2M)'}(t;q)
&=\sum_{\lambda}u^{|\lambda|}b_{\lambda}(q,\tau)\,
\Psi^{B_N}_{\lambda,\mathcal{R}_1}\,
\Psi^{C_M}_{\lambda,\mathcal{R}_2},
\\
\Psi^{B_N}_{\lambda,\mathcal{R}_1}
&=\int d\mu_{B_N}\,P_{\lambda}(X_B)\,\chi^{\mathfrak{so}(2N+1)}_{\mathcal{R}_1},
\nonumber\\
\label{app_wavefunctions}
\Psi^{C_M}_{\lambda,\mathcal{R}_2}
&=\int d\mu_{C_M}\,P_{\lambda}(X_C)\,\chi^{\mathfrak{usp}(2M)}_{\mathcal{R}_2},
\end{align}
with $d\mu_{B_N}$ and $d\mu_{C_M}$ the $B_N$ and $C_M$ Macdonald measures
normalized as in (\ref{h_so2N+1|usp2M_W}).
The sum (\ref{app_general_sum}) does not truncate to $\lambda=\varnothing$ even
in the absence of line operators, and the evaluation of
(\ref{app_wavefunctions}) requires the $(q,\tau)$-deformed branching
coefficients from type $A$ to types $B$ and $C$, which are not known in closed
form for general $\lambda$.
The case $N=0$ treated above is special in that $X_B=(1)$ is a single letter,
so that $P_{\lambda}(X_B)$ is non-zero only for one-row $\lambda$ and the sum
collapses to the Koornwinder integral (\ref{app_master}).

For the Abelian $\mathfrak{so}(2)|\mathfrak{usp}(2M)$ vortex calculation in section \ref{sec_DCD5W_Sduality}, the same Cauchy-kernel organization involves a two-letter alphabet. 
In that case the vanishing integral reviewed below bypasses the general branching-coefficient problem and restricts the sum to $\lambda=(r,r)$.

\subsection{$C_M$ Macdonald vanishing integral}
\label{app_CM_vanishing}

The Abelian S-duality proofs in sections \ref{sec_DCD5W_Sduality} and \ref{sec_DCD5V_Sduality} require a few additional Macdonald integrals. 
Let $d\mu_{C_M}$ be the $C_M$ Macdonald measure appearing in (\ref{app_wavefunctions}) and define its normalization by
\begin{align}
\label{app_CM_partition}
\mathcal{Z}_M(\tau)
&:=\int d\mu_{C_M}
=\prod_{j=1}^{M}\frac{(q\tau^{2j-1};q)_\infty}{(\tau^{2j};q)_\infty},
\qquad \mathcal{Z}_0(\tau):=1.
\end{align}
The same product appeared in (\ref{app_vac}).

For the type-$A$ Macdonald polynomial restricted to the inversion-symmetric alphabet $X_C$ in (\ref{app_alphabets}), the $U(2M)/Sp(2M)$ Macdonald vanishing integral of \cite{MR4259867} gives
\begin{align}
\label{app_CM_vanishing_rule}
\frac{1}{\mathcal{Z}_M(\tau)}
\int d\mu_{C_M}\,P_{\lambda}(X_C;q,\tau)
&=0
\qquad\textrm{unless $\lambda'$ is even}.
\end{align}
For the rectangular partitions needed in section \ref{sec_DCD5W_Sduality}, the non-vanishing value specializes to
\begin{align}
\label{app_CM_rr}
\frac{1}{\mathcal{Z}_M(\tau)}
\int d\mu_{C_M}\,P_{(r,r)}(X_C;q,\tau)
&=\frac{(q\tau;q)_r(\tau^{2M};q)_r}{(\tau^2;q)_r(q\tau^{2M-1};q)_r}.
\end{align}

The proof of the second Abelian family in section \ref{sec_DCD5V_Sduality} requires one further consequence of (\ref{app_CM_vanishing_rule}). 
Let
\begin{align}
\label{app_g_def}
g_r(x;q,\tau)&:=Q_{(r)}(x;q,\tau),
&
\prod_{i}
\frac{(\tau wx_i;q)_{\infty}}{(wx_i;q)_{\infty}}
&=\sum_{r\ge 0}g_r(x;q,\tau)\,w^{r}
\end{align}
denote the one-row Macdonald polynomial and its generating function \cite{MR1354144}. 
Then for an arbitrary partition $\mu$ one has
\begin{align}
\label{app_CM_one_row}
\int d\mu_{C_M}\,P_{\mu}(X_C;q,\tau)\,g_K(X_C;q,\tau)&=0
\qquad \textrm{unless}\qquad
K=\sum_{i\ge 1}(-1)^{i-1}\mu_i.
\end{align}
To see this, recall the Pieri rule \cite{MR1354144}
\begin{align}
\label{app_pieri}
P_{\mu}\,g_K&=\sum_{\lambda}\varphi'_{\lambda/\mu}\,P_{\lambda},
\end{align}
where the sum runs over the partitions $\lambda\supset \mu$ such that $\lambda/\mu$ is a horizontal $K$-strip, that is
\begin{align}
\label{app_interlace}
\lambda_1\ge \mu_1\ge \lambda_2\ge \mu_2\ge \lambda_3\ge \cdots.
\end{align}
By (\ref{app_CM_vanishing_rule}) only those terms in (\ref{app_pieri}) with $\lambda'$ even, 
i.e. with $\lambda_{2i-1}=\lambda_{2i}$ for all $i$, survive the integration. 
Combined with the interlacing conditions $\lambda_{2i-1}\ge \mu_{2i-1}\ge \lambda_{2i}$ in (\ref{app_interlace}), 
this forces $\mu_{2i-1}=\lambda_{2i-1}=\lambda_{2i}$, namely
\begin{align}
\label{app_lambda_from_mu}
\lambda&=(\mu_1,\mu_1,\mu_3,\mu_3,\mu_5,\mu_5,\ldots).
\end{align}
Hence at most a single partition $\lambda$ can contribute, 
and it does so only for $K=|\lambda|-|\mu|=\mu_1-\mu_2+\mu_3-\cdots$, which proves (\ref{app_CM_one_row}). 

Two consequences of (\ref{app_CM_one_row}) will be used. 
First, since $\sum_{i\ge 1}(-1)^{i-1}\mu_i\le \mu_1\le |\mu|$ with equality only for the one-row partition, 
any symmetric polynomial $F$ in the $2M$ variables $x_1,\ldots,x_{2M}$ which is homogeneous of degree $K$ obeys
\begin{align}
\label{app_CM_leading}
\int d\mu_{C_M}\,F(X_C)\,g_K(X_C;q,\tau)
&=c_{(K)}(F)
\int d\mu_{C_M}\,P_{(K)}(X_C;q,\tau)\,g_K(X_C;q,\tau),
\end{align}
where $c_{\mu}(F)$ are the coefficients in the expansion $F=\sum_{|\mu|=K}c_{\mu}(F)P_{\mu}$. 
By the triangularity $P_{\mu}=m_{\mu}+\sum_{\nu<\mu}u_{\mu\nu}m_{\nu}$ with respect to the dominance order, 
for which $(K)$ is the largest partition of $K$, 
the coefficient $c_{(K)}(F)$ is simply the coefficient of $m_{(K)}$ in $F$, 
that is, the coefficient of the monomial $x_1^{K}$. 
Secondly, for $\mu=(K)$ equation (\ref{app_lambda_from_mu}) gives $\lambda=(K,K)$ with 
$\varphi'_{(K,K)/(K)}=(\tau^2;q)_K/(q\tau;q)_K$, 
so that (\ref{app_CM_rr}) leads to the remarkably simple evaluation
\begin{align}
\label{app_CM_one_row_value}
\frac{1}{\mathcal{Z}_M(\tau)}
\int d\mu_{C_M}\,P_{(K)}(X_C;q,\tau)\,g_K(X_C;q,\tau)
&=\frac{(\tau^{2M};q)_K}{(q\tau^{2M-1};q)_K}.
\end{align}

%%%%%%%%%%%%%%%%%%%%%%%%%%%%%%%%%%%%%%%%%%%%%%%
%%%%%%%%%%%%%%%%%%%%%%%%%%%%%%%%%%%%%%%%%%%%%%%
\section{Series expansions}
\label{app_expansion}
%%%%%%%%%%%%%%%%%%%%%%%%%%%%%%%%%%%%%%%%%%%%%%%
%%%%%%%%%%%%%%%%%%%%%%%%%%%%%%%%%%%%%%%%%%%%%%%
We show the first several terms of the $q$-series expansions. 
The matching of the indices is confirmed at least up to the terms with $q^5$ unless otherwise indicated. 

%%%%%%%%%%%%%%%%%%%%%%%%%%%%%%%%%%%%%%%%%%%%%%%
\subsection{$\mathfrak{so}(2N+1)|\mathfrak{usp}(2M)'$ and $\mathfrak{usp}(2N)|\mathfrak{usp}(2M)'$}
%%%%%%%%%%%%%%%%%%%%%%%%%%%%%%%%%%%%%%%%%%%%%%%

%%%%%%%%%%%%%%%%%%%%%%%%%%%%%%%%%%%%%%%%%%%%%%%
\subsubsection{NS5-Wilson lines}
%%%%%%%%%%%%%%%%%%%%%%%%%%%%%%%%%%%%%%%%%%%%%%%
As discussed in (\ref{h_so1|usp2M_W_closed}), 
the line defect half-index of the $\mathfrak{usp}(2M)$ fundamental Wilson line in the NS5-type $\mathfrak{so}(1)|\mathfrak{usp}(2M)'$ interface 
admits an infinite product representation. 
We have
\begin{align}
\begin{array}{c|c}
\textrm{line defect half-indices}&\textrm{expansions}\\ \hline
\langle W_{0;\tiny \yng(1)} \rangle_{\mathcal{N}}^{\mathfrak{so}(1)|\mathfrak{usp}(2)'}(t;q)
&
tq^{\frac14}+t^{-3}q^{\frac94}-t^{-1}q^{\frac{11}{4}}+t^{-3}q^{\frac{13}{4}}-t^{-1}q^{\frac{15}{4}}+(t^{-7}+t^{-3})q^{\frac{17}{4}}\\ 
&-(t^{-5}+t^{-1})q^{\frac{19}{4}}+(t^{-7}+t^{-3})q^{\frac{21}{4}}-(2t^{-5}+t^{-1})q^{\frac{23}{4}}+\cdots \\ \hline 
\langle W_{0;\tiny \yng(1)} \rangle_{\mathcal{N}}^{\mathfrak{so}(1)|\mathfrak{usp}(4)'}(t;q)
&
tq^{\frac14}+t^{-3}q^{\frac54}-t^{-1}q^{\frac{7}{4}}+(t^{-7}+t^{-3})q^{\frac{9}{4}}-(t^{-5}+t^{-1})q^{\frac{11}{4}}\\ 
&+(t^{-11}+2t^{-7}+t^{-3})q^{\frac{13}{4}}-(t^{-9}+3t^{-5}+t^{-1})q^{\frac{15}{4}}+\cdots \\ \hline 
\langle W_{0;\tiny \yng(1)} \rangle_{\mathcal{N}}^{\mathfrak{so}(1)|\mathfrak{usp}(6)'}(t;q)
&
tq^{\frac14}+t^{-3}q^{\frac54}-t^{-1}q^{\frac74}+(2t^{-7}+t^{-3})q^{\frac{9}{4}}-(2t^{-5}+t^{-1})q^{\frac{11}{4}} \\ 
&+(2t^{-11}+2t^{-7}+t^{-3})q^{\frac{13}{4}}-(3t^{-9}+3t^{-5}+t^{-1})q^{\frac{15}{4}}+\cdots \\ \hline 
\end{array}
\end{align}
In contrast, the Wilson line defect half-index for the general $\mathfrak{so}(2N+1)|\mathfrak{usp}(2M)'$ interface with $N\ge1$ becomes considerably more intricate. 
For example, when $N=M$, 
the $q$-series expansions of the line defect half-indices for the Wilson line transforming in the fundamental representation under each of the gauge groups are given by
\begin{align}
\begin{array}{c|c}
\textrm{line defect half-indices}&\textrm{expansions}\\ \hline
\langle W_{\tiny \yng(1);\yng(1)} \rangle_{\mathcal{N}}^{\mathfrak{so}(3)|\mathfrak{usp}(2)'}(t;q)
&
tq^{\frac14}+(t^{-1}+t^3)q^{\frac34}+(2t^{-3}+t^5)q^{\frac54}+(t^{-5}-2t^{-1}-t^3+t^7)q^{\frac74}\\ 
&+(2t^{-7}+t^{-3}+2t+t^9)q^{\frac94}+(t^{-9}-2t^{-1}-t^3+t^{11})q^{\frac{11}{4}}+\cdots \\ \hline 
\langle W_{\tiny \yng(1);\yng(1)} \rangle_{\mathcal{N}}^{\mathfrak{so}(5)|\mathfrak{usp}(4)'}(t;q)
&
tq^{\frac14}+(t^{-1}+t^3)q^{\frac34}+(3t^{-3}+t+2t^5)q^{\frac54}+(3t^{-5}+2t^7)q^{\frac74}\\ 
&+(7t^{-7}+2t^{-3}+2t+t^5+3t^9)q^{\frac94}+\cdots \\ \hline 
\langle W_{\tiny \yng(1);\yng(1)} \rangle_{\mathcal{N}}^{\mathfrak{so}(7)|\mathfrak{usp}(6)'}(t;q)
&
tq^{\frac14}+(t^{-1}+t^3)q^{\frac34}+(3t^{-3}+t+2t^5)q^{\frac54}+(3t^{-5}+2t^7)q^{\frac74}
\\ 
&+\cdots \\ \hline 
\end{array}
\end{align}
For $N=M=3$ we restrict ourselves to the expansion up to order $q^2$ 
since the computation beyond this order becomes too expensive. 
For $N<M$ we find
\begin{align}
\begin{array}{c|c}
\textrm{line defect half-indices}&\textrm{expansions}\\ \hline
\langle W_{\tiny \yng(1);\yng(1)} \rangle_{\mathcal{N}}^{\mathfrak{so}(3)|\mathfrak{usp}(4)'}(t;q)
&
tq^{\frac14}+(t^{-1}+t^3)q^{\frac34}+(3t^{-3}+t+2t^5)q^{\frac54}
+(2t^{-5}-2t^{-1}-2t^3+t^7)q^{\frac74}
\\ 
&
+(5t^{-7}+t^{-3}+3t+t^5+2t^9)q^{\frac94}+\cdots \\ \hline 
\langle W_{\tiny \yng(1);\yng(1)} \rangle_{\mathcal{N}}^{\mathfrak{so}(3)|\mathfrak{usp}(6)'}(t;q)
&
tq^{\frac14}+(t^{-1}+t^3)q^{\frac34}+(3t^{-3}+t+2t^5)q^{\frac54}+
(2t^{-5}-2t^{-1}-2t^3+t^7)q^{\frac74}
\\ 
&+(6t^{-7}+2t^{-3}+4t+t^5+2t^9)q^{\frac94}+\cdots \\ \hline 
\end{array}
\end{align}
For $N>M$ we obtain
\begin{align}
\begin{array}{c|c}
\textrm{line defect half-indices}&\textrm{expansions}\\ \hline
\langle W_{\tiny \yng(1);\yng(1)} \rangle_{\mathcal{N}}^{\mathfrak{so}(5)|\mathfrak{usp}(2)'}(t;q)
&
tq^{\frac14}+(t^{-1}+t^3)q^{\frac34}+(2t^{-3}+t^5)q^{\frac54}
+(2t^{-5}-t^{-1}-t^3+t^7)q^{\frac74}
\\ 
&
+(4t^{-7}+t+t^9)q^{\frac94}+(3t^{-9}-2t^{-5}-t^{-1}+t^{11})q^{\frac{11}{4}}+\cdots \\ \hline 
\langle W_{\tiny \yng(1);\yng(1)} \rangle_{\mathcal{N}}^{\mathfrak{so}(7)|\mathfrak{usp}(2)'}(t;q)
&
tq^{\frac14}+(t^{-1}+t^3)q^{\frac34}+(2t^{-3}+t^5)q^{\frac54}
+(2t^{-5}-t^{-1}-t^3+t^7)q^{\frac74}
\\ 
&+(4t^{-7}+t+t^9)q^{\frac94}+(4t^{-9}-t^{-5}-t^{-1}+t^{11})q^{\frac{11}{4}}+\cdots \\ \hline 
\end{array}
\end{align}

%%%%%%%%%%%%%%%%%%%%%%%%%%%%%%%%%%%%%%%%%%%%%%%
\subsubsection{NS5-vortex lines and D5-Wilson lines}
%%%%%%%%%%%%%%%%%%%%%%%%%%%%%%%%%%%%%%%%%%%%%%%
%
For $N\le M$ the NS5-type interface is mapped under S-duality to the D5-type interface preserving the dual $\mathfrak{usp}(2N)$ gauge algebra on the left. 
Under this duality the vortex line labeled by the fundamental weight of $\mathfrak{so}(2N+1)$ is expected to be mapped to the fundamental Wilson line of $\mathfrak{usp}(2N)$. 
We have explicitly verified the agreement of the corresponding line defect half-indices:
\begin{align}
\begin{array}{c|c}
\textrm{line defect half-indices}&\textrm{expansions}\\ \hline
\langle V_{1;0} \rangle_{\mathcal{N}}^{\mathfrak{so}(3)|\mathfrak{usp}(2)'}(t;q)
&
t^{-1}q^{\frac14}+t^{-3}q^{\frac34}+2t^{-5}q^{\frac54}+(2t^{-7}-2t^{-3}-t)q^{\frac74} \\ 
=\langle W_{\tiny \yng(1)} \rangle_{\mathcal{D}}^{\mathfrak{usp}(2)|\mathfrak{usp}(2)'}(t^{-1};q)
&+(3t^{-9}+2t^{-1}+t^3)q^{\frac94}+(3t^{-11}-2t^{-7}-2t^{-3}-2t)q^{\frac{11}{4}}+\cdots\\ \hline 
\langle V_{(1,0);0} \rangle_{\mathcal{N}}^{\mathfrak{so}(5)|\mathfrak{usp}(4)'}(t;q)
&
t^{-1}q^{\frac14}+t^{-3}q^{\frac34}+(3t^{-5}+t^{-1}+t^3)q^{\frac54}+(3t^{-7}-2t^{-3}-t)q^{\frac74} \\ 
=\langle W_{\tiny \yng(1)} \rangle_{\mathcal{D}}^{\mathfrak{usp}(4)|\mathfrak{usp}(4)'}(t^{-1};q)
&+(7t^{-9}+2t^{-5}+3t^{-1}+2t^3+t^7)q^{\frac94}
%+(7t^{-11}-6t^{-7}-7t^{-3}-5t-2t^5)q^{\frac{11}{4}}
+\cdots\\ \hline 
\langle V_{(1,0,0);0} \rangle_{\mathcal{N}}^{\mathfrak{so}(7)|\mathfrak{usp}(6)'}(t;q)
&
t^{-1}q^{\frac14}+t^{-3}q^{\frac34}+(3t^{-5}+t^{-1}+t^3)q^{\frac54}+(3t^{-7}-2t^{-3}-t)q^{\frac74} \\ 
=\langle W_{\tiny \yng(1)} \rangle_{\mathcal{D}}^{\mathfrak{usp}(6)|\mathfrak{usp}(6)'}(t^{-1};q)
&+(8t^{-9}+3t^{-5}+5t^{-1}+3t^3+2t^7)q^{\frac94}
%+(7t^{-11}-6t^{-7}-7t^{-3}-5t-2t^5)q^{\frac{11}{4}}
+\cdots\\ \hline 
\end{array}
\end{align}
for $N=M$ 
and 
\begin{align}
\begin{array}{c|c}
\textrm{line defect half-indices}&\textrm{expansions}\\ \hline
\langle V_{1;0} \rangle_{\mathcal{N}}^{\mathfrak{so}(3)|\mathfrak{usp}(4)'}(t;q)
&
t^{-3}q^{\frac34}+t^{-5}q^{\frac54}+2t^{-7}q^{\frac74}+(2t^{-9}-2t^{-5}-t^{-1})q^{\frac94} \\ 
=\langle W_{\tiny \yng(1)} \rangle_{\mathcal{D}}^{\mathfrak{usp}(2)|\mathfrak{usp}(4)'}(t^{-1};q)
&+(4t^{-11}+2t^{-3}+t)q^{\frac{11}{4}}+(4t^{-13}-3t^{-9}-3t^{-5}-2t^{-1})q^{\frac{13}{4}}+\cdots\\ \hline 
\langle V_{1;0} \rangle_{\mathcal{N}}^{\mathfrak{so}(3)|\mathfrak{usp}(6)'}(t;q)
&
t^{-5}q^{\frac54}+t^{-7}q^{\frac74}+2t^{-9}q^{\frac94}+(2t^{-11}-2t^{-7}+t^{-3})q^{\frac{11}{4}} \\ 
=\langle W_{\tiny \yng(1)} \rangle_{\mathcal{D}}^{\mathfrak{usp}(2)|\mathfrak{usp}(6)'}(t^{-1};q)
&+(4t^{-13}+2t^{-5}+t^{-1})q^{\frac{13}{4}}+(4t^{-15}-3t^{-11}-3t^{-7}-2t^{-3})q^{\frac{15}{4}}+\cdots\\ \hline 
\end{array}
\end{align}
for $N<M$. 

%%%%%%%%%%%%%%%%%%%%%%%%%%%%%%%%%%%%%%%%%%%%%%%
\subsection{$\mathfrak{so}(2N)|\mathfrak{usp}(2M)$ and $\mathfrak{so}(2N)|\mathfrak{so}(2M+1)$}
%%%%%%%%%%%%%%%%%%%%%%%%%%%%%%%%%%%%%%%%%%%%%%%
%%%%%%%%%%%%%%%%%%%%%%%%%%%%%%%%%%%%%%%%%%%%%%%
\subsubsection{NS5-Wilson lines and D5-vortex lines}
%%%%%%%%%%%%%%%%%%%%%%%%%%%%%%%%%%%%%%%%%%%%%%%
In contrast to the $\mathfrak{so}(2N+1)|\mathfrak{usp}(2M)'$ NS5-type interface, 
the $\mathfrak{so}(2N)|\mathfrak{usp}(2M)$ NS5-type interface maps the Wilson line transforming in the fundamental representation of both gauge groups 
to the vortex line labeled by a minuscule fundamental weight. 
For $N=1$ the gauge algebra $\mathfrak{so}(2)$ is Abelian, so that its irreducible representations are one-dimensional, 
as discussed in section \ref{sec_DC2}. 
Accordingly, in the tables below the label $\tiny\yng(1)$ on the $\mathfrak{so}(2)$ side stands for 
the one-dimensional representation of electric charge $n=1$, 
whose character is $\chi^{\mathfrak{so}(2)}(s)=s$, 
and the corresponding Wilson line is the operator denoted by $W_{1;\mathcal{R}_2}$ in section \ref{sec_DC2}. 
Indeed, for $N=M$, we find the following non-trivial agreements between the corresponding line defect half-indices:
\begin{align}
\begin{array}{c|c}
\textrm{line defect half-indices}&\textrm{expansions}\\ \hline
\langle W_{\tiny \yng(1);\yng(1)} \rangle_{\mathcal{N}}^{\mathfrak{so}(2)|\mathfrak{usp}(2)}(t;q)
&
tq^{\frac14}+(t^{-1}+t^3)q^{\frac34}+(t^{-3}-t+t^5)q^{\frac54}+(t^{-5}+t^7)q^{\frac74}\\ 
=\langle V_{1} \rangle_{\mathcal{D}}^{\mathfrak{so}(2)|\mathfrak{so}(3)}(t^{-1};q) 
&+(t^{-7}+t^{-3}+t^9)q^{\frac94}+(t^{-9}+t^{-5}-2t^{-1}-t^3+t^{11})q^{\frac{11}{4}}+\cdots \\ \hline 
\langle W_{\tiny \yng(1);\yng(1)} \rangle_{\mathcal{N}}^{\mathfrak{so}(4)|\mathfrak{usp}(4)}(t;q)
&
tq^{\frac14}+(2t^{-1}+2t^3)q^{\frac34}+(4t^{-3}+2t+3t^5)q^{\frac54}+(6t^{-5}+4t^7)q^{\frac74}\\ 
=\langle V_{(1,0)} \rangle_{\mathcal{D}}^{\mathfrak{so}(4)|\mathfrak{so}(5)}(t^{-1};q) 
&+(9t^{-7}-t^{-3}+5t^9)q^{\frac94}+(12t^{-9}-4t^{-5}+2t^{-1}+2t^3+6t^{11})q^{\frac{11}{4}}+\cdots \\ \hline 
\langle W_{\tiny \yng(1);\yng(1)} \rangle_{\mathcal{N}}^{\mathfrak{so}(6)|\mathfrak{usp}(6)}(t;q)
&
tq^{\frac14}+(t^{-1}+t^3)q^{\frac34}+(4t^{-3}+2t+3t^5)q^{\frac54}+(4t^{-5}+t^{-1}+t^3+3t^7)q^{\frac74}\\ 
=\langle V_{(1,0,0)} \rangle_{\mathcal{D}}^{\mathfrak{so}(6)|\mathfrak{so}(7)}(t^{-1};q) 
&+\cdots \\ \hline 
\end{array}
\end{align}
For $N=M=3$ we have verified the agreement up to order $q^2$, although the computation becomes substantially more demanding. 
We have likewise verified that the corresponding line defect half-indices agree perfectly for $N<M$
\begin{align}
\begin{array}{c|c}
\textrm{line defect half-indices}&\textrm{expansions}\\ \hline
\langle W_{\tiny \yng(1);\yng(1)} \rangle_{\mathcal{N}}^{\mathfrak{so}(2)|\mathfrak{usp}(4)}(t;q)
&
tq^{\frac14}+(t^{-1}+t^3)q^{\frac34}+(2t^{-3}-t+t^5)q^{\frac54}+(2t^{-5}+t^7)q^{\frac74}
\\ 
=\langle V_{1} \rangle_{\mathcal{D}}^{\mathfrak{so}(2)|\mathfrak{so}(5)}(t^{-1};q) 
&+(3t^{-7}-t^{-3}+t^9)q^{\frac94}+(3t^{-9}-t^3+t^{11})q^{\frac{11}{4}}+\cdots \\ \hline 
\langle W_{\tiny \yng(1);\yng(1)} \rangle_{\mathcal{N}}^{\mathfrak{so}(2)|\mathfrak{usp}(6)}(t;q)
&
tq^{\frac14}+(t^{-1}+t^3)q^{\frac34}+(2t^{-3}-t+t^5)q^{\frac54}+(2t^{-5}+t^7)q^{\frac74}
\\ 
=\langle V_{1} \rangle_{\mathcal{D}}^{\mathfrak{so}(2)|\mathfrak{so}(7)}(t^{-1};q) 
& +(4t^{-7}-t^{-3}+t^{9})q^{\frac94}+(4t^{-9}-t^3+t^{11})q^{\frac{11}{4}}+\cdots \\ \hline
\langle W_{\tiny \yng(1);\yng(1)} \rangle_{\mathcal{N}}^{\mathfrak{so}(4)|\mathfrak{usp}(6)}(t;q)
&
tq^{\frac14}+(2t^{-1}+2t^3)q^{\frac34}+(4t^{-3}+2t+3t^5)q^{\frac54}+(6t^{-5}+t^{-1}+4t^7)q^{\frac74}
\\ 
=\langle V_{1} \rangle_{\mathcal{D}}^{\mathfrak{so}(4)|\mathfrak{so}(7)}(t^{-1};q) 
&
+(10t^{-7}+t^{-3}+t+5t^9)q^{\frac{9}{4}}+(14t^{-9}+t^{-1}+3t^{3}+6t^{11})q^{\frac{11}{4}}+\cdots \\ \hline 
\end{array}
\end{align}
Similarly, for $N>M$, we find the following non-trivial agreement between the corresponding line defect half-indices:
\begin{align}
\begin{array}{c|c}
\textrm{line defect half-indices}&\textrm{expansions}\\ \hline
\langle W_{\tiny \yng(1);\yng(1)} \rangle_{\mathcal{N}}^{\mathfrak{so}(4)|\mathfrak{usp}(2)}(t;q)
&
tq^{\frac14}+(2t^{-1}+t^3)q^{\frac34}+(3t^{-3}+t^5)q^{\frac54}+(4t^{-5}-3t^{-1}-t^3+t^7)q^{\frac74}
\\ 
=\langle V_{1} \rangle_{\mathcal{D}}^{\mathfrak{so}(4)|\mathfrak{so}(3)}(t^{-1};q) 
&+(5t^{-7}-2t^{-3}+2t+t^9)q^{\frac94}+(6t^{-9}-t^{-5}+3t^{-1}+t^3+t^{11})q^{\frac{11}{4}}+\cdots \\ \hline 
\langle W_{\tiny \yng(1);\yng(1)} \rangle_{\mathcal{N}}^{\mathfrak{so}(6)|\mathfrak{usp}(2)}(t;q)
&
tq^{\frac14}+(t^{-1}+t^3)q^{\frac34}+(3t^{-3}+t^5)q^{\frac54}+(3t^{-5}-t^{-1}-t^3+t^7)q^{\frac74}
\\ 
=\langle V_{1} \rangle_{\mathcal{D}}^{\mathfrak{so}(6)|\mathfrak{so}(3)}(t^{-1};q) 
&
+(6t^{-7}-2t^{-3}+t+t^9)q^{\frac94}+(6t^{-9}-3t^{-5}+t^{11})q^{\frac{11}{4}}+\cdots \\ \hline 
\end{array}
\end{align}

%ADDED
For $N=1$ the preserved $\mathfrak{so}(2)$ gauge algebra of the D5-type interface is Abelian 
and the vortex line of an arbitrary magnetic charge $n$ is free from the monopole bubbling effects. 
Its line defect half-index (\ref{h_so2|so2M+1_V}) is matched 
with the half-index of the Wilson line $W_{n;(n)}$ at the dual NS5-type $\mathfrak{so}(2)|\mathfrak{usp}(2M)$ interface, 
which carries the $\mathfrak{so}(2)$ electric charge $n$ 
and transforms in the rank-$n$ symmetric representation of $\mathfrak{usp}(2M)$. 
For $n=2$ we find
\begin{align}
\begin{array}{c|c}
\textrm{line defect half-indices}&\textrm{expansions}\\ \hline
\langle W_{2;(2)} \rangle_{\mathcal{N}}^{\mathfrak{so}(2)|\mathfrak{usp}(2)}(t;q)
&
t^2q^{\frac12}+(1+t^4)q+(t^{-2}-t^2+t^6)q^{\frac32}+(t^{-4}+t^8)q^2
\\
=\langle V_{2} \rangle_{\mathcal{D}}^{\mathfrak{so}(2)|\mathfrak{so}(3)}(t^{-1};q)
&
+(t^{-6}+t^{10})q^{\frac52}+(t^{-8}-1-t^4+t^{12})q^3+\cdots
\\ \hline 
\langle W_{2;(2)} \rangle_{\mathcal{N}}^{\mathfrak{so}(2)|\mathfrak{usp}(4)}(t;q)
&
t^2q^{\frac12}+(1+t^4)q+(2t^{-2}-t^2+t^6)q^{\frac32}+(2t^{-4}+t^8)q^2
\\
=\langle V_{2} \rangle_{\mathcal{D}}^{\mathfrak{so}(2)|\mathfrak{so}(5)}(t^{-1};q)
&
+(3t^{-6}-t^{-2}+t^{10})q^{\frac52}+(3t^{-8}-t^4+t^{12})q^3+\cdots
\\ \hline 
\end{array}
\end{align}
whereas for $n=3$ we obtain
\begin{align}
\begin{array}{c|c}
\textrm{line defect half-indices}&\textrm{expansions}\\ \hline
\langle W_{3;(3)} \rangle_{\mathcal{N}}^{\mathfrak{so}(2)|\mathfrak{usp}(2)}(t;q)
&
t^3q^{\frac34}+(t+t^5)q^{\frac54}+(t^{-1}-t^3+t^7)q^{\frac74}+(t^{-3}+t^9)q^{\frac94}
\\
=\langle V_{3} \rangle_{\mathcal{D}}^{\mathfrak{so}(2)|\mathfrak{so}(3)}(t^{-1};q)
&
+(t^{-5}+t^{11})q^{\frac{11}{4}}+(t^{-7}-t-t^5+t^{13})q^{\frac{13}{4}}+\cdots
\\ \hline 
\end{array}
\end{align}
Here the $\mathfrak{usp}(2M)$ character of the rank-$n$ symmetric representation is 
the complete homogeneous symmetric polynomial $h_n$ of degree $n$ in the $2M$ variables $s_i^{(2)\pm}$. 

%%%%%%%%%%%%%%%%%%%%%%%%%%%%%%%%%%%%%%%%%%%%%%%
\subsubsection{NS5-vortex lines and D5-Wilson lines}
%%%%%%%%%%%%%%%%%%%%%%%%%%%%%%%%%%%%%%%%%%%%%%%
%Vortex
We again consider the case $N\le M$. 
Under S-duality, the vortex line labeled by the fundamental weight of the left $\mathfrak{so}(2N)$ gauge algebra of the NS5-type interface 
is expected to be mapped to the Wilson line transforming in the fundamental representation of the preserved $\mathfrak{so}(2N)$ gauge algebra of the D5-type interface. 
We have explicitly verified the agreement of the corresponding line defect half-indices. 
For the case $N=M$ we obtain
\begin{align}
\begin{array}{c|c}
\textrm{line defect half-indices}&\textrm{expansions}\\ \hline
\langle V_{1;0} \rangle_{\mathcal{N}}^{\mathfrak{so}(2)|\mathfrak{usp}(2)}(t;q)
&
t^{-2}q^{\frac12}+t^{-4}q+(2t^{-6}-t^{-2})q^{\frac32}+(1+2t^{-8}-t^{-4})q^2\\ 
=\langle W_{\tiny \yng(1)} \rangle_{\mathcal{D}}^{\mathfrak{so}(2)|\mathfrak{so}(3)}(t^{-1};q)
&+(3t^{-10}-t^{-6}-t^{-2})q^{\frac52}+(3t^{-12}-t^{-8}-t^{-4})q^3
%+(4t^{-14}-t^{-10}-2t^{-6}+t^{-2}+t^2)q^{\frac72}
+\cdots\\ \hline 
\langle V_{(1,0);0} \rangle_{\mathcal{N}}^{\mathfrak{so}(4)|\mathfrak{usp}(4)}(t;q)
&
t^{-2}q^{\frac12}+(1+2t^{-4})q+(4t^{-6}+t^{-2}+t^2)q^{\frac32}+(-1+6t^{-8}-t^{-4}+t^4)q^2
 \\ 
=\langle W_{\tiny \yng(1)} \rangle_{\mathcal{D}}^{\mathfrak{so}(4)|\mathfrak{so}(5)}(t^{-1};q)
&+(10t^{-10}-t^{-6}+t^{-2}+t^2+t^6)q^{\frac52}+\cdots\\ \hline 
\end{array}
\end{align}
For $N<M$ we find 
\begin{align}
\begin{array}{c|c}
\textrm{line defect half-indices}&\textrm{expansions}\\ \hline
\langle V_{1;0} \rangle_{\mathcal{N}}^{\mathfrak{so}(2)|\mathfrak{usp}(4)}(t;q)
&
t^{-4}q+t^{-6}q^{\frac32}+(2t^{-8}-t^{-4})q^2+(2t^{-10}-t^{-6}+t^{-2})q^{\frac52}
\\ 
=\langle W_{\tiny \yng(1)} \rangle_{\mathcal{D}}^{\mathfrak{so}(2)|\mathfrak{so}(5)}(t^{-1};q)
&
+(4t^{-12}-t^{-8}-t^{-4})q^3+(4t^{-14}-2t^{-10})q^{\frac72}+\cdots\\ \hline 
\langle V_{1;0} \rangle_{\mathcal{N}}^{\mathfrak{so}(2)|\mathfrak{usp}(6)}(t;q)
&
t^{-6}q^{\frac32}+t^{-8}q^2+(2t^{-10}-t^{-6})q^{\frac52}+(2t^{-12}-t^{-8}+t^{-4})q^3
 \\ 
=\langle W_{\tiny \yng(1)} \rangle_{\mathcal{D}}^{\mathfrak{so}(2)|\mathfrak{so}(7)}(t^{-1};q)
&+(4t^{-14}-t^{-10}-t^{-6})q^{\frac72}+(4t^{-16}-2t^{-12})q^4+\cdots\\ \hline 
\end{array}
\end{align}

%ADDED
For $N=1$ the preserved $\mathfrak{so}(2)$ gauge algebra is Abelian, 
so that the vortex line of an arbitrary magnetic charge $n$ is free from the monopole bubbling effects 
and its line defect half-index is given by (\ref{h_so2|usp2M_V}). 
It matches the half-index of the Wilson line of electric charge $n$ at the dual D5-type interface. 
For $n=2$ we find
\begin{align}
\begin{array}{c|c}
\textrm{line defect half-indices}&\textrm{expansions}\\ \hline
\langle V_{2;0} \rangle_{\mathcal{N}}^{\mathfrak{so}(2)|\mathfrak{usp}(2)}(t;q)
&
t^{-4}q+t^{-6}q^{\frac32}+(2t^{-8}-t^{-4})q^2+(2t^{-10}-t^{-6})q^{\frac52}
\\
=\langle W_{2} \rangle_{\mathcal{D}}^{\mathfrak{so}(2)|\mathfrak{so}(3)}(t^{-1};q)
&
+(3t^{-12}-t^{-8}-t^{-4})q^3+(3t^{-14}-t^{-10}-t^{-6}+t^{-2})q^{\frac72}+\cdots
\\ \hline 
\langle V_{2;0} \rangle_{\mathcal{N}}^{\mathfrak{so}(2)|\mathfrak{usp}(4)}(t;q)
&
t^{-8}q^2+t^{-10}q^{\frac52}+(2t^{-12}-t^{-8})q^3+(2t^{-14}-t^{-10})q^{\frac72}
\\ 
=\langle W_{2} \rangle_{\mathcal{D}}^{\mathfrak{so}(2)|\mathfrak{so}(5)}(t^{-1};q)
&
+(4t^{-16}-t^{-12}-t^{-8})q^4+(4t^{-18}-2t^{-14}-t^{-10}+t^{-6})q^{\frac92}+\cdots
\\ \hline 
\end{array}
\end{align}
while for $n=3$ we obtain
\begin{align}
\begin{array}{c|c}
\textrm{line defect half-indices}&\textrm{expansions}\\ \hline
\langle V_{3;0} \rangle_{\mathcal{N}}^{\mathfrak{so}(2)|\mathfrak{usp}(2)}(t;q)
&
t^{-6}q^{\frac32}+t^{-8}q^{2}+(2t^{-10}-t^{-6})q^{\frac52}+(2t^{-12}-t^{-8})q^{3}
\\ 
=\langle W_{3} \rangle_{\mathcal{D}}^{\mathfrak{so}(2)|\mathfrak{so}(3)}(t^{-1};q)
&
+(3t^{-14}-t^{-10}-t^{-6})q^{\frac72}+(3t^{-16}-t^{-12}-t^{-8})q^4+\cdots
\\ \hline 
\end{array}
\end{align}
Here $W_{n}$ stands for the Wilson line of electric charge $n$, 
whose character is $\chi^{\mathfrak{so}(2)}(s)=s^n$.

%%%%%%%%%%%%%%%%%%%%%%%%%%%%%%%%%%%%%%%%%%%%%%%
\subsection{$\mathfrak{o}(2N)^+|\mathfrak{usp}(2M)$ and $\mathfrak{o}(2N)^+|\mathfrak{so}(2M+1)$}
%%%%%%%%%%%%%%%%%%%%%%%%%%%%%%%%%%%%%%%%%%%%%%%
%%%%%%%%%%%%%%%%%%%%%%%%%%%%%%%%%%%%%%%%%%%%%%%
\subsubsection{NS5-Wilson lines and D5-vortex lines}
%%%%%%%%%%%%%%%%%%%%%%%%%%%%%%%%%%%%%%%%%%%%%%%
The same agreements of the corresponding line defect half-indices are obtained by replacing $\mathfrak{so}(2N)$ with $\mathfrak{o}(2N)$. 
For the Wilson lines at the NS5-type interface and the vortex lines at the D5-type interface we find
\begin{align}
\begin{array}{c|c}
\textrm{line defect half-indices}&\textrm{expansions}\\ \hline
\langle W_{\tiny \yng(1);\yng(1)} \rangle_{\mathcal{N}}^{\mathfrak{o}(4)^+|\mathfrak{usp}(4)}(t;q)
&
tq^{\frac14}+(t^{-1}+t^3)q^{\frac34}+(3t^{-3}+t+2t^5)q^{\frac54}+(3t^{-5}+2t^7)q^{\frac74}
\\ 
=\langle V_{(1,0)} \rangle_{\mathcal{D}}^{\mathfrak{o}(4)^+|\mathfrak{so}(5)}(t^{-1};q) 
&+(6t^{-7}+3t^9)q^{\frac94}+(6t^{-9}-2t^{-5}+t^3+3t^{11})q^{\frac{11}{4}}+\cdots \\ \hline 
\end{array}
\end{align}
for $N=M=2$ 
and 
\begin{align}
\begin{array}{c|c}
\textrm{line defect half-indices}&\textrm{expansions}\\ \hline
\langle W_{\tiny \yng(1);\yng(1)} \rangle_{\mathcal{N}}^{\mathfrak{o}(2)^+|\mathfrak{usp}(4)}(t;q)
&
tq^{\frac14}+(t^{-1}+t^3)q^{\frac34}+(2t^{-3}+t^5)q^{\frac54}+(2t^{-5}-t^{-1}-t^3+t^7)q^{\frac74}
\\ 
=\langle V_{1} \rangle_{\mathcal{D}}^{\mathfrak{o}(2)^+|\mathfrak{so}(5)}(t^{-1};q) 
&+(3t^{-7}-t^{-3}+t+t^9)q^{\frac94}+(3t^{-9}+t^{-1}+t^{11})q^{\frac{11}{4}}+\cdots \\ \hline 
\langle W_{\tiny \yng(1);\yng(1)} \rangle_{\mathcal{N}}^{\mathfrak{o}(6)^+|\mathfrak{usp}(2)}(t;q)
&
tq^{\frac14}+(t^{-1}+t^3)q^{\frac34}+(2t^{-3}+t^5)q^{\frac54}+(2t^{-5}-t^{-1}-t^3+t^7)q^{\frac74}
\\ 
=\langle V_{(1,0)} \rangle_{\mathcal{D}}^{\mathfrak{o}(6)^+|\mathfrak{so}(3)}(t^{-1};q) 
&+(4t^{-7}+t+t^9)q^{\frac94}+(4t^{-9}-t^{-5}-t^{-1}+t^{11})q^{\frac{11}{4}}+\cdots \\ \hline 
\end{array}
\end{align}
for $N\neq M$. 

%%%%%%%%%%%%%%%%%%%%%%%%%%%%%%%%%%%%%%%%%%%%%%%
\subsubsection{NS5-vortex lines and D5-Wilson lines}
%%%%%%%%%%%%%%%%%%%%%%%%%%%%%%%%%%%%%%%%%%%%%%%
%Vortex
For the vortex lines at the NS5-type interface and the Wilson lines at the D5-type interface we have confirmed the agreement for $N=M=2$ 
\begin{align}
\begin{array}{c|c}
\textrm{line defect half-indices}&\textrm{expansions}\\ \hline
\langle V_{(1,0);0} \rangle_{\mathcal{N}}^{\mathfrak{o}(4)^+|\mathfrak{usp}(4)}(t;q)
&
t^{-2}q^{\frac12}+t^{-4}q+(3t^{-6}+t^{-2}+t^2)q^{\frac32}+(-1+3t^{-8}-2t^{-4})q^2
 \\ 
=\langle W_{\tiny \yng(1)} \rangle_{\mathcal{D}}^{\mathfrak{o}(4)^+|\mathfrak{so}(5)}(t^{-1};q)
&+(7t^{-10}+t^{-6}+2t^{-2}+t^2+t^6)q^{\frac52}
%+(-1+7t^{-12}-5t^{-8}-3t^{-4})q^3
+\cdots\\ \hline 
\end{array}
\end{align}

%%%%%%%%%%%%%%%%%%%%%%%%%%%%%%%%%%%%%%%%%%%%%%%
%%%%%%%%%%%%%%%%%%%%%%%%%%%%%%%%%%%%%%%%%%%%%%%
\section{Line defect Witten indices and residual gauge groups}
\label{app_residual}
%%%%%%%%%%%%%%%%%%%%%%%%%%%%%%%%%%%%%%%%%%%%%%%
%%%%%%%%%%%%%%%%%%%%%%%%%%%%%%%%%%%%%%%%%%%%%%%
In this appendix we show that the line defect Witten indices are determined by the residual gauge groups 
at generic points of the Coulomb and Higgs branches, and we summarize our checks. 
The same mechanism governs the line defect Witten indices of 3d $\mathcal{N}\ge 4$ supersymmetric gauge theories \cite{Hayashi:2026ngm}, 
where the average of the inserted characters over the residual gauge group, 
the principal isotropy subgroup of the gauge group acting on the branch, 
likewise reproduces the Coulomb and Higgs limits of the normalized line defect indices. 
The argument applies to the NS5-type and to the D5-type interfaces alike, 
and we first fix the notation common to the two limits. 

%%%%%%%%%%%%%%%%%%%%%%%%%%%%%%%%%%%%%%%%%%%%%%%
\subsection{Setup and notation}
\label{app_setup}
%%%%%%%%%%%%%%%%%%%%%%%%%%%%%%%%%%%%%%%%%%%%%%%
We write $G$ for the gauge group whose holonomy is integrated over in the limit under consideration, 
$r$ for its rank, $W_G$ for its Weyl group and $\Phi_G$ for its root system. 
For the NS5-type interfaces both vector multiplets obey the Neumann boundary condition and $G=G_1\times G_2$, 
while for the D5-type interfaces the Nahm pole leaves $G=G_{\min}$, 
the gauge group associated with $\mathfrak{g}_{\min}$. 
We write $\mathcal{R}$ for the tensor product of the inserted representations, 
that is, $\mathcal{R}=\mathcal{R}_1\otimes \mathcal{R}_2$ as a representation of $G_1\times G_2$ for the Wilson line $\mathcal{W}_{\mathcal{R}_1;\mathcal{R}_2}$ 
at the NS5-type interfaces and $\mathcal{R}=\mathcal{R}_1\otimes \cdots \otimes \mathcal{R}_k$ as a representation of $G_{\min}$ 
for the $k$ Wilson lines at the D5-type ones, 
$\chi_{\mathcal{R}}$ for its character and $\mathcal{R}[0]$ for its zero weight subspace. 
In both limits the line defect half-index becomes a Molien-Weyl integral 
for $G$ acting on a representation $V=\bigoplus_a V_a$, 
built out of the chiral fields which survive the limit and graded by $\mathfrak{q}^{d_a}$, 
times an $s$-independent factor $c(\mathfrak{q})$, which collects the Cartan and Nahm pole contributions 
and cancels in the normalized index, so that it plays no role below. 
The four cases are listed in Table \ref{Table_app_setup}. 
In the Coulomb limit the hypermultiplets and the Nahm pole drop out and $V$ is the adjoint representation of weight $\mathfrak{q}^2$, 
whereas in the Higgs limit the Neumann vector multiplet only yields the Haar measure, 
so that $V$ consists of the interface hypermultiplets alone, 
while the vector multiplet preserved at a D5-type interface still contributes its adjoint chiral field of weight $\mathfrak{q}^2$ 
descending from the bulk hypermultiplet. 

\begin{table}[t]
\centering
\begin{tabular}{c|c|c|c}
limit & interface & $G$ & $V$ with the weights $\mathfrak{q}^{d_a}$ \\ \hline
Coulomb & NS5-type & $G_1\times G_2$ & $\mathfrak{g}_1\oplus \mathfrak{g}_2$ with $\mathfrak{q}^2$ \\
 & D5-type & $G_{\min}$ & $\mathfrak{g}_{\min}$ with $\mathfrak{q}^2$ \\ \hline
Higgs & NS5-type & $G_1\times G_2$ & bifundamental half-hypermultiplet with $\mathfrak{q}$ \\
 & D5-type & $G_{\min}$ & $\mathfrak{g}_{\min}$ with $\mathfrak{q}^2$, fundamental with $\mathfrak{q}^{d_{\mathrm{f}}}$ \\
\end{tabular}
\caption{
The gauge group $G$ and the representation $V$ of the chiral fields entering the Molien-Weyl integrals 
(\ref{app_HLmeasure}) and (\ref{app_MW}) in the two limits. 
The weight $d_{\mathrm{f}}$ of the fundamental hypermultiplet at a D5-type interface depends on the interface, 
being $1+2|N-M|$ for $\mathfrak{usp}(2N)|\mathfrak{usp}(2M)'$ in (\ref{hH_usp2N|usp2M_W}) 
and $2+2(M-N)$ or $2+2(N-M-1)$ for $\mathfrak{so}(2N)|\mathfrak{so}(2M+1)$ with $N\le M$ or $N>M$ 
in (\ref{hH_so2N|so2M+1_Wa}) and (\ref{hH_so2N|so2M+1_Wb}). 
}
\label{Table_app_setup}
\end{table}

%%%%%%%%%%%%%%%%%%%%%%%%%%%%%%%%%%%%%%%%%%%%%%%
\subsection{Coulomb limit}
%%%%%%%%%%%%%%%%%%%%%%%%%%%%%%%%%%%%%%%%%%%%%%%
In the Coulomb limit each vector multiplet contributes 
\begin{align}
\label{app_HLmeasure}
\frac{1}{|W_G|}\frac{1}{(1-\mathfrak{q}^2)^r}\oint \prod_{i=1}^r\frac{ds_i}{2\pi is_i}
\prod_{\alpha\in \Phi_G}\frac{1-s^{\alpha}}{1-\mathfrak{q}^2s^{\alpha}}, 
\end{align}
where $s^{\alpha}$ denotes the holonomy evaluated on the root $\alpha$. 
This is the Molien-Weyl integral for $G$ acting on its complexified Lie algebra, which we denote by $\mathfrak{g}$, with weight $\mathfrak{q}^2$, 
which is the first block of Table \ref{Table_app_setup}, the factor $c(\mathfrak{q})$ being trivial here. 
Hence the Coulomb limit of the Wilson line defect half-index is the Hilbert series of the module of covariants 
$(\mathbb{C}[\mathfrak{g}]\otimes \mathcal{R})^G$. 
According to Kostant \cite{MR158024}, $\mathbb{C}[\mathfrak{g}]$ is a free module over the ring of invariants, 
$\mathbb{C}[\mathfrak{g}]\cong \mathbb{C}[\mathfrak{g}]^G\otimes \mathrm{Harm}(\mathfrak{g})$, 
where $\mathrm{Harm}(\mathfrak{g})$ is the space of harmonic polynomials, 
and an irreducible representation $\mathcal{R}$ appears in $\mathrm{Harm}(\mathfrak{g})$ with multiplicity $\dim \mathcal{R}[0]$. 
Therefore the normalized index is the graded multiplicity of $\mathcal{R}$ in $\mathrm{Harm}(\mathfrak{g})$, 
which is given by (\ref{halfBPS_K}) \cite{MR737932,MR593631}, and its value at $\mathfrak{q}=1$ is $\dim \mathcal{R}[0]$. 

The same result can be seen directly from the matrix integral. 
In the normalized index the prefactor $1/(1-\mathfrak{q}^2)^r$ cancels out. 
For $|x|=1$ we have $|1-x|/|1-\mathfrak{q}^2x|\le \mathfrak{q}^{-1}$ and the ratio approaches $1$ as $\mathfrak{q}\rightarrow 1$. 
Thus the limit can be taken inside the integral and the Hall-Littlewood measure becomes unity 
\begin{align}
\label{app_WI_Coulomb}
\lim_{\mathfrak{q}\rightarrow 1}
\frac{\oint d\mu(s) \prod_{\alpha\in \Phi_G}\frac{1-s^{\alpha}}{1-\mathfrak{q}^2s^{\alpha}}\,\chi_{\mathcal{R}}(s)}
{\oint d\mu(s) \prod_{\alpha\in \Phi_G}\frac{1-s^{\alpha}}{1-\mathfrak{q}^2s^{\alpha}}}
&=\oint d\mu(s)\, \chi_{\mathcal{R}}(s)=\dim \mathcal{R}[0], 
\end{align}
where $d\mu(s)=\prod_{i=1}^r \frac{ds_i}{2\pi is_i}$. 
Physically, the W-bosons and the off-diagonal components of the adjoint scalar field pair up and decouple, 
leaving the maximal torus $T\subset G$ as the residual gauge group at a generic point of the Coulomb branch. 
For the NS5-type interfaces the two vector multiplets factorize and we obtain (\ref{WI_NS5_C_intro}), 
while for the D5-type interfaces only the preserved gauge group of rank $\mathrm{rank}(\mathfrak{g}_{\min})$ remains, 
which reproduces the conjectured formulae in section \ref{sec_BC2} and section \ref{sec_DC2}. 

%%%%%%%%%%%%%%%%%%%%%%%%%%%%%%%%%%%%%%%%%%%%%%%
\subsection{Higgs limit}
%%%%%%%%%%%%%%%%%%%%%%%%%%%%%%%%%%%%%%%%%%%%%%%
With the second block of Table \ref{Table_app_setup} 
the Higgs limit of the line defect half-index takes the form 
\begin{align}
\label{app_MW}
{\langle W_{\mathcal{R}}\rangle}^{(H)}(\mathfrak{q})
&=\frac{c(\mathfrak{q})}{|W_G|}\oint \prod_{i=1}^r\frac{ds_i}{2\pi is_i}\prod_{\alpha\in \Phi_G}(1-s^{\alpha})\,
\mathrm{PE}\Big[\sum_a \mathfrak{q}^{d_a}\chi_{V_a}(s)\Big]\chi_{\mathcal{R}}(s), 
\end{align}
where $\mathrm{PE}$ is the plethystic exponential and the Haar measure $\prod_{\alpha\in \Phi_G}(1-s^{\alpha})$ 
is all that the Neumann vector multiplet leaves behind, 
the adjoint chiral field of the D5-type interfaces being included in $V$ instead. 
This is the Molien-Weyl integral computing the Hilbert series of the module of covariants 
$M_{\mathcal{R}}=(\mathbb{C}[V]\otimes \mathcal{R})^G$, 
while ${\mathbb{II}}^{(H)}$ is that of the ring of invariants $A=\mathbb{C}[V]^G$. 
Since $M_{\mathcal{R}}$ is a finitely generated graded $A$-module 
and $A$ is a polynomial ring generated by homogeneous invariants of degrees $e_1,\cdots, e_d$ in the cases at hand, 
the Hilbert series of $M_{\mathcal{R}}$ takes the form $P_{\mathcal{R}}(\mathfrak{q})/\prod_{k=1}^{d}(1-\mathfrak{q}^{e_k})$ 
with $P_{\mathcal{R}}(\mathfrak{q})\in \mathbb{Z}[\mathfrak{q}]$. 
Therefore the normalized line defect Higgs index is the polynomial $P_{\mathcal{R}}(\mathfrak{q})$, 
and its value at $\mathfrak{q}=1$ is the rank of $M_{\mathcal{R}}$ over $A$. 

The rank can be computed at a generic point of the quotient $V/\!/G$. 
Since $G$ is reductive, taking invariants commutes with the base change, 
so that the rank is $\dim (\mathbb{C}[F]\otimes \mathcal{R})^G$, where $F$ is the generic fiber of the quotient map $V\rightarrow V/\!/G$. 
By Luna's slice theorem \cite{MR342523}, 
$F\cong G\times_{H_0} \mathcal{N}(S)$, 
where $H_0$ is the stabilizer of a closed orbit of generic type, 
$S$ is the slice representation of $H_0$ with the trivial summands removed 
and $\mathcal{N}(S)$ is its null cone. 
Then the Frobenius reciprocity leads to 
\begin{align}
\label{app_WI_rank}
\lim_{\mathfrak{q}\rightarrow 1}
\frac{{\langle W_{\mathcal{R}}\rangle}^{(H)}(\mathfrak{q})}{{\mathbb{II}}^{(H)}(\mathfrak{q})}
&=\dim \mathrm{Hom}_{H_0}\left(\mathbf{1}, \mathcal{R}\otimes \mathbb{C}[\mathcal{N}(S)]\right). 
\end{align}
The closed orbits correspond to the solutions of the D-term equations \cite{MR555701}, 
so that $H_0$ is the gauge group which remains unbroken at a generic point of the Higgs branch 
and $S$ describes the charged hypermultiplets which stay massless there. 
When the action is stable, i.e. the generic orbits are closed, $S$ is trivial 
and (\ref{app_WI_rank}) reduces to (\ref{WI_residual_intro}). 
When $\mathcal{N}(S)=S$, which is the case for the only non-trivial $S$ below, 
(\ref{app_WI_rank}) can be written as 
$\lim_{\mathfrak{q}\rightarrow 1}\int_{H_0} d\mu_{H_0}\,\chi_{\mathcal{R}}\,\mathrm{PE}[\mathfrak{q}\chi_S]$, 
which is (\ref{WI_NS5_H_intro}) once $H_0$ and $S$ are taken from Table \ref{Table_residual}. 
The three formulae (\ref{WI_so2N+1|usp2M_H1}), (\ref{WI_so2N+1|usp2M_H2}) and (\ref{WI_so2N|usp2M_H}) 
are special cases of it, 
and the D5-type interfaces, for which $H_0$ is trivial as shown below, give $\dim \mathcal{R}$. 
An immediate consequence is that any central element of $G$ acting trivially on $V$ belongs to $H_0$, 
so that the Witten index vanishes unless $\mathcal{R}$ has a trivial charge under it, 
which is consistent with the selection rules for the spinor Wilson lines. 

%%%%%%%%%%%%%%%%%%%%%%%%%%%%%%%%%%%%%%%%%%%%%%%
\subsection{Residual gauge groups}
%%%%%%%%%%%%%%%%%%%%%%%%%%%%%%%%%%%%%%%%%%%%%%%
For the NS5-type interfaces $V$ is the bifundamental half-hypermultiplet. 
We regard it as a linear map $X:\mathbb{C}^{2M}\rightarrow \mathbb{C}^{L}$ with $L=2N+1$ or $2N$, 
where $\mathbb{C}^{L}$ and $\mathbb{C}^{2M}$ are endowed with the symmetric and the symplectic forms respectively, 
and we let $X^{\dagger}:\mathbb{C}^{L}\rightarrow \mathbb{C}^{2M}$ be its adjoint with respect to these forms. 
Then $XX^{\dagger}\in \mathfrak{so}(L)$ and $X^{\dagger}X\in \mathfrak{usp}(2M)$ are the complex moment maps, 
whose non-zero eigenvalues coincide. 
The invariants are generated by $\mathrm{Tr}(X^{\dagger}X)^{2k}$ with $k=1,\cdots,\min(N,M)$, 
where for $L=2N\le 2M$ the top one is replaced by the Pfaffian of $XX^{\dagger}$. 
On a closed orbit of generic type the eigenvalues $\pm a_i$ with $i=1,\cdots,\min(N,M)$ are distinct and non-zero, and 
\begin{align}
\mathbb{C}^{L}&=\bigoplus_{i}(U_{a_i}\oplus U_{-a_i})\oplus U_0, &
\mathbb{C}^{2M}&=\bigoplus_{i}(W_{a_i}\oplus W_{-a_i})\oplus W_0, 
\end{align}
where $X$ maps $W_{\pm a_i}$ isomorphically onto $U_{\pm a_i}$ and $X|_{W_0}=0$. 
The stabilizer consists of $U(1)^{\min(N,M)}$ acting as $\mathrm{diag}(u_i,u_i^{-1})$ on $U_{a_i}\oplus U_{-a_i}$ and $W_{a_i}\oplus W_{-a_i}$ simultaneously, 
and of $SO(U_0)\times USp(W_0)$ with $\dim U_0=L-2\min(N,M)$ and $\dim W_0=2(M-\min(N,M))$. 
At most one of $U_0$ and $W_0$ is non-trivial except for $L=2N+1$ with $N<M$, 
where $\dim U_0=1$ and the slice contains $\mathrm{Hom}(W_0,U_0)\cong \mathbb{C}^{2(M-N)}$, 
the fundamental representation of $USp(2(M-N))$, whose null cone is $\mathbb{C}^{2(M-N)}$ itself. 
One can check that $\dim V-\dim G+\dim H_0=\min(N,M)+\dim S$ in all cases, 
in agreement with the number of generators of the ring of invariants. 
The results are summarized in Table \ref{Table_residual}. 
In the brane picture, $\min(N,M)$ D3-branes reconnect across the half NS5-brane and move away from the O3-plane along the $789$ directions. 
Each of them gives rise to a $U(1)$ gauge group, which is the diagonal subgroup of the gauge groups on both sides. 
The remaining D3-branes end on the half NS5-brane from one side, 
and the massless half-hypermultiplet for the $\mathfrak{so}(2N+1)|\mathfrak{usp}(2M)'$ interface with $N<M$ 
is attributed to the stuck half D3-brane of the $\widetilde{\textrm{O3}}^-$-plane. 

For the D5-type interfaces $V$ consists of the adjoint representation of the preserved gauge group $G_{\min}$ with weight $\mathfrak{q}^2$ 
and the fundamental representation. 
A generic pair $(\sigma,v)$ with $\sigma$ regular semisimple and all the weight components of $v$ non-zero has a closed orbit. Indeed, 
$\sigma$ breaks $G_{\min}$ to the maximal torus, whose orbit of $v$ is closed, 
and the stabilizer of $v$ in the maximal torus is trivial. 
Since $\dim V-\dim G_{\min}$ is equal to the dimension of the fundamental representation, 
which coincides with the number of generators of the ring of invariants, 
the slice representation is trivial and the action is stable. 
Hence $H_0$ is trivial and (\ref{WI_residual_intro}) gives $\dim \mathcal{R}$, 
in accordance with the conjectures in section \ref{sec_BC2} and section \ref{sec_DC2}. 

\begin{table}[t]
\centering
\begin{tabular}{c|c|c|c}
interface & ranks & $H_0$ & $S$ \\ \hline
$\mathfrak{so}(2N+1)|\mathfrak{usp}(2M)'$ & $N\ge M$ & $U(1)^M\times SO(2(N-M)+1)$ & -- \\
 & $N<M$ & $U(1)^N\times USp(2(M-N))$ & $\mathbb{C}^{2(M-N)}$ \\ \hline
$\mathfrak{so}(2N)|\mathfrak{usp}(2M)$ & $N\le M$ & $U(1)^N\times USp(2(M-N))$ & -- \\
 & $N>M$ & $U(1)^M\times SO(2(N-M))$ & -- \\ \hline
D5-type & & $\{1\}$ & -- \\
\end{tabular}
\caption{
The residual gauge groups $H_0$ and the massless charged half-hypermultiplets $S$ at a generic point of the Higgs branch. 
The factor $U(1)^{\min(N,M)}$ is embedded diagonally into both gauge groups as $s_i^{(1)}=s_i^{(2)}=u_i$. 
In the Coulomb limit $H_0$ is the maximal torus of $G_1\times G_2$ for the NS5-type interfaces and that of $G_{\min}$ for the D5-type interfaces. 
}
\label{Table_residual}
\end{table}

%%%%%%%%%%%%%%%%%%%%%%%%%%%%%%%%%%%%%%%%%%%%%%%
\subsection{Checks}
%%%%%%%%%%%%%%%%%%%%%%%%%%%%%%%%%%%%%%%%%%%%%%%
We have evaluated the Higgs limit matrix integrals (\ref{hH_so2N+1|usp2M_W}) and (\ref{hH_so2N|usp2M_W}) exactly as $\mathfrak{q}$-series, 
by expanding the plethystic exponential and extracting the multiplicities by means of the Weyl character formula, 
and compared the normalized indices at $\mathfrak{q}=1$ with (\ref{app_WI_rank}). 
We have examined all pairs of irreducible representations $(\lambda_1;\lambda_2)$ with $|\lambda_1|+|\lambda_2|\le 3$, 
and up to $4$ or $5$ for lower ranks, 
for the NS5-type $\mathfrak{so}(2N+1)|\mathfrak{usp}(2M)'$ interfaces with $(N,M)=(0,3)$, $(1,1)$, $(1,2)$, $(1,3)$, $(2,1)$, $(2,2)$ 
and the NS5-type $\mathfrak{so}(2N)|\mathfrak{usp}(2M)$ interfaces with $(N,M)=(1,1)$, $(1,2)$, $(1,3)$, $(2,1)$, $(2,2)$, $(3,1)$. 
In all the cases we have examined, the normalized indices are polynomials in $\mathfrak{q}$ and their values at $\mathfrak{q}=1$ agree with (\ref{app_WI_rank}). 
Table \ref{Table_check} shows several examples for the $\mathfrak{so}(3)|\mathfrak{usp}(4)'$ interface, 
where the residual half-hypermultiplet is essential. 
We have also checked that the Coulomb limit (\ref{app_HLmeasure}) yields $K_{\lambda,0}(\mathfrak{q}^2)$ with $K_{\lambda,0}(1)=\dim \mathcal{R}[0]$ 
for low-rank examples, 
and that the Higgs limits of the D5-type interfaces give $\dim \mathcal{R}$ for non-minuscule representations 
as shown in section \ref{sec_BC2} and section \ref{sec_DC2}. 

\begin{table}[t]
\centering
\begin{tabular}{c|c|c|c}
$(\mathcal{R}_1;\mathcal{R}_2)$ & normalized Higgs index & Witten index & compact $H_0$ only \\ \hline
$(0;\tiny \yng(1))$ & $\mathfrak{q}^3$ & $1$ & $0$ \\
$(0;\tiny \yng(2))$ & $\mathfrak{q}^2+\mathfrak{q}^6$ & $2$ & $1$ \\
$(0;\tiny \yng(3,1))$ & $\mathfrak{q}^6+\mathfrak{q}^8+\mathfrak{q}^{10}$ & $3$ & $1$ \\
$(\tiny \yng(1);\yng(1))$ & $\mathfrak{q}+\mathfrak{q}^3+\mathfrak{q}^5$ & $3$ & $2$ \\
$(\tiny \yng(1);\yng(2,1))$ & $\mathfrak{q}^3+2\mathfrak{q}^5+2\mathfrak{q}^7+\mathfrak{q}^9$ & $6$ & $2$ \\
\end{tabular}
\caption{
The normalized Wilson line defect Higgs half-indices of the NS5-type $\mathfrak{so}(3)|\mathfrak{usp}(4)'$ interface. 
The Witten indices agree with (\ref{WI_so2N+1|usp2M_H2}), 
while the integral over the compact residual gauge group $U(1)\times USp(2)$ alone, shown in the last column, fails to reproduce them. 
}
\label{Table_check}
\end{table}

%%%%%%%%%%%%%%%%%%%%%%%%%%%%%%%%%%%%%%%%%%%%%%%
\subsection{Range of validity}
%%%%%%%%%%%%%%%%%%%%%%%%%%%%%%%%%%%%%%%%%%%%%%%
Two conditions are implicit in the argument above. 
The first is that the ring of invariants has the expected dimension, 
so that ${\mathbb{II}}^{(H)}\neq 0$ and the rank is defined relative to it. 
This fails for the NS5-type $\mathfrak{o}(2N)^-|\mathfrak{usp}(2M)$ interface with $N>M$, 
where the twisted sector cancels the untwisted one and the Higgs half-index vanishes identically, 
with or without line defect insertions, 
since a $\mathbb{Z}_2$-odd invariant of $SO(2N)$ requires the epsilon tensor contracted with $2N$ vectors 
while only $2M<2N$ of them are available. 
The normalized index is then $0/0$ and there is nothing to compare (\ref{app_WI_rank}) with. 
For $N\le M$ instead the $\mathbb{Z}_2$ acts trivially on $H_0$, 
the covariants odd under it are generated by the Pfaffian of $XX^{\dagger}$, 
and we find 
\begin{align}
{\mathbb{II}^{\mathfrak{o}(2N)^+|\mathfrak{usp}(2M)}_{\mathcal{N}}}^{(H)}(\mathfrak{q})
&=\prod_{n=1}^{N}\frac{1}{1-\mathfrak{q}^{4n}}, &
{\mathbb{II}^{\mathfrak{o}(2N)^-|\mathfrak{usp}(2M)}_{\mathcal{N}}}^{(H)}(\mathfrak{q})
&=\mathfrak{q}^{2N}\prod_{n=1}^{N}\frac{1}{1-\mathfrak{q}^{4n}}, 
\end{align}
in agreement with the Coulomb limits (\ref{hC_o2N+|so2M+1}) and (\ref{hC_o2N-|so2M+1}) 
at the S-dual D5-type interface. 

The second is that the Wilson line in question is a genuine line operator. 
For $N<M$ the residual $USp(2(M-N))$ gauge group carries a single fundamental half-hypermultiplet, 
that is, an odd number of them, so that its discrete theta angle matters 
and a Wilson line charged under its center need not be genuine by itself. 
The matrix integral still produces the values quoted above, 
but their reading as a count of the junction states requires care in that case. 

Similar limitations are encountered in 3d $\mathcal{N}\ge 4$ supersymmetric gauge theories \cite{Hayashi:2026ngm}. 
There the Coulomb limit formula presupposes that $H_0$ is the maximal torus 
and the Higgs limit formula that $H_0$ is trivial, 
and both are modified whenever a larger subgroup survives, 
as happens in the presence of Chern-Simons terms, when $H_0$ acquires a disconnected piece 
and in the rank deformed ABJ theories. 
The average over $H_0$ in (\ref{WI_residual_intro}) remains the appropriate statement in those cases as well. 

\bibliographystyle{utphys}
\bibliography{ref}

\end{document}